\documentclass[10pt,aps,prc,amsmath,amssymb,twocolumn,floatfix,nofootinbib,showpacs,superscriptaddress]{revtex4-1}

\usepackage[utf8]{inputenc}

\usepackage{amssymb,amsthm,amsmath,amstext,amsbsy}
\usepackage{amsfonts}
\usepackage{verbatim}
\usepackage[labelfont={small},subrefformat=parens,caption=false]{subfig}

\usepackage{graphicx}
\usepackage{bbm}
\usepackage{url}
\usepackage{nicefrac}
\usepackage{hyperref}

\graphicspath{{./figures_updated/}}

\newcommand{\beq}{\begin{equation}}
\newcommand{\eeq}{\end{equation}}

\newcommand{\bseq}{\begin{subequations}}
\newcommand{\eseq}{\end{subequations}}

\newcommand{\ts}{\textstyle}

\newcommand{\bi}{\begin{itemize}}
\newcommand{\ei}{\end{itemize}}
\newcommand{\I}{\item}
\newcommand{\be}{\begin{enumerate}}
\newcommand{\ee}{\end{enumerate}}
\newcommand{\bc}{\begin{center}}
\newcommand{\ec}{\end{center}}

\newcommand{\pr}{\text{pr}}
\newcommand{\avec}{\mathbf{a}}

\newcommand{\amarg}{\mathbf{a}_{\text{marg}}}

\newcommand{\gth}{g_{\text{th}}}

\newcommand{\ktilde}{\tilde{k}}

\newcommand{\bnorm}[0]{\begin{large}}
\newcommand{\enorm}[0]{\end{large}}
\newcommand{\chiEFT}{$\chi$EFT}
\newcommand{\chiPT}{$\chi$PT}

\newcommand{\Order}{\mathcal{O}}

\newcommand{\expm}{\text{exp}}
\newcommand{\kmax}{k_{\text{max}}}
\newcommand{\kord}{k}
\newcommand{\xmax}{x_{\text{max}}}
\newcommand{\abar}{\bar{a}}
\newcommand{\abarmin}{\abar_{<}}
\newcommand{\abarmax}{\abar_{>}}

\newcommand{\pdf}{\pr}  

\newcommand{\Q}{x}  

\newcommand{\dataset}[3]{\ensuremath{{\textrm #1}#2_{(#3\%)}}}
\newcommand{\datasetmn}[4]{\ensuremath{{\textrm #1}{\textrm #2}#3_{(#4\%)}}}
\newcommand{\datasetnonlin}[4]{\ensuremath{{\textrm #1}[\textrm{#2}]#3_{(#4\%)}}}

\newcommand{\Aprime}{A$'$}
\newcommand{\Cprime}{C$'$}
\newcommand{\emcee}{emcee}
\newcommand{\Dtilde}{\ensuremath{\widetilde{\mathrm{D}}}}
\newcommand{\cvec}{\mathbf{c}}
\newcommand{\cmarg}{\cvec_{\mathrm{marg}}}

\newcommand{\pord}{\ensuremath{p}}
\newcommand{\pmax}{\ensuremath{p_{\text{max}}}}
\newcommand{\abarzero}{\abar_{\rm fix}}  
\newcommand{\mmax}{\ensuremath{m_{\mathrm{max}}}}
\newcommand{\ktildevec}{\mathbf{\ktilde}}

\begin{document}

\title{Bayesian parameter estimation for effective field theories}

\author{S. Wesolowski}
\email{wesolowski.14@osu.edu}
\affiliation{Department of Physics, The Ohio State University, Columbus, OH 43210, USA}

\author{N. Klco}
\email{nk405210@ohio.edu}
\affiliation{Institute of Nuclear and Particle Physics and Department of Physics and
Astronomy, Ohio University, Athens, OH 45701, USA}
\affiliation{Department of Physics, University of Washington, Seattle, WA 98195, USA}

\author{R.J. Furnstahl}
\email{furnstahl.1@osu.edu}
\affiliation{Department of Physics, The Ohio State University, Columbus, OH 43210, USA}

\author{D.R. Phillips}
\email{phillips@phy.ohiou.edu}
\author{A. Thapaliya}
\email{at311509@ohio.edu}
\affiliation{Institute of Nuclear and Particle Physics and Department of Physics and
Astronomy, Ohio University, Athens, OH 45701, USA}

\date{\today}

\begin{abstract}

We present procedures based on Bayesian statistics for estimating, from data,
the parameters of effective field theories
(EFTs).
The extraction of low-energy constants (LECs) is guided by
theoretical expectations
in a quantifiable way through the specification of Bayesian priors.
A prior for natural-sized LECs
reduces the possibility of overfitting,
and leads to a consistent accounting of different sources of uncertainty.
A set of diagnostic tools are developed that analyze the fit and
ensure that the priors do not bias the EFT parameter estimation.
The procedures are illustrated using representative model problems, including 
the extraction of LECs for the nucleon mass expansion
in SU(2) chiral perturbation theory from synthetic lattice data.
\end{abstract}

\smallskip
  \pacs{02.50.-r, 11.10.Ef, 21.45.-v, 21.60.-n}

\maketitle

\section{Introduction} \label{sec:intro}

Effective field theories (EFTs) describe physics in the presence of
a separation of scales.  They exploit known symmetries to
construct general interactions at the lower momentum scale that do not
depend on unresolved details of the
physics from the higher scale~\cite{Kaplan:1995uv,Phillips:2002da,Burgess:2007pt,Epelbaum:2010nr}.
A ratio (or, in the case of multiple scales, ratios) between the separated scales is formed, enabling a power-counting
scheme that organizes the (renormalized) contributions of different operators to
observables in an order-by-order expansion in the ratio(s)~\cite{Epelbaum:2010nr}.
The effects of unresolved physics are parametrized
by the coefficients of these operators, called low-energy constants (LECs).
Estimating these LECs using experimental data or numerical simulations of
the underlying theory is essential if the EFT is to be used to make
quantitative predictions.
But, since the EFT expansion is written in terms of a ratio of disparate scales, 
even in the absence of such data, we expect the LECs in the EFT should be of ``natural" size,
that is, of order unity when expressed using the appropriate scales.

In this paper we lay out a framework for EFT practitioners to obtain estimates
of the LECs in their EFT while incorporating \emph{all} the information at
their disposal. A Bayesian formulation of the problem is ideal for this, as it facilitates folding
theoretical expectations---such as that the LECs should be natural---into parameter estimation.
Using only data (together with their statistical errors) to constrain the LECs
can lead to distortions in the fits; this is already recognized by EFT practitioners
who adapt their fitting procedures to include additional information (e.g., in
fitting NN scattering data, see Sec.~\ref{subsec:least-squares}). The use of priors is thus very well motivated
in the EFT context. Even though it is still controversial in other contexts, such Bayesian reasoning has
recently found several significant applications in EFTs for nuclear physics~\cite{DescotesGenon:2003cg,Ledwig:2014cla,Zhang:2015ajn,Furnstahl:2015rha,Perez:2015ufa,Griesshammer:2015ahu}. But priors can bias the parameter extraction if not used carefully.
Validating the prior is a key element of robust parameter estimation.

The naturalness priors and Bayesian framework also allow us to examine the impact
of higher-order terms in
the EFT. Those terms can be incorporated into the fit, either through a known
functional form or via the assumed EFT expansion for observables
(see Ref.~\cite{Furnstahl:2015rha}), and then accounted for using the rule of
 marginalization.
This allows the parameter estimation to account for all three sources of
uncertainty in the extraction of LECs~\cite{Furnstahl:2014xsa}:
\be
  \I Data uncertainties from experiment or numerical simulations.
  \I Systematic errors from truncating the EFT.
  \I Errors from methods used to compute observables.
\ee
These uncertainties are often intertwined and must be dealt with consistently, and each
is nontrivial to estimate. For example, even when there is a large amount of high-quality
data available,
as in the case of NN scattering, that data must be carefully analyzed for consistency,
as in the self-consistent database of Ref.~\cite{Perez:2013mwa}.
But, while errors of the first type are often included in LEC extractions---and errors of the 
third type are sometimes also accounted for---it is far rarer for such extractions to 
consider the systematic errors from truncation of the EFT. Yet neglecting those
means  we fail to incorporate all the information we have about the 
EFT when determining the best values for LECs.

Using information on LEC naturalness also ameliorates
the well-known phenomena of ``overfitting'' and ``underfitting''
within the EFT:
\bi
  \I[] \textbf{Overfitting.} The parameters get fine-tuned to the data.
  A naturalness prior reduces the risk of this, since it
  restricts degenerate directions in the EFT parameter space.
  Together with marginalization over higher orders, the prior additionally
  ensures that the LECs are not tuned to reproduce the data better than expected from
  the theoretical truncation error.

  \I[] \textbf{Underfitting.} The model provides an inadequate representation
  of at least part of the data being fit (e.g., not high enough order). 
  Marginalization over higher EFT orders
  ensures that the combined data and truncation uncertainties---which together
  determine the total uncertainty of the LECs---grows as the EFT parameter increases.
  Without this feature, blind application of least-squares fitting can lead to low-order
  LECs being overly influenced by data in an energy domain where the corresponding 
  EFT expression has large errors.
\ei
When conducting EFT parameter estimation using data
over a large range of the expansion parameter, it is possible to 
experience tendencies toward both overfitting and underfitting.
In the past these biases have usually been decoupled in 
EFT fits by limiting the parameter range over which data are fitted. But,
ideally, all available data would be used to
inform results. A well-constructed prior
protects against these biases while simultaneously relieving
the scientist of the responsibility of justifying an
arbitrarily chosen cutoff of data. We will show that our
Bayesian EFT parameter estimation converges
simply and intuitively, and that it does so even for data sets that 
include regions where the EFT converges poorly, or not at all.

This is achieved by computing so-called posterior probability distributions (pdfs)
for LECs with all assumptions, such as naturalness, made
explicit. We develop a system of diagnostics that
ensure consistent, reproducible posteriors, with all
assumptions manifested through the use of Bayesian priors.
The diagnostics also allow tests of the sensitivity to particular prior prescriptions.
Such a framework avoids both overfitting and underfitting
in a statistically well-defined way, thereby maximizing the information from data that
enters the LEC determination. 

The probability distributions yield estimates of the parameters of
the effective theory that are derived from both the data considered and
underlying knowledge of the theory itself.
Propagating the resulting uncertainties to the EFT's predictions
displays the impact that uncertainties in the data used to fix the LECs
have on those predictions.
By combining those results with our prescription
for truncation errors from Ref.~\cite{Furnstahl:2015rha} we can fully account
for the uncertainties in the EFT (we do not address here the uncertainties
from calculational methods, although their inclusion is well defined),
thereby providing reliable guidance as to the power
of new experimental or lattice data to improve theoretical predictions.

Following Refs.~\cite{Schindler:2008fh, Furnstahl:2014xsa}, where a model problem
was introduced to simulate the systematic behavior of an EFT expansion, we will
explore representative examples to demonstrate issues that occur in EFT parameter
estimation. To simulate the extraction of EFT parameters from data, we contrive
functions of a single variable, say, $g(x)$. This function is designed to have a
Taylor series whose coefficients are order unity within a radius of convergence that
is equal, or close, to 1:
\begin{equation}
g(x)=a_0 + a_1 x + a_2 x^2 + \ldots; \quad |x| < 1.
\end{equation}
The $a_i$s play the role of LECs in this example. They are extracted from synthetic data
to which Gaussian noise is added
to simulate experimental errors.
In this case, the data are generated by choosing a set of $N_d$---typically
evenly spaced---$x$ points up to a maximum value of $x=\xmax$;
given the $j^{\rm th}$ point $x_j$, the data point $d(x_j) \equiv d_j$ and error
$\sigma_j$ are
\beq
   d_j = g(x_j) (1 + c\eta_j)
    \quad \Longrightarrow \quad \sigma_j = c \, d_j
   \;,
   \label{eq:pseudodata}
\eeq
where $\eta_j$ is normally distributed with mean 0 and standard deviation 1. $c$ is a specified relative error.

The goal of the model
problems is then to reliably estimate as many EFT parameters as possible
given certain synthetic data.
We do this by choosing a truncation order $\kord$ for the ``theory" 
(which is really just a polynomial) and
determining the posterior pdfs of the coefficients $a_i$  in
\beq
  g_{\textrm{th}}(x) \equiv \sum_{i=0}^{\kord} a_i x^i
    \label{eq:model-th-expansion}
\eeq
from the data.%
\footnote{The extension to an EFT with multiple coefficients at a given order 
and/or more than one expansion parameter because of multiple low-energy scales follows
by generalizing the vector of coefficients 
(e.g., see Sec.~\ref{sec:nucleonmass}).}
To ensure that the extraction is robust and reliable,
one includes by a marginalization integral the contributions 
from a (hopefully finite) number of influential higher-order coefficients
$\{a_{\kord+1}, \ldots, a_{\kmax} \}$ up to some order $\kmax$. 
This accounts for the uncertainties in the posterior 
that arise as an artifact of finite-order truncation of the EFT, as presented in 
Ref.~\cite{Furnstahl:2015rha}. 
For the simple models here, this marginalization is the same as doing parameter
estimation up to truncation order $\kmax$.

These model problems allow us to control the ``natural'' size
of the coefficients being determined, the distribution and range of the
``experimental" data, and its precision.
We can explore pertinent EFT issues such as the presence of unnatural coefficients
or improvements gained by adding new data, while avoiding the computational
cost of computing observables from EFTs order-by-order.

However, such models do not include certain generic aspects of
actual EFTs. An obvious extension of Eq.~(\ref{eq:model-th-expansion}) is
to a problem with two light scales, as occurs in chiral perturbation theory
($\chi$PT), where the pion mass and the (assumed small) external momentum 
can be used to form two independent expansion parameters.
Here we deal only with series in one variable; the extension to additional variables
should be straightforward. 

Another aspect of EFTs which is not reflected in Eq.~(\ref{eq:model-th-expansion})
is that the coefficient of the term of $\mathcal{O}(x^i)$ will typically
be a non-analytic function of $x$, ${\cal A}_i(x)$, of which the LEC $a_i$ forms only the constant 
part. For the parameter-estimation problems being discussed here it is crucial that the 
non-analytic parts of ${\cal A}_i$---i.e. all but the LEC that appears in it---are predicted by the EFT.  (Indeed, at 
certain orders in the EFT expansion ${\cal A}_i$ contains no LEC and is completely predicted.) Thus, 
up to the order to which the EFT series has been computed, these non-analytic pieces  can be removed from 
the parameter-estimation problem by simply subtracting them from the data. Of course, such a subtraction 
is only feasible for orders where the ${\cal A}_i$ are known. But, at all orders in the EFT, 
the ${\cal A}_i$ should be $\mathcal{O}(1)$ for $x$ in the domain where the EFT applies. This allows us to deal with higher-order 
terms in the EFT series by imposing a naturalness prior on
unknown higher-order coefficients---even in this general case where the coefficient is, in fact, a function---thereby
facilitating marginalization over the higher-order terms.

Such non-analytic terms occur in one of our examples: 
the estimation of LECs for the two-flavor \chiPT~expansion of nucleon mass up to sixth chiral order.
In this application, which was previously considered in Ref.~\cite{Schindler:2008fh}, the LECs are linearly related to the data used for the
extractions, just like our generic polynomial model.
This example is difficult because it involves extracting a relatively large
number of LECs from lattice data in a region where contributions
at different orders are not easily distinguished.

In anticipation of later applications to chiral EFT for nuclei, we also examine 
the impact on our 
procedures of situations where the quantity for which we have data is a \emph{nonlinear} function 
of the LECs $a_i$. 
In addition, we will need to account for higher-order contributions to the 
posterior when they have an unknown
form or are impractical to compute. 
Our strategy is to build upon 
Ref.~\cite{Furnstahl:2015rha} and use the expected
naturalness of the expansion coefficients $\{c_{\kord+1},  \ldots, c_{\kmax} \}$
for the fit observables rather than the LECs themselves 
above the order $\kord$.
This alternative approach is discussed in Sec.~\ref{subsec:setup}.

In Sec.~\ref{sec:stat-methods}, we first briefly review alternative procedures used
to supplement standard least-squares methods when fitting LECs for chiral EFT, 
and then
outline our Bayesian framework for EFT
parameter estimation.
Sec.~\ref{sec:procedures} gives an overview of the procedures and
diagnostics for LEC estimation, illustrated with a particular model problem.
A flowchart for using the diagnostic framework is presented in
Sec.~\ref{sec:model-problems} and the major steps are illustrated with
examples using a different model.
Additional case studies, including a nonlinear model and
the \chiPT\ mass expansion, are briefly explored in Sec.~\ref{sec:case_studies}.
In Sec.~\ref{sec:conclusion} we summarize and
present plans for future analyses with the diagnostic framework
by the BUQEYE%
\footnote{Bayesian Uncertainty Quantification: Errors for Your EFT.}
collaboration.


\section{Statistical Methods} \label{sec:stat-methods}

\subsection{Least-squares minimization} \label{subsec:least-squares}

The underlying basis for extracting LECs is generally minimization
of a least-squares objective function for a set of observables
(e.g., see~\cite{Ekstrom:2013kea}):
\beq
    \chi^2 = \sum_{i=1}^{N_d} \biggl(
    \frac{d_i - \gth(x_i)}{\sigma_{i}} \biggr)^2 \;,
    \label{eq:objective-func}
\eeq
where  $x_i$ is the value of the EFT expansion
parameter at which the observable was measured, $N_d$ is the total number of data,
$d_i$ is the experimental measurement of the observable,
$\gth(x_i)$ is the corresponding EFT prediction at $x_i$ given
a set of LECs (which we denote collectively as the vector $\avec$), 
and $\sigma_{i}$ is the uncertainty associated with the observable.
A conventional least-squares minimization would take $\sigma_{i}$ to be the
experimental error, but, as already noted, in practice
the procedure is modified to take account of additional elements, such as
theoretical errors. Recent fits made of LECs for nucleon-nucleon scattering in chiral EFT
in Ref.~\cite{Carlsson:2015vda} and Ref.~\cite{Epelbaum:2014efa},
which apply different regularization schemes, provide
representative examples of the modifications.

Carlsson et al.\ add the different sources of uncertainty in quadrature for
each observable~\cite{Carlsson:2015vda},
\beq
  \sigma^2 = \sigma^2_{\textrm{exp}} + \sigma^2_{\textrm{numerical}}
      + \sigma^2_{\textrm{method}} + \sigma^2_{\textrm{model}} \;,
      \label{eq:sigma_quadrature}
\eeq
where $\sigma_{\textrm{exp}}$ is the experimental uncertainty,
 $\sigma_{\textrm{numerical}}$ and $\sigma_{\textrm{method}}$
are computational uncertainties (both included as the third type of
uncertainty listed in the Introduction),
and $\sigma_{\textrm{model}}$ is the systematic EFT truncation error.
The latter is estimated for scattering at momentum $p$ as a constant
times the expected expansion parameter raised to a power given by the
first omitted order ($k+1$), namely $(p/\Lambda_b)^{k+1}$, with $\Lambda_b$ the breakdown
scale of chiral EFT.  The
constant is determined self-consistently by requiring that $\chi^2$
including the full $\sigma$ of Eq.~\eqref{eq:sigma_quadrature}
equals the number of degrees of freedom,
as advocated in Ref.~\cite{Dobaczewski:2014jga}.

Epelbaum et al.~\cite{Epelbaum:2014efa} use $\sigma = \sigma_{\textrm{exp}}$
but  address the danger of underfitting
by limiting the region of fit data according to estimates of the domain of
validity at each EFT order.
They also augment the $\chi^2$ function with penalty terms to enforce an expected
relation between two LECs (Wigner symmetry) and to limit the allowed range of
the D-state probability of the deuteron.
These serve to restrict the parameter space for the least-squares
fit, which limits overfitting.
As part of the uncertainty quantification (but not the fit itself), 
error estimates are made based on expectations for truncation errors.
In Ref.~\cite{Furnstahl:2015rha}, these estimates were shown to be justified 
semi-quantitatively by Bayesian arguments and a naturalness
prescription.

The Bayesian approach we advocate for parameter estimation
has a different structure to the procedures of Ref.~\cite{Carlsson:2015vda,Epelbaum:2014efa}; it is an interesting and relevant
question whether those procedures can also be derived or motivated by a Bayesian framework
under prescribed conditions.  For example, in Ref.~\cite{Furnstahl:2014xsa},
it was noted (drawing on Stump et al.~\cite{Stump:2001gu}) that when Gaussian priors are used
one can account for missing higher-order contributions by a 
modified, augmented $\chi^2$.
However, this does not take the form of
adding experimental and theory errors in quadrature as in
Eq.~\eqref{eq:sigma_quadrature}.
We do not intend to pursue these connections further here.
Rather we have sketched these examples to highlight that the use of
\textit{a priori} supplementary information is standard practice in
state-of-the-art LEC optimization.

\subsection{Bayesian parameter estimation} \label{subsec:Bayesian_parameter_est}

We generically use the notation $\pr(x|y)$ to denote the pdf of
$x$ given that $y$ is true. The Bayesian framework for parameter estimation
determines the \emph{posterior} $\pr(\avec|D,I)$,
which is the joint probability distribution for the full set of
LECs $\avec$ given the data $D$ (including their errors) and any other information $I$.
$I$ includes the form of the EFT expansion and the  order at which it is truncated, together with 
additional knowledge such as the expected naturalness of the LECs.
We emphasize that the posterior distribution contains much greater content than
just the ``best-fit'' values for the coefficients (which might, but need not necessarily,
be identified as the vector $\avec$ that maximizes the posterior).
When a Gaussian approximation to the posterior is appropriate, the full content
may be effectively summarized by values for the maximum and a covariance matrix.
But, for determining how the uncertainties propagate,
it is generally necessary to start with the full posterior.

To make progress, we express the desired pdf $\pr(\avec|D,I)$ in terms
of other pdfs that we can calculate or use to implement prior information, e.g., about
naturalness.  This is done using the basic rules of probabilistic inference,
the sum and product rules, and their immediate consequences such as Bayes'
theorem and marginalization~\cite{Gregory:2005,Sivia:2006}.
It was demonstrated by Cox that this is the unique set of rules
that follow from basic postulates of
consistency~\cite{Cox:1946,Cox:1961,Jaynes:2003}.

The product rule dictates two ways to decompose a joint probability:
\beq
  \pdf(x,y|I) = \pdf(x|y,I)\,\pdf(y|I)
              = \pdf(y|x,I)\,\pdf(x|I)
              \;.
       \label{eq:product}
\eeq
A simple rearrangement of the two decompositions yields
Bayes' theorem, which when applied to the posterior for parameter
estimation yields
\beq
	\pr(\avec|D,I)  = \frac{\pr(D|\avec,I) \pr(\avec|I)}{\pr(D|I)} \;.
	\label{eq:bayes-thm}
\eeq
The pdf in the denominator, $\pr(D|I)$, which is called the evidence for $D$,
has no dependence on $\avec$ and is therefore 
determined here by normalization. (In this context the normalization follows from the sum rule for
probabilities, but the unnormalized magnitude of the resulting integral is useful
in Bayesian model selection, see Sec.~\ref{subsec:bayes_factor} below.)
Rewriting Eq.~\eqref{eq:bayes-thm}
as a proportionality, we have
\beq
	\pr(\avec|D,I) \propto \pr(D|\avec,I) \pr(\avec|I) \;.
	\label{eq:bayes_proportionality}
\eeq
The posterior is therefore determined by the product of two probabilities,
the \emph{likelihood} $\pr(D|\avec,I)$ and the \emph{prior} $\pr(\avec|I)$.
The Bayesian procedure for estimating LECs $\avec$ proceeds by identifying
a likelihood based on the available experimental data, assigning an
appropriate prior, and finally analyzing the properties of the posterior for
$\avec$, including the dependence on the choice of prior.
From the computed posterior, we can propagate the LEC uncertainties to the
predictions of the EFT. We describe the practical implementation of this
procedure in Sec.~\ref{sec:procedures}. We pay particular attention to problems 
that can occur if the prior is not well 
chosen and develop a set of diagnostics by which such problems may be identified.

The unadorned least-squares minimization
procedure with $\sigma_j = \sigma_{j,\expm}$ follows from particular choices
for the likelihood and prior.
The pdf for $N_d$ independent data $D = \{d_i\}$ given the LECs $\avec$ and relevant
information $I$ including the form and truncation order of the EFT is given by
\beq
	\pr(D| \avec, I) = \prod_{j=1}^{N_d}
	\left( \frac{1}{\sqrt{2 \pi}\sigma_{j,\expm}} \right)
	e^{-\chi^2/2} \;,
	\label{eq:least-squares-likelihood}
\eeq
with the $\chi^2$ objective function defined in Eq.~\eqref{eq:objective-func}.%
\footnote{The principle of maximum entropy (MaxEnt)~\cite{Jaynes:1957}
can be used to derive this likelihood in the Bayesian framework~\cite{Gregory:2005,Sivia:2006}.
In general we need a covariance matrix to account for correlated data.}
Note that the theory itself, including the LECs and all other
pertinent information, is on the right-hand side of the conditional here,
i.e., it is given information.
If we identify the ``best-fit'' parameters with the maximum of the posterior
from Eq.~\eqref{eq:bayes_proportionality} (maximum likelihood),
then the Bayesian procedure is equivalent to least-squares minimization
\emph{if} we choose the prior $\pr(\avec|I)$ to be uniform (i.e., independent
of $\avec$) in regions where the likelihood is nonzero (or non-negligible).
This equivalence does, though, blur the significant interpretative difference between
	$\pr(D| \avec, I)$  and $\pr(\avec|D, I)$. This difference in frequentist and Bayesian perspectives will be 
	discussed in Sec.~\ref{subsec:interpretation}

One might think that a uniform prior is a natural, unbiased choice, but there are
subtleties (e.g., in what variable it should be uniform and whether uniformity accounts for
all available constraints)~\cite{Sivia:2006}. Furthermore, in the context of EFTs
we know that the LECs should not have any unrestricted value, but instead can be
expected to be distributed in some ``natural'' range around zero. 
Thus a uniform prior encodes information that is \emph{not} expected, namely that
parameters in any interval are equally likely.
But this prior is the baseline for many analyses, and so we use it throughout
as a benchmark.
We will use the term ``least-squares fit'' interchangeably with 
maximizing the posterior of $\avec$ using a uniform prior.
However, we stress again that in practice, especially in the EFT context,
least-squares is often augmented with other 
criteria, such as those reviewed in Sec.~\ref{subsec:least-squares}.

\subsection{Marginalization}  \label{subsec:marginalization}

The sum rule of probabilities says
a pdf should be normalized, that is,
\beq
	\int dx \, \pr(x|I) = 1\;,
\eeq
where the integration is over the appropriate range
of $x$. The sum rule implies that for a joint pdf $\pr(x,y|I)$ we find
\begin{align}
	\pr(x|I) & = \int dy \, \pr(x,y|I) \nonumber \\
	         & = \int dy \, \pr(x|y,I)\, \pr(y|I)
	         \;,
\end{align}
which is marginalization, with the product rule applied to obtain the second
form. 

This rule allows us to ``integrate out'' parameters from a joint distribution to obtain a pdf
which depends only on the parameter(s) of interest. One can also ``integrate in'' parameters,
expressing a pdf for some parameters in terms of integrals over joint distributions
with other parameters. This is especially useful when the form of a pdf of interest
is not known, but joint distributions with other parameters are known.

\begin{figure*}[tbh]
  \subfloat{%
    \label{fig:triangle_D1_uniform}%
  \includegraphics[width=0.48\textwidth]{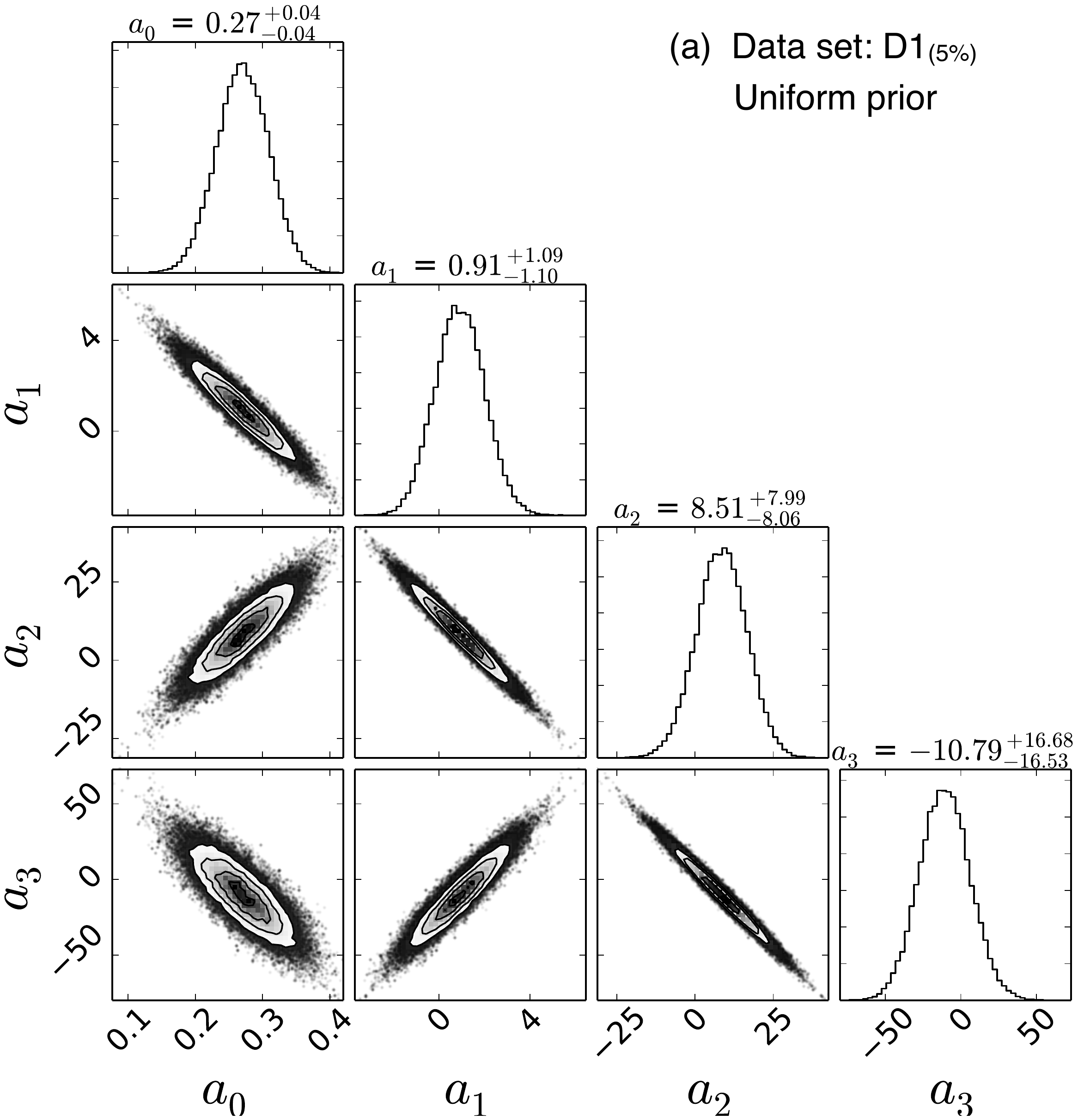}%
  }
  \hspace*{0.02\textwidth}  
  \subfloat{%
    \label{fig:triangle_D1_Gaussian}%
  \includegraphics[width=0.48\textwidth]{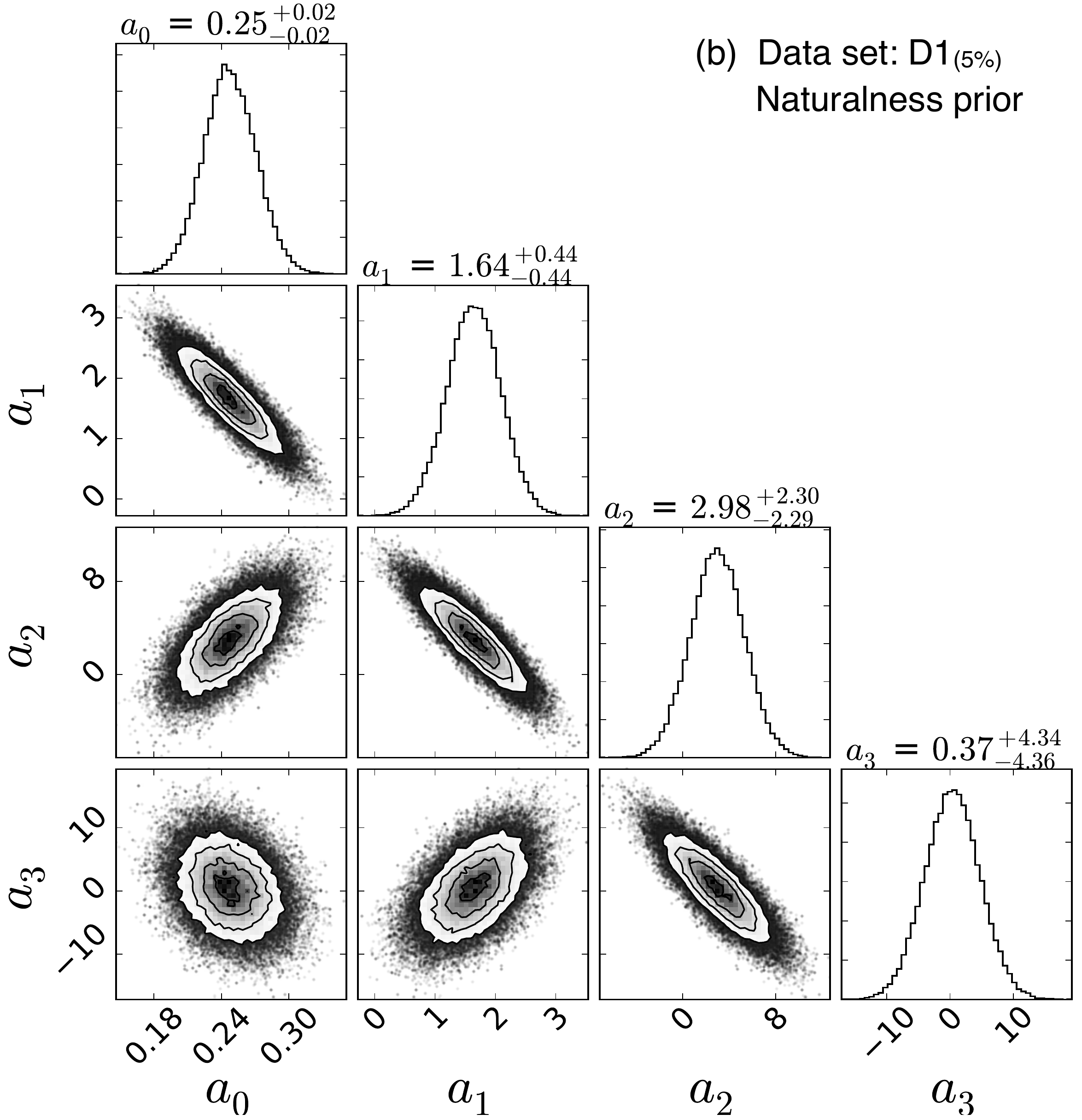}%
  } 
  \caption{
  Projected posterior plots~\cite{dan_foreman_mackey_2014_11020} for Model~D 
  [the function defined in Eq.~\eqref{eq:toy_func_SP}] calculated at order
  $\kord=3$, $\kmax=3$ given data set
   \dataset{D}{1}{5}, as described in Sec.~\ref{sec:procedures}, with (a) a uniform prior,
  and (b) the Set~\Cprime\ prior with $\abarzero=5$. The full posterior was calculated using
  Eq.~\eqref{eq:post-marg-abar0}.
  The plots on the diagonal are marginalized pdfs $\pr(a_i|D,I)$ for the 
  coefficient $a_i$ on the $x$-axis, calculated
  according to Eq.~\eqref{eq:2dto1d}. 
  The lower triangle plots are two-dimensional pdfs $\pr(a_i,a_j|D,I)$,
  calculated according to Eq.~\eqref{eq:projected-post}.
  The contours indicate 0.5, 1.0, 1.5, 2.0 sigma in a Gaussian approximation.
  The mean and 68\% DoB interval are also given for each coefficient. Note the
  change of scale between (a) and (b).
  \label{fig:triangle_D1}}
\end{figure*}

We use the rule of marginalization extensively in several different contexts.
\be
  \I \textbf{Projection:} to quote LEC results and plot pdfs. 
     We derive two-dimensional pdfs $\pr(a_i,a_j|D,I)$ from the full 
     joint posterior $\pr(\avec|D,I)$ by integrating over all the other
     parameters:
      \begin{multline}
      \qquad\pr(a_i,a_j|D,I)  \\
        = \int\! {\prod_{m=0, m \neq i,j}^{\kord}} da_m \, \pr(\avec|D,I)
      \;.
      \label{eq:projected-post}
      \end{multline}
    Such marginalized posteriors for the model used in 
    Refs.~\cite{Schindler:2008fh,Furnstahl:2014xsa} 
    are shown in the plots below the diagonal in Fig.~\ref{fig:triangle_D1}, where
    the $x$-axis scale is at the bottom.   (This model will be used in
    Sec.~\ref{sec:procedures} to illustrate the general procedures and diagnostics.) The $\pr(a_i,a_j|D,I)$s are displayed  as intensity plots, 
    and so reveal the regions of high joint probability for pairs of LECs.  Such 
    projected posterior plots 	
    are therefore useful for diagnosing correlations between the LECs obtained from
    the analysis.  
    Further details of their interpretation will be discussed in
    Sec.~\ref{subsec:triangleplots}.
    Note that in this case, the posteriors are two-dimensional Gaussians, but
    in general they can be more complicated and even multi-modal. 
 
 One further marginalization step reduces the two-dimensional pdfs to one-dimensional pdfs for a particular LEC, i.e.
   \begin{equation}
    \qquad\pr(a_i|D,I)
    = \int da_j \, \pr(a_i,a_j|D,I) \;.
    \label{eq:2dto1d}
  \end{equation}
  These pdfs appear on the diagonal in Fig.~\ref{fig:triangle_D1}.  Here, the individual 
  posteriors are simple Gaussians, but again, in general, they can be
  highly non-Gaussian.

  \I \textbf{Higher orders:} to account for higher-order terms in the EFT expansion and avoid
  underfitting when extracting low-order LECs. Marginalization is a straightforward method to account for
  such higher-order terms whose coefficients may not
  be well-constrained by the data but can still be assumed to be natural.
  In particular, if we need to determine coefficients up to order $\kord$ but wish to
  account for contributions up to order $\kmax$, we marginalize according to
  \begin{multline}
    \qquad\pr(a_0, \ldots, a_\kord|D,I) \\ 
    = \int da_{\kord+1} \ldots da_{\kmax} \,\pr(\avec|D,I) \;.
    \label{eq:marginpdf}
  \end{multline}

  \I \textbf{Integrating in:} to describe pdfs of interest in terms of pdfs with known analytic forms. For
  example, the prior distribution $\pr(\avec|I)$ can be expressed in
  terms of a naturalness parameter, $\abar$
  \beq
    \qquad\pr(\avec|I) = \int d\abar \, \pr(\avec|\abar,I) \pr(\abar|I) \;,
    \label{eq:marginabar}
  \eeq
  where $\abar$ parameterizes the width of prior pdf for the LECs $\pr(\avec|\abar,I)$.
  By introducing the marginalization integral over $\abar$, the prior for the LECs
  is specified quantitatively by its width $\abar$ and can be calculated.
  \ee

\subsection{Interpretation}

\label{subsec:interpretation}

In Sec.~\ref{subsec:Bayesian_parameter_est} we noted how the Bayesian
procedure can yield the same maximum likelihood result obtained from a
frequentist procedure.
However, the interpretations are quite different.
In the strict frequentist view, the only valid interpretation of probability
is as the relative frequency of an event out of
a large ensemble of repeatable experiments~\cite{Agashe:2014kda}.
Thus the outcomes of experiments are treated as random variables.
Model parameters (e.g., the LECs), on the other hand, are fixed --- they do
not have a probability distribution, so the Bayesian posterior we seek has
no meaning.  The complete focus is on the likelihood. Thus, strictly speaking, a frequentist confidence interval (CI) exists only for
data. It gives the probability that, for fixed parameters, a large set of experiments
will yield results that fall in the interval $p$\% of the time.

\begin{figure*}[tbh]
  \subfloat{%
    \label{fig:likelihood-win}%
  \includegraphics[width=0.45\textwidth]{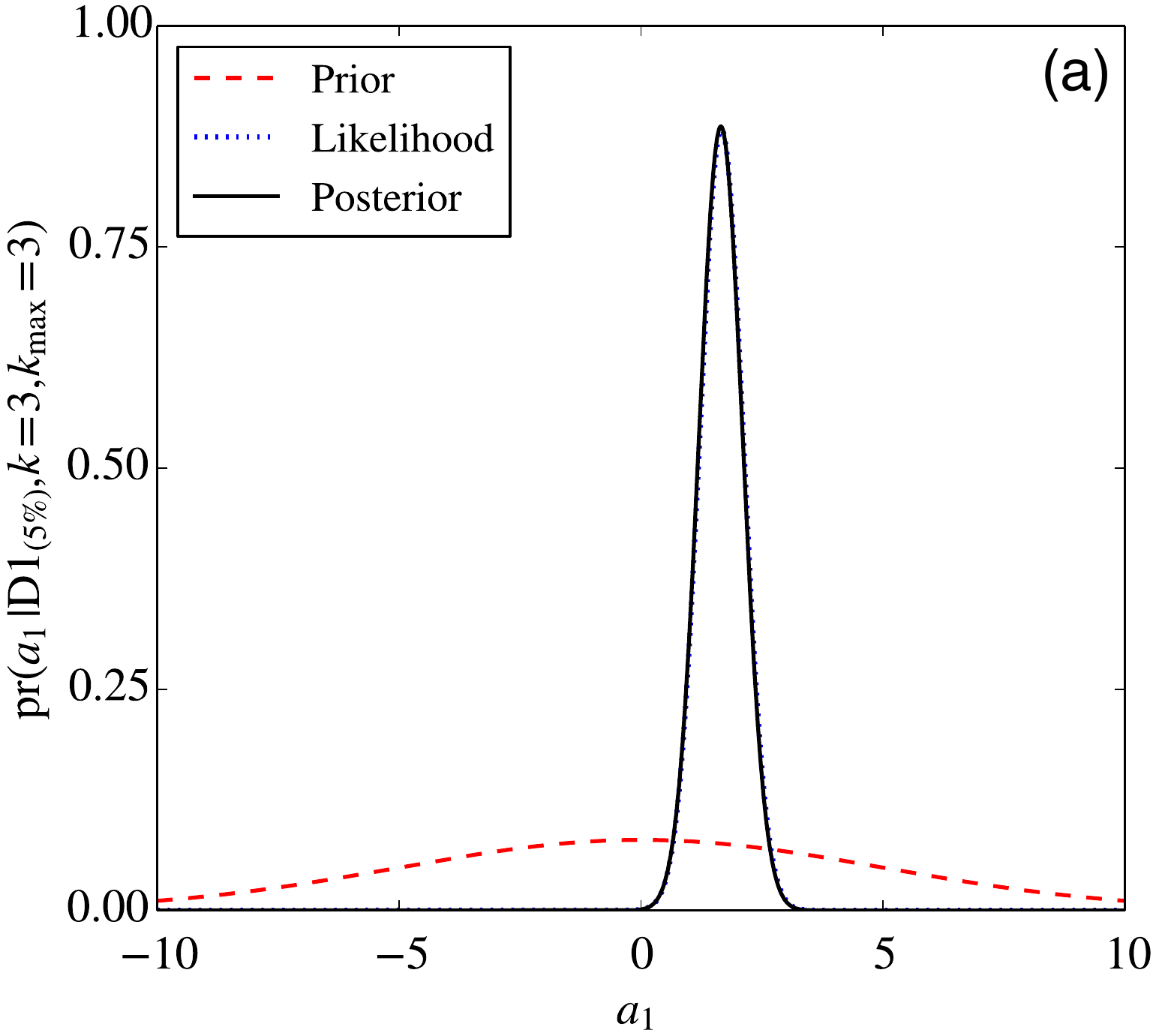}%
  }
  \hspace*{0.05\textwidth}  
  \subfloat{%
    \label{fig:prior-win}%
  \includegraphics[width=0.45\textwidth]{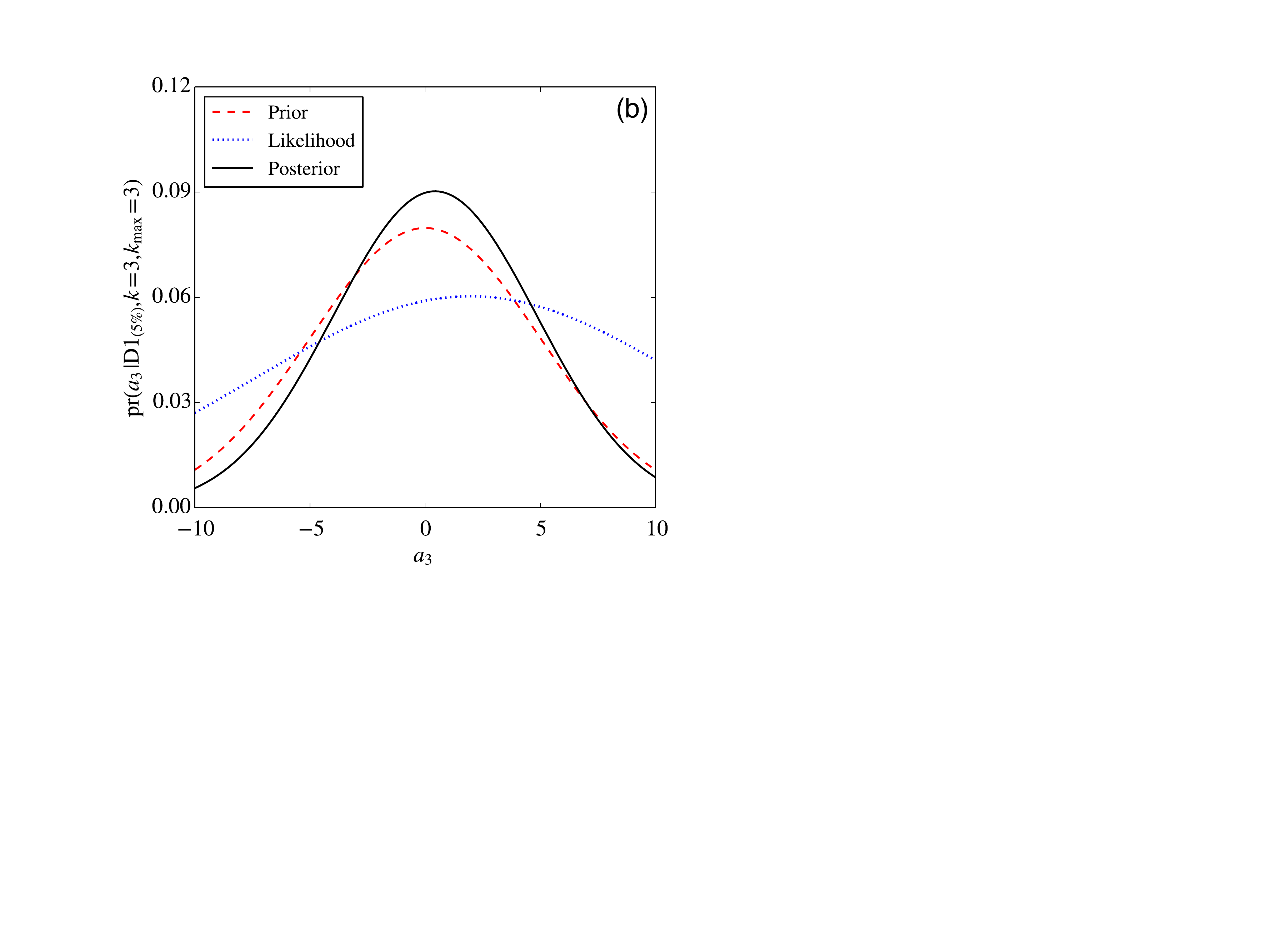}%
  } 
  \caption{(color online)
  Marginalized posteriors for two coefficients calculated at $\kord=3$, $\kmax=3$
  given data set \dataset{D}{1}{5}, 
  which is described in Sec.~\ref{sec:procedures},  
  along with the likelihood and prior pdfs in each case.
  For parameter $a_1$ in (a), the posterior is dominated by the likelihood (causing the two
  to overlap in the plot) and
  the prior is essentially uniform. 
  For parameter $a_3$ in (b), the likelihood is widely spread and the posterior
  is mostly given by the prior.
  \label{fig:likelihood-prior-compare}}
\end{figure*}

In the Bayesian interpretation, probabilities express the current state of
knowledge about
the parameters, which allows us to calculate Bayesian degree-of-belief (DoB)
intervals for the true values of the parameters.
These are established by integrations over posteriors for $\avec$ (or some subset)
to determine the desired intervals.
For example, a 68\% DoB interval for a parameter $a_i$ means that the particular input
data and information lead to a probability of 68\% that the
true value of $a_i$ is in that interval.
The data are fixed while the parameters have a pdf.

The use of the word ``belief'' here is somewhat unfortunate, because it might suggest to some
readers that results are subjective. And indeed, different priors can lead to different outcomes for the posterior pdfs---particularly
in the case where data are sparse or of low quality. But, the Bayesian framework is completely rigorous
once the prior is specified. It permits a careful tracing of assumptions through to probabilistic consequences for parameters.
In the case of EFT 
the assignment and interpretation of the prior should not be a source of controversy:
it merely encodes \emph{a priori} information about the parameter of interest. Furthermore, the prior can also represent
knowledge from previous determinations or measurements.
The use of Bayesian priors thus provides a
quantifiable way to include guidance from both theory and previous experiments. It makes explicit the assumptions that are
inherent in EFT parameter estimation, but are usually left largely implicit.  We also mention in this context that, given specified pieces of ``testable
information'' for the parameters, the Method of Maximum Entropy (MaxEnt)~\cite{Jaynes:1957} can
be used to derive the least informative priors based on the known information.

The interplay of likelihood and prior in limiting cases embodies desired outcomes
that otherwise have to be imposed by hand.
If the data are numerous and of good quality, a highly peaked likelihood will
be the dominant
contribution to the posterior if the prior is not too restrictive~\cite{Sivia:2006}.
On the other hand, if the likelihood for a parameter is not well-constrained by
the data, the prior will have the dominant effect. In the case of LEC parameter
 estimates, a naturalness prior can restrict the effect of parameters that
are not well-constrained by data by suppressing contributions from regions
of parameter space driven by spurious correlations.

An example of a parameter that is well-determined by data is shown in
Fig.~\subref*{fig:likelihood-win},
where the prior does not affect the posterior for that parameter because
it is essentially uniform in the region where the likelihood has support.
Figure~\subref*{fig:prior-win}
shows an example where the posterior largely reflects the input prior, since the
parameter is not strongly localized by the slowly varying likelihood. We refer to
this latter situation as the posterior ``returning the prior''.
These are marginalized posterior distributions from the same joint pdf $\pr(\avec|D,I)$,
which means that even though the prior does not seem to play a role in
Fig.~\subref*{fig:likelihood-win}, this is because it has restricted the
domain of the parameter in Fig.~\subref*{fig:prior-win}.
Thus the first parameter is insensitive, but only because the prior prevented
other, higher-order, parameters from contributing in unnatural regions of parameter
space. An example of a spurious correlation can be seen in Fig.~\subref*{fig:triangle_D1_uniform},
where $a_0$ is correlated with the ill-determined cubic coefficient $a_3$. The amelioration of this problem is then seen in Fig.~\ref{fig:triangle_D1}\subref{fig:triangle_D1_Gaussian},
where the correlation is almost completely removed when an appropriate naturalness prior is applied.

\subsection{Bayes evidence}  \label{subsec:bayes_factor}

In this section we explain how Bayesian methods facilitate a quantitative approach to 
\emph{model selection}: which model from a set of models is preferred by the data?
Our presentation draws on material from Refs.~\cite{Sivia:2006,Gregory:2005,Trotta:2008qt}. 

We should first clarify what we mean by ``model'' in this context.
After all,
an EFT is said to be model-independent because the theory is constructed in
the most general form consistent with the symmetries
of the underlying interaction.
We use ``model'' here to mean
any theoretical construct for computing an observable.
Examples of quantitative Bayesian comparisons between EFT models could be the
predictive power of EFTs with different degrees of freedom for the same data set,
or an analysis of the increase in predictive power for observables using
different orders in the same EFT. 

For two models $M_1$ and $M_2$ and the same data set $D$, the relevant pdfs to compare
are the posteriors for each model given the data, called the model evidence: $\pr(M_1|D,I)$
and $\pr(M_2|D,I)$.
Note that there is no reference to a particular parameter set here; the comparison
is not between two fits but between two models.
We can use Bayes' theorem
to express the evidence ratio for the two models:
\beq
	\frac{\pr(M_2|D,I)}{\pr(M_1|D,I)}
	  = \frac{\pr(D|M_2,I)\,\pr(M_2|I)}{\pr(D|M_1,I)\,\pr(M_1|I)}
	 \;,
\eeq
where the common factor of $\pr(D|I)$ cancels in the ratio.
If neither model is \emph{a priori} more likely,
the ratio $\pr(M_2|I)/\pr(M_1|I) = 1$
so that the evidence ratio becomes equal to the \emph{Bayes factor}
\beq
	\frac{\pr(M_2|D,I)}{\pr(M_1|D,I)}
	\longrightarrow \frac{\pr(D|M_2,I)}{\pr(D|M_1,I)}
	\;.
  \label{eq:bayes-factor}
\eeq
To express the right-hand-side in terms of the parameters of each
model, we apply the marginalization rule so that
\beq
	\frac{\pr(D|M_2,I)}{\pr(D|M_1,I)}
	= \frac{\int d\avec_2 \, \pr(D|\avec_2,M_2,I)\,\pr(\avec_2|M_2,I)}
	  {\int d\avec_1 \, \pr(D|\avec_1,M_1,I)\,\pr(\avec_1|M_1,I)} \;.
	\label{eq:evi-ratio}
\eeq
Thus we integrate over the entire parameter space.
The evidence ratio provides a quantitative diagnostic to assess
the effectiveness of EFTs with different degrees of freedom
(e.g., comparing \chiEFT\ with and without explicit
$\Delta$-resonances~\cite{Epelbaum:2007sq}) or distinguishing
between different power-counting schemes.

Here we focus on using the evidence ratio to assess whether adding another
order to our EFT is favored by the given data.
Thus we consider the case where 
$M_1 \rightarrow M_\kord$ and $M_2 \rightarrow M_{\kord+1}$ are the same EFT but evaluated
at successive orders in the EFT expansion, and that neither is more likely \emph{a priori}
so that the evidence ratio for the two models is equal to the Bayes factor as in Eq.~\eqref{eq:bayes-factor}.
(We will continue to use $\avec_1$ and $\avec_2$ to denote the corresponding
vectors of coefficients.)

For simplicity, let $M_{\kord+1}$ have one additional parameter $a'$ and assume that
the priors for the parameters factor into individual priors, so
\begin{align}
  \pr(\avec_2|M_{\kord+1},I) &\equiv \pr(\avec_1,a'|M_{\kord+1},I)
   \nonumber \\
    &= \pr(\avec_1|M_{\kord+1},I)\,\pr(a'|M_{\kord+1},I)
    \;.
\end{align}
If the values of $a'$ that contribute to the integrand in the numerator of
Eq.~(\ref{eq:evi-ratio}) are determined predominantly by the likelihood (for 
the values of $\avec_1$ where the integrand is large), and that 
likelihood is sharply peaked, then we can 
 approximate the impact of the prior by taking it to be roughly constant with an effective width $\Delta a'$,
so $\pr(a'|M_{\kord+1},I) \approx 1/\Delta a'$.
We  then approximate the integral over $a'$ as the value at the likelihood peak $\widehat a'$
times the width of that peak, $\delta a'$. This yields the Bayes factor
\begin{align}
  \frac{\pr(D|M_{\kord+1},I)}{\pr(D|M_\kord,I)}
  \approx& \frac{\delta a'}{\Delta a'}
       \int\!d\avec_1\, \pr(D|\avec_1,\widehat a',M_{\kord+1},I)                
                 \nonumber \\
          & \null  
             \qquad\qquad\times \pr(\avec_1|M_{\kord+1},I)  
          \nonumber \\
    & \hspace*{-0.4in} \null\times \Bigl[
        \int\!d\avec_1\, \pr(D|\avec_1,M_\kord,I) \, \pr(\avec_1|M_\kord,I)
        \Bigr]^{-1}
     \;.
     \label{eq:bayes_factor1}
\end{align}
The ratio of $\avec_1$ integrals represents
the gain in the likelihood from having this extra parameter with value
$\widehat a'$. 
But the ratio is also multiplied by 
the ``Occam factor'' or ``Occam penalty'' $\delta a'/\Delta a'$,
which is the factor by which the parameter space collapses when the
data are taken into account (this is large when the parameters can range
over a large domain compared to the restriction imposed by the data).

Equation~\eqref{eq:bayes_factor1}
tells us that we will favor the more complicated model only if the
likelihood gain is larger than this factor.
We are assuming that $M_\kord$ is contained within $M_{\kord+1}$, i.e.,
$M_\kord$ is $M_{\kord+1}$ with $a' = 0$, which means this likelihood gain
is always greater than or equal to one.
In a situation such as Fig.~\subref*{fig:likelihood-win}, we have a peaked 
likelihood compared to the prior pdf, which yields a substantial Occam penalty.
However, the likelihood gain here is much greater than that penalty, because it involves
the 
likelihood peak at $\widehat a' > 0$ in comparison to the likelihood for $a' = 0$, and the
latter is very small.
That means that the evidence ratio for the addition of $a'$ is $\gg 1$, and
the inclusion of this parameter is therefore highly favored.
This leads us to expect that if we calculate
 the evidence for the data 
$\pr(D|M_\kord,I)$ as a function of the order $k$ (which we will henceforth 
call the ``evidence'' for brevity) that function will increase as long as
additional parameters are leading to a marked decrease in the $\chi^2$.
However, if the likelihood function at $a'=0$ is not well below the peak value (e.g., its width is greater
than in this example, 
or the peak is shifted toward zero),
than the Occam penalty can dominate and disfavor an additional parameter,
which is an implementation of Occam's razor.
In such a situation 
the evidence will decrease from order $k$ to order $k+1$.
We would then expect the evidence function $\pr(D|M_\kord,I)$ to show a peak.

A different behavior can be expected if the comparison of likelihood and
prior as in 
Fig.~\subref*{fig:prior-win}
holds, in which case the analysis of Eq.~\eqref{eq:evi-ratio} is inverted.
That is, we now say that the
dependence on $a'$ in the likelihood can be replaced by a constant $\widehat a'$
and the likelihood pulled out of the integral of $a'$:
\begin{align}
  \frac{\pr(D|M_{\kord+1},I)}{\pr(D|M_\kord,I)}
  \approx& \int\!d\avec_1\, \pr(D|\avec_1,\widehat a',M_{\kord+1},I)                
                 \nonumber \\
          & \null  
             \times \pr(\avec_1|M_{\kord+1},I)  \int\!da'\, \pr(a'|M_{\kord+1},I)
          \nonumber \\
    & \hspace*{-0.4in} \null\times \Bigl[
        \int\!d\avec_1\, \pr(D|\avec_1,M_\kord,I) \, \pr(\avec_1|M_\kord,I)
        \Bigr]^{-1}
     \;.
     \label{eq:bayes_factor2}
\end{align}
The integral over $a'$ is now just a normalization integral, equal to one.
We are dominated by the prior, so
$\widehat a' \approx 0$ and the Bayes ratio is one, because $M_{\kord+1}$ with the last
coefficient equal to zero is simply $M_\kord$.
The same argument goes through for each higher value of $k$,
meaning that we have \emph{saturation}
rather than a peak for $\pr(D|M_{k+1},I)$ as a function of the order $k$~\cite{Trotta:2008qt}.
In summary, the naturalness prior cuts down the ``wasted'' parameter space
that might be ruled out by the data and which leads to an Occam penalty.
This means that it limits the ``phase space'' of an EFT,
which is therefore a simpler model (in the model selection sense) than the
same functional form with unconstrained or only weakly constrained LECs.

\subsection{Sampling with MCMC}\label{sec:MCMCsampling}

While in some situations
it is possible to evaluate posterior pdfs analytically, in
many cases we must resort to numerical methods.  
Although many of the integrals we confront in this present work do not require its use, 
we employ Markov Chain Monte Carlo (MCMC)
sampling methods to obtain posterior pdfs, since they are easily generalized to cases with
more complicated probability distributions. 

The MCMC algorithm generates
$N$ samples $\{a_j\}$ according to the posterior probability distribution 
$ \pr(\avec|D,I)$. 
Expectation integrals may then be performed using 
those samples:
\beq
  \langle f(\avec) \rangle = \int d \avec \, \pr(\avec|D,I) f(\avec) \approx \frac{1}{N}
  \sum_{j=1}^N f(\avec_j) \;.
  \label{eq:ilovemc}
\eeq
The result accounts for all correlations between the parameters. The propagation 
of those correlations to the function $f$ can be simplified if the posterior is well
approximated as a correlated Gaussian; in that case the correlation matrix
can be employed for this purpose. But Eq.~(\ref{eq:ilovemc}) achieves
this task independent of whether $ \pr(\avec|D,I)$ is Gaussian or not. 
Marginalization integrals over parameters [see, e.g., Eq.~\eqref{eq:projected-post}] can also
be performed trivially by retaining only samples in the parameters of interest. 
The samples $\{a_j\}$ are constructed via MCMC, in a particular implementation called
\emcee~\cite{Foreman_Mackey:2013aa}. Implementation details can be found in 
Appendix~\ref{app:emcee}.


\section{Diagnostics and procedures for parameter estimation}
 \label{sec:procedures}

\subsection{Set up of a test case}\label{subsec:setup}

In this section we present a suite of diagnostics and associated procedures
for EFT parameter estimation.
For continuity, we use for illustration  a model problem explored in 
previous work~\cite{Schindler:2008fh,Furnstahl:2014xsa}, which is to
use a given data set to 
extract as many coefficients in a particular function's
Taylor expansion as possible. For the function we choose, in this section~\cite{Schindler:2008fh}:
\begin{align}
   g(x) &= \left( \frac12 + \tan(\frac{\pi}{2}x) \right)^2  \nonumber \\
         &= 0.25 + 1.57x + 2.47x^2 + 1.29x^3
           + \cdots
       \;.
       \label{eq:toy_func_SP}
\end{align}
The function is designed to have coefficients 
that are $\Order{(1)}$ up to about tenth order, 
and the singularity at $x=1$ means this is where 
the simulated EFT expansion breaks down. 
We follow the previous work and consider 10 equally spaced data points covering the range $0 < x \leq 1/\pi$,
each with a data error of 5\% [thus $c=0.05$ in Eq.~\eqref{eq:pseudodata}]. The data set is
enumerated in Appendix~C of Ref.~\cite{Schindler:2008fh} and is also available
as supplementary material to this paper.
To conform to later usage, we refer to this as data set \dataset{D}{1}{5}, 
where ``D'' indicates the function
defined by Eq.~\eqref{eq:toy_func_SP}, the ``1'' labels a choice of $x$ points where
the function was sampled, and the
subscript indicates the data error.
Other models and the nucleon mass expansion, with varied data ranges, precision,
and numbers of points, will be considered in 
Secs.~\ref{sec:model-problems} and~\ref{sec:case_studies}.

\begin{table}[bh]
  \caption{Examples of prior pdfs encoding various naturalness assumptions. 
  \label{tab:priors}}
  \begin{tabular}{c|cc}
  set  &   $\pdf(a_i| \abar)$   &   $\pdf(\abar)$  \\
  \hline
  A &
  $\frac{\ts 1}{\ts 2\abar}\,\theta(\abar-|a_i|)$
    & $\frac{\ts 1}{\ts \ln \abarmax/\abarmin}
   \frac{\ts 1}{\ts\mathstrut\abar}
    \theta ( \abar - \abarmin) \theta( \abarmax - \abar)$
   \\[8pt]
  \Aprime &
  $\frac{\ts 1}{\ts 2\abar}\,\theta(\abar-|a_i|)$
  & $\delta(\abar - \abarzero)$
   \\[8pt]
  B &
  $\frac{\ts 1}{\ts 2\abar}\,\theta(\abar-|a_i|)$
  &  $\frac{\ts 1}{\ts\sqrt{2\pi}\abar\sigma} e^{-(\log\abar)^2/2\sigma^2}$
   \\[8pt]
  C &
  $\frac{\ts 1}{\ts\sqrt{2\pi}\abar} e^{-a_i^2/2\abar^2}$
    & $\frac{\ts 1}{\ts \ln \abarmax/\abarmin}
   \frac{\ts 1}{\ts\mathstrut\abar}
    \theta ( \abar - \abarmin) \theta( \abarmax - \abar)$
  \\[8pt]
  \Cprime &
  $\frac{\ts 1}{\ts\sqrt{2\pi}\abar} e^{-a_i^2/2\abar^2}$
  & $\delta(\abar - \abarzero)$
   \\[5pt]
  \hline
  \end{tabular}
\end{table}

\begin{table*}[t]
  \caption{\label{tab:diagnostic_list}Diagnostic tools for parameter estimation.
  Figure numbers in Secs.~\ref{sec:stat-methods} and \ref{sec:procedures} are given for each type of plot.  
  Additional 
  examples of these figures can be found in Secs.~\ref{sec:model-problems} 
  and \ref{sec:case_studies}.}
  \begin{ruledtabular}
  \begin{tabular}{c|c|p{0.75\linewidth}}
     \textbf{Name} & \textbf{Fig.} &
          \multicolumn{1}{c}{\textbf{Description and uses}} \\ \hline
     \parbox[t]{0.15\linewidth}{projected posterior plot} &  \ref{fig:triangle_D1}  &
      Matrix of subplots with posteriors from marginalizing
      $\pr(\avec|D,I)$ over all but one $a_i$ (diagonal) or all but a pair
      $a_i, a_j$ (lower triangle). Compare different priors; visualize correlations; identify
      when posterior $\approx$ prior.
      \\ \hline

      $\abar$ posterior & \ref{fig:D15-abar-post-k3-kmax3}  &  $\pr(\abar|D,\kord,\kmax)$.
       Show weighting of $\abar$ values when marginalized;
       identify appropriate marginalization range for $\abar$;
       signals unnatural coefficients.
      \\ \hline

     $\abar$ relaxation plot & \ref{fig:D1_abar_relaxation}  &
      Evolution of marginalized posterior for single coefficients for full
      range of $\abar$. Check whether $\abar$ is too restrictive (rapid changes
      with $\abar$);
      identify regions insensitive to $\abar$ to help
      identify marginalization range; show transition to least-squares result.
      \\ \hline

     evidence vs.\ $k$ &  \ref{fig:D15_evidence}  &
      $\pr(D|k,\abar) \propto \pr(k|D,\abar)$ (if uniform
     prior on $k$).  Show transition from dominance by likelihood
     [e.g., Fig.~\subref*{fig:likelihood-win}]
     to dominance by prior [e.g., Fig.~\subref*{fig:prior-win}].
      \\ \hline

     $\xmax$ plot & \ref{fig:xmax_D15}  &
      Evolution of marginalized posterior for single coefficients, 
      $\pr(a_i|D,I)$, as data
      range specified by $\xmax$ is increased. Check for stability with
      data range.
      \\ \hline

     multi-set plot & \subref*{fig:D15_datasets}  &
      Evolution of marginalized posterior for single coefficients using multiple
      equal-sized sets of data, each over the same range.
      \\ \hline

     accumulation plot & \subref*{fig:D15_accumulated}   &
      Comparison of marginalized posteriors for single coefficients as data are
      accumulated from combining multiple equal-sized sets of data.
      \\ \hline

     residual plot & \ref{fig:D15-Lepage-kmax3-k0-to-2}  &
      Log-log plot of residuals versus expansion parameter $x$.  
      Check for power-law behavior at different
      orders.  Test for data or theory error dominance in
      different regions of $\Q$.
  \end{tabular}
  \end{ruledtabular}
\end{table*}

Now that we have data for a simulated EFT, the first step in our analysis is to select 
prior pdfs for the coefficients.
Some possible sets of priors encoding naturalness assumptions are listed in 
Table~\ref{tab:priors}.  Prior Sets A, B, and C were used in 
Ref.~\cite{Furnstahl:2015rha}.  The assumption for the coefficients 
themselves (the $a_i$s) in Sets~A and B
correspond to a flat distribution bounded by a maximum, $\abar$.
For Set~A, this naturalness parameter $\abar$ is itself subject to a 
scale-invariant uniform (Jeffreys) prior~\cite{Schindler:2008fh} 
between the limits $\bar{a}_<$ and $\bar{a}_>$, while Set~B has 
the naturalness parameter $\abar$ following a log-normal distribution of 
width $\sigma$. Set~C assumes a Gaussian prior distribution for the $a_i$s, with the
width $\abar$ distributed according to the Jeffreys prior.
In general, arguments can be made to support any of these assumptions, which means we need to 
carefully check the sensitivity of our results to different choices.

For most of our examples, we choose a simpler variation of Set~C, 
labeled Set~\Cprime, in which we adopt the Gaussian prior on the coefficients 
$\avec \equiv \{a_0,\cdots,a_k\}$,
\beq
    \pr(\avec|\abar,I) = \left(\frac{1}{\sqrt{2\pi}\abar} \right)^{\kord+1}
        \exp\left(-\frac{\avec^2}{2 \abar^2}\right) \;,
        \label{eq:Gaussian-prior}
\eeq
where $I$ will include the order of the model and the marginalization order, 
and fix the value of the naturalness parameter at $\abarzero$:
\beq
    \pr(\abar) = \delta(\abar - \abarzero) \;,
    \label{eq:delta-func-abar}
\eeq
so that the marginalization over $\abar$ as in Eq.~\eqref{eq:marginabar}
picks out this value. The impact of the choice of a particular $\abarzero$ can be anticipated by consulting
the $\abar$ posterior plot described below.

We seek the posterior for the parameters $\avec$ for the model EFT expansion
up to order $\kord$, but we will marginalize over coefficients as in 
Sec.~\ref{subsec:marginalization}, up to
order $\kmax$ where we define the coefficients $\amarg \equiv \{a_{k+1},\cdots,a_{\kmax}\}$;
we designate this posterior as $\pr(\avec|D,\kord,\kmax)$.
We can calculate this using Eqs.~\eqref{eq:marginpdf}--\eqref{eq:marginabar}, and Bayes' theorem:
\begin{widetext}
\beq
    \begin{split}
    \pr(\avec|D,\kord,\kmax) = & 
    \int da_{\kord+1} \cdots da_{\kmax} \, 
       \pr(\avec, a_{\kord+1}, \cdots, a_{\kmax} | D, \kord,\kmax) \\
    = & \int d\amarg \ \frac{\pr(D|\avec,\amarg,\kord,\kmax) \pr(\avec,\amarg|\kord,\kmax)}
    {\pr(D|\kord,\kmax)} \\
    = & \int d\amarg \int d\abar \ \frac{\pr(D|\avec,\amarg,\kord,\kmax) 
    \pr(\avec,\amarg|\abar,\kord,\kmax) \pr(\abar)}
    {\pr(D|\kord,\kmax)}
    \;,
    \end{split}
    \label{eq:avec-post-general-exp}
\eeq
where we have assumed that the prior for 
the naturalness parameter is independent of the truncation and marginalization orders.
Substituting the prior from Eq.~\eqref{eq:delta-func-abar}, we have
\beq
    \pr(\avec|D,\kord, \kmax)  = \int d\amarg \
    \frac{\pr(D|\avec,\amarg,\kord, \kmax) \pr(\avec,\amarg|\abarzero,\kord,\kmax)}
    {\pr(D|\kord,\kmax)} \;,
    \label{eq:post-marg-abar0}
\eeq
The likelihood $\pr(D|\avec,\amarg,\kord,\kmax)$ for this
problem is simply the least-squares likelihood from Eq.~\eqref{eq:least-squares-likelihood} 
calculated with \emph{all} the coefficients $\{\avec,\amarg\}$ up to order $\kmax$. For Set~\Cprime\ 
the joint prior
$\pr(\avec,\amarg|\abarzero,\kord,\kmax)$ is simply the product of Gaussian priors with width
$\abarzero$ in Eq.~\eqref{eq:Gaussian-prior} for all the coefficients up to order $\kmax$.

\textbf{Note on $\kord$ and $\kmax$.}
The posterior pdf in Eq.~\eqref{eq:avec-post-general-exp}
is the pdf for the coefficients $\avec$ up to some order $\kord$, having accounted 
for the higher-order coefficients $\{a_{k+1},\cdots,a_{\kmax}\}$. Had we computed 
the posterior for \emph{all} the coefficients up to order $\kmax$ 
and then performed the marginalization integral to generate the projected 
posterior plot up to order $\kord$, the
results would be the same. 
Thus in the examples in this paper, 
where the posterior has the form of Eq.~\eqref{eq:avec-post-general-exp},
the parameter estimation 
will be completely controlled by $\kmax$.
The order $\kord\leq\kmax$ controls the actual calculation of observables,
for example as in Fig.~\ref{fig:D15-Lepage-kmax3-k0-to-2}.

\textbf{Alternative accounting of higher-order effects.}
As noted in Sec.~\ref{sec:intro}, explicitly calculating the observables appearing in
the likelihood up to order $\kmax$ may not be feasible---yet we still need to account 
for these contributions. 
To do so, we build on Ref.~\cite{Furnstahl:2015rha} and use the expected naturalness
of the expansion coefficients for observables at higher order.
That is, the contribution to an observable beyond order $\kord$ is given by
the truncation error (which may also have an overall scale~\cite{Furnstahl:2015rha}) 
\beq
  \Delta_k(x) = \sum_{i=\kord+1}^{\kmax} c_i x^i \;,
  \label{eq:delta-cmarg}
\eeq
and we introduce a naturalness prior for the $c_i$s. For the linear model observables
introduced in Eq.~\eqref{eq:model-th-expansion}, the $c_i$s here are just the $a_i$s.
The generalization of Eq.~\eqref{eq:avec-post-general-exp} where we marginalize over 
$\cmarg \equiv \{c_{\kord+1}, \ldots, c_{\kmax}\}$ is

\beq
  \pr(\avec|D,\kord,\kmax) = \int d \cmarg \, \frac{\pr(D|\avec, \cmarg, \kord, \kmax) \,
  \pr(\avec|\kord,\kmax) \, \pr(\cmarg|\kord,\kmax)}{\pr(D|\kord,\kmax)} \;,
  \label{eq:cmarg-avec-post-pdf}
\eeq
where we have assumed statistical independence of the coefficients $\avec$
from each other and from higher-order corrections, so that
the prior pdf $\pr(\avec|\cmarg,\kord,\kmax)=\pr(\avec|\kord,\kmax)$.
The posterior now depends on where the line is drawn (at $\kord$) between higher-order corrections
and the LECs we want to compute. 
Thus, in contrast to the case just described for Eq.~\eqref{eq:avec-post-general-exp}, 
it is possible to have posteriors for $\avec$ up to order $\kord$ that are not 
controlled just by $\kmax$. 
The two prior pdfs for the $\cmarg$ and the $\avec$ coefficients will now encode our naturalness
assumptions. Reference~\cite{Furnstahl:2015rha} contains an extensive analysis of different naturalness priors
for observable coefficients. The likelihood in Eq.~\eqref{eq:cmarg-avec-post-pdf} will not 
in general be a simple
least-squares likelihood, because the systematic error in the calculation given in Eq.~\eqref{eq:delta-cmarg}
cannot simply be added in quadrature to the experimental error \cite{Stump:2001gu,Furnstahl:2014xsa}.
This alternative
accounting will not be used in this paper, but will be explored in future work.

\end{widetext}

When we are not marginalizing over higher orders and $\kmax=\kord$, 
Eq.~\eqref{eq:post-marg-abar0} can be computed analytically
as an augmented $\chi^2$, which was derived by Schindler and Phillips in Ref.~\cite{Schindler:2008fh}.
When $\kmax > \kord$, the marginalization over the higher-order coefficients can be
understood as accounting for the correlated systematic errors from leaving out
higher-order terms. The posterior in this case is then a \emph{modified}, augmented $\chi^2$
\cite{Furnstahl:2014xsa} which was derived in Refs.~\cite{Stump:2001gu,Schindler:2008fh}.
As a benchmark, we compare results obtained for estimates of $\avec$ using prior Set \Cprime\
from Eq.\eqref{eq:post-marg-abar0} with the results of least-squares fits (equivalent to choosing
a uniform prior instead) using our test data set \dataset{D}{1}{5}.

\begin{figure}[tbh]
    \includegraphics[width=0.45\textwidth]{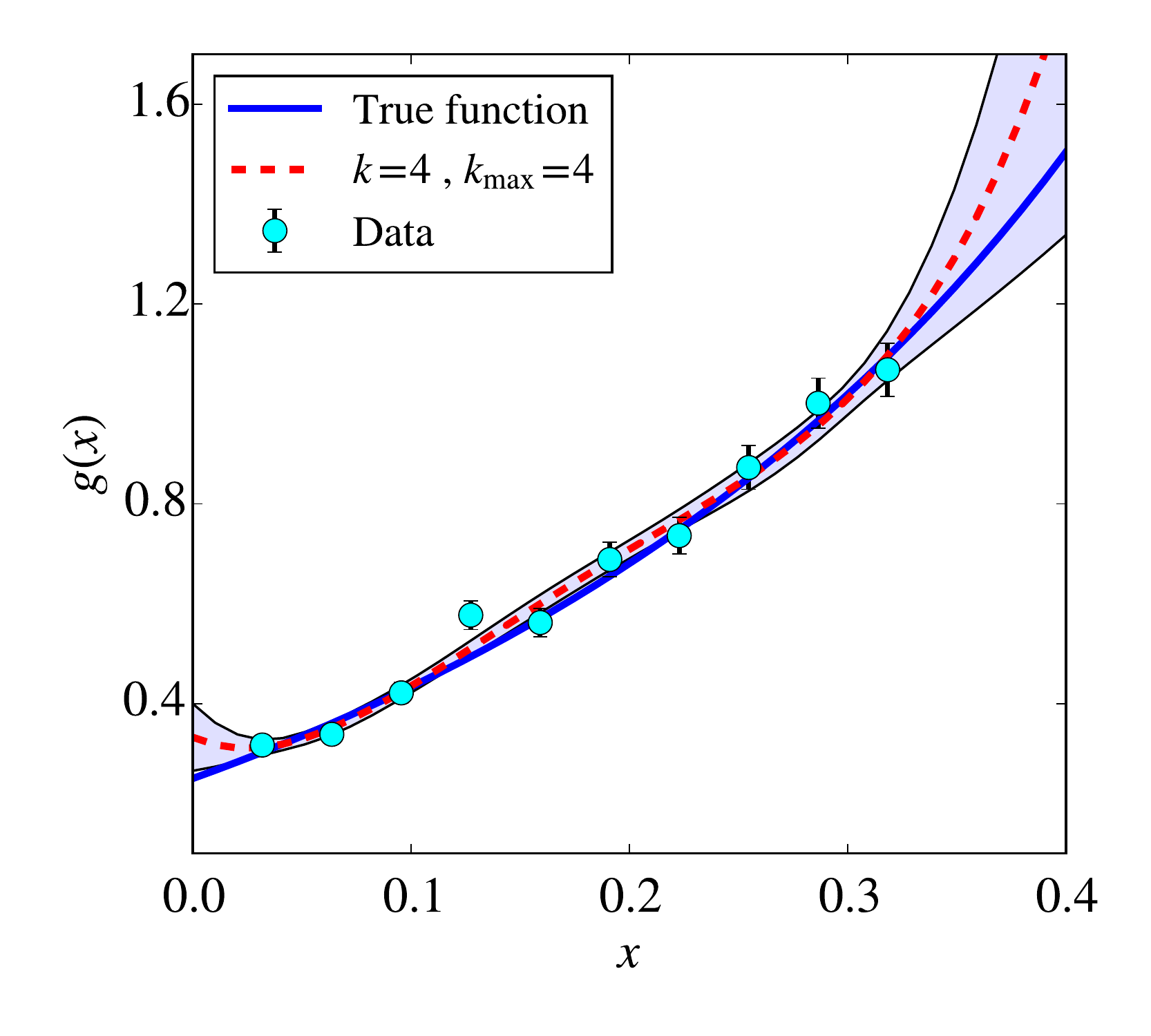}
    \caption{(color online) 
    Comparison of data set \dataset{D}{1}{5}, the underlying function for Model~D from
    Eq.~\eqref{eq:toy_func_SP}, and a least-squares prediction calculated at order
    $\kord=4$, $\kmax=4$ from that data set. The error bands represent 1-$\sigma$ (68\% DoBs).
     \label{fig:D1-ls-k4-fit}}
\end{figure}

\begin{figure}[tbh]
    \includegraphics[width=0.45\textwidth]{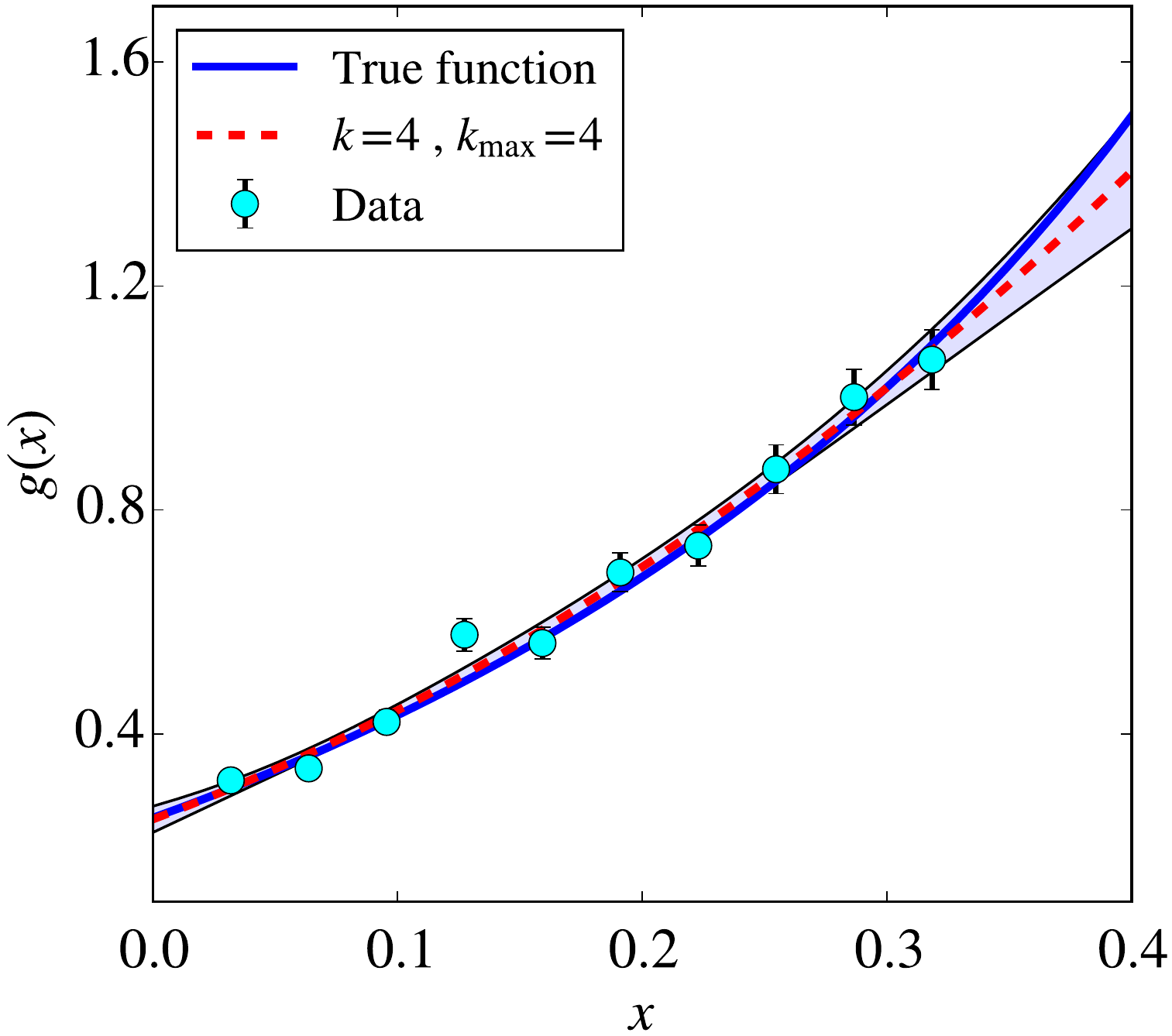}
    \caption{(color online)
     Comparison of data set \dataset{D}{1}{5}, the underlying function for Model~D from
    Eq.~\eqref{eq:toy_func_SP}, and a Bayesian prediction calculated at order
    $\kord=4$, $\kmax=4$ using prior Set \Cprime with
    $\abarzero=5$, from that data set. The error bands represent 1-$\sigma$ (68\% DoBs).
    \label{fig:D1-bayes-abar-5-k4-fit} }
\end{figure}

\begin{table*}[tbh]
   \caption{Coefficient estimates from sampling of $\pr(\avec|\dataset{D}{1}{5},\kord,\kmax)$
   given the expansion from Eq.~\eqref{eq:model-th-expansion} at different orders
   (these results are controlled by $\kmax$ only, see Sec.~\ref{subsec:setup}).
   The left side of the table is for a uniform prior, 
   which is equivalent to a least-squares fit, and includes the $\chi^2$/dof values.
   The right side of the table is using prior Set~\Cprime\ 
   from Table~\ref{tab:priors} with $\abarzero = 5$, and includes the evidence
   $\pr(\dataset{D}{1}{5}|\kord,\kmax)$.
   For both priors the posterior pdf is a multi-dimensional
   Gaussian.}
    \label{tab:D1-results}
  \begin{tabular}{|c|c||c|c|c|c||c|c|c|c|}
    \hline
     \multicolumn{2}{|c||}{} & \multicolumn{4}{|c||}{Uniform prior} &  \multicolumn{4}{|c|}{Gaussian prior} \\
    \hline
      $\kord$ & $\kmax$ & $\chi^2/$dof & $a_0$ & $a_1$ & $a_2$ 
      & Evidence & $a_0$ & $a_1$ & $a_2$ \\ 
    \hline
        0 & 0 & 67 & 0.48$\pm$0.01 &  &  & $\sim 0$ & 0.48$\pm$0.01 &  &  \\
        1 & 1 & 2.2 & 0.20$\pm$0.01 & 2.6$\pm$0.1 &  &$6.0\times10^{2}$ & 0.20$\pm$0.01 & 2.6$\pm$0.1 &  \\
        2 & 2 & 1.6 & 0.25$\pm$0.02 & 1.6$\pm$0.4 & 3.3$\pm$1.3 & $3.3\times10^{3}$ & 0.25$\pm$0.02 & 1.6$\pm$0.4 & 3.1$\pm$1 \\ 
        2 & 3 & 1.9 & 0.27$\pm$0.04 & 1.0$\pm$1 & 8.1$\pm$8.0 & $2.9\times10^{3}$ & 0.25$\pm$0.02 & 1.7$\pm$0.5 & 3.0$\pm$2 \\ 
        2 & 4 & 2.0 & 0.33$\pm$0.07 & $-$1.9$\pm$3 & 45$\pm$30 & $2.8\times10^{3}$ & 0.25$\pm$0.02 & 1.7$\pm$0.5 & 3.0$\pm$2 \\ 
        2 & 5 & 1.4 & 0.57$\pm$0.1 & $-$15$\pm$7 & 280$\pm$100 & $2.8\times10^{3}$ & 0.25$\pm$0.02 & 1.7$\pm$0.5 & 3.0$\pm$2 \\ 
        2 & 6 & 1.9 & 0.59$\pm$0.3 & $-$16$\pm$20 & 310$\pm$400 & $2.8\times10^{3}$ & 0.25$\pm$0.02 & 1.7$\pm$0.5 & 3.0$\pm$2 \\ 
    \hline
        \multicolumn{3}{|c|}{True values}  & 0.25 & 1.57 & 2.47 & & 0.25 & 1.57 & 2.47 \\
    \hline
\end{tabular}
\end{table*}

In the next subsections we analyze and interpret $\pr(\avec|D,\kord,\kmax)$ using
the set of diagnostics summarized in Table~\ref{tab:diagnostic_list}.
Since the posterior is simple to compute for this linear model problem and 
prior Set~\Cprime, we can easily check computations using MCMC sampling against the
analytic results.
Marginalization in more complicated situations such as
nonlinear models will be discussed in Sec.~\ref{sec:case_studies}.
For Model~D and both priors in Fig.~\ref{fig:triangle_D1} we are sampling from a
posterior which is a ($k+1$)-dimensional Gaussian parametrized by the LEC
means and covariance matrix. We therefore use the LEC means and 1-$\sigma$
error bands (which are equivalent to 68\% DoB intervals since the posterior is Gaussian) 
in the subsequent diagnostic plots, but remind the reader that in general 
the posterior will not be Gaussian, and that many distributions will need to be
treated more carefully before resorting to a covariance analysis.

\subsection{Coefficient estimates and correlations} \label{subsec:triangleplots}

The marginalized posterior pdfs computed for $\kord=3$, $\kmax=3$, 
are shown in Figs.~\subref*{fig:triangle_D1_uniform} 
and \subref*{fig:triangle_D1_Gaussian} respectively for the uniform prior and for the prior Set~\Cprime\ 
with $\abarzero = 5$.
Projected posterior plots as in Fig.~\ref{fig:triangle_D1} 
help to visualize the correlations
among the extracted parameters, and are particularly useful for comparing the
effects of different priors (including a uniform prior). 
For example, it is evident that for $a_3$ in Fig.~\subref*{fig:triangle_D1_Gaussian}
 the marginalized one-dimensional posterior (on the diagonal) is simply returning the
Gaussian prior of width $\abarzero$ [the case illustrated in 
Fig.~\subref*{fig:prior-win}]; in Fig.~\subref*{fig:triangle_D1_uniform} the same posterior,
while still Gaussian, is very wide.  This leads to overfitting in the 
uniform-prior case, as $a_3$ is able to play off against other coefficients to push the
maximum of the likelihood to unnatural regions of parameter space.%
\footnote{Indeed, we generically
find problems with the least-squares extraction of a $a_\kord$ in a $\kord$th-order fit.
Presumably this is because $a_{\kord+1}$, with which $a_\kord$ is highly
anti-correlated, is artificially forced to zero in such a fit.}  
Note also that the parameters $a_0$ and $a_3$ are
strongly correlated without the prior but become largely uncorrelated with the
prior. This trend of decoupling low-order coefficients from poorly determined 
high-order ones via application of a prior continues as $k$ is increased. In the case
of the uniform prior, the overfitting becomes more pronounced as $k$ is increased.

Figure~\ref{fig:D1-ls-k4-fit} shows the results of a least-squares prediction 
where overfitting occurs at $\kord=\kmax=4$. 
The 1-$\sigma$ error bands indicate that the leading behavior is
poorly estimated, and that the fit is fine-tuned to reproduce the data where the
error band is smaller. Figure \ref{fig:D1-bayes-abar-5-k4-fit} shows the same prediction made using
Bayesian parameter estimates with the Gaussian naturalness prior.
The true leading-order behavior (manifested at low $x$) and the prediction for
$x$ above the last data point are much better reproduced, indicating
that overfitting is avoided.

Table~\ref{tab:D1-results} shows the central values and 68\% intervals obtained
for the three leading coefficients, $a_0$, $a_1$, and $a_2$ from both the
least-squares posterior pdf (left side) and the posterior pdf corresponding to the naturalness prior (right side).
Results are shown for fits at different $\kmax$ up to $\kmax=6$. 
The parameter estimates are obtained by MCMC integration and are consistent with
  the exact results in Tables 1 and 4 in Ref.~\cite{Schindler:2008fh}. The 
   details of the exact calculations for these simple priors can be found in Ref.~\cite{Schindler:2008fh}. 
   When computing the parameter estimates with MCMC, the error bars for the estimates were obtained
   from the ensemble of samples (see Sec.~\ref{sec:MCMCsampling}). These results for the parameter
   estimates with their error bars are presented with a precision that is a conservative estimate of the
   error of the MCMC method itself; by increasing the number of samples in these calculations, the
   estimates will approach the exact ones in Ref.~\cite{Schindler:2008fh}.
   We note that Table~4 of Ref.~\cite{Schindler:2008fh} contains an error:
   the logarithm of the posterior at the maximum parameter values quoted there for the Bayesian fit should be shifted
   down by one row. This mistake is corrected in the calculation that produced Table~\ref{tab:D1-results}.   
   
Our results confirm the finding of Ref.~\cite{Schindler:2008fh}: as the
number of parameters is increased, the least-squares estimates of the leading coefficients
degrade. As seen in Fig.~\ref{fig:D1-ls-k4-fit} for the $\kord=\kmax=4$ prediction, overfitting
occurs as the coefficients become correlated in order to reproduce the data.
This leads to large errors outside the fit region: compare the quality of the predictions there
in Figs.~\ref{fig:D1-ls-k4-fit} and \ref{fig:D1-bayes-abar-5-k4-fit}.

The left side of the table uses the $\chi^2/\textrm{dof}$ to indicate the quality of the fit, while
the right side gives the evidence $\pr(D|\kord,\kmax)$ to show the relative
extent to which the
model describes the data
(see Sec.~\ref{subsec:quality}). 
The evidence here is controlled by $\kmax$, as discussed in Sec.~\ref{subsec:quality}, and as 
we increase $\kmax$,  the naturalness prior
prevents fine-tuning by restricting the correlations between the leading
coefficients and higher-order ones as we saw in the lower left of
Fig.~\subref*{fig:triangle_D1_Gaussian}. For any value of $\kmax$, the first two
coefficients ($\kord=1$) are reliably extracted, with the posterior indicating when a
coefficient is not well-determined by the data.

\begin{figure}[tb]
  \includegraphics[width=0.99\columnwidth]{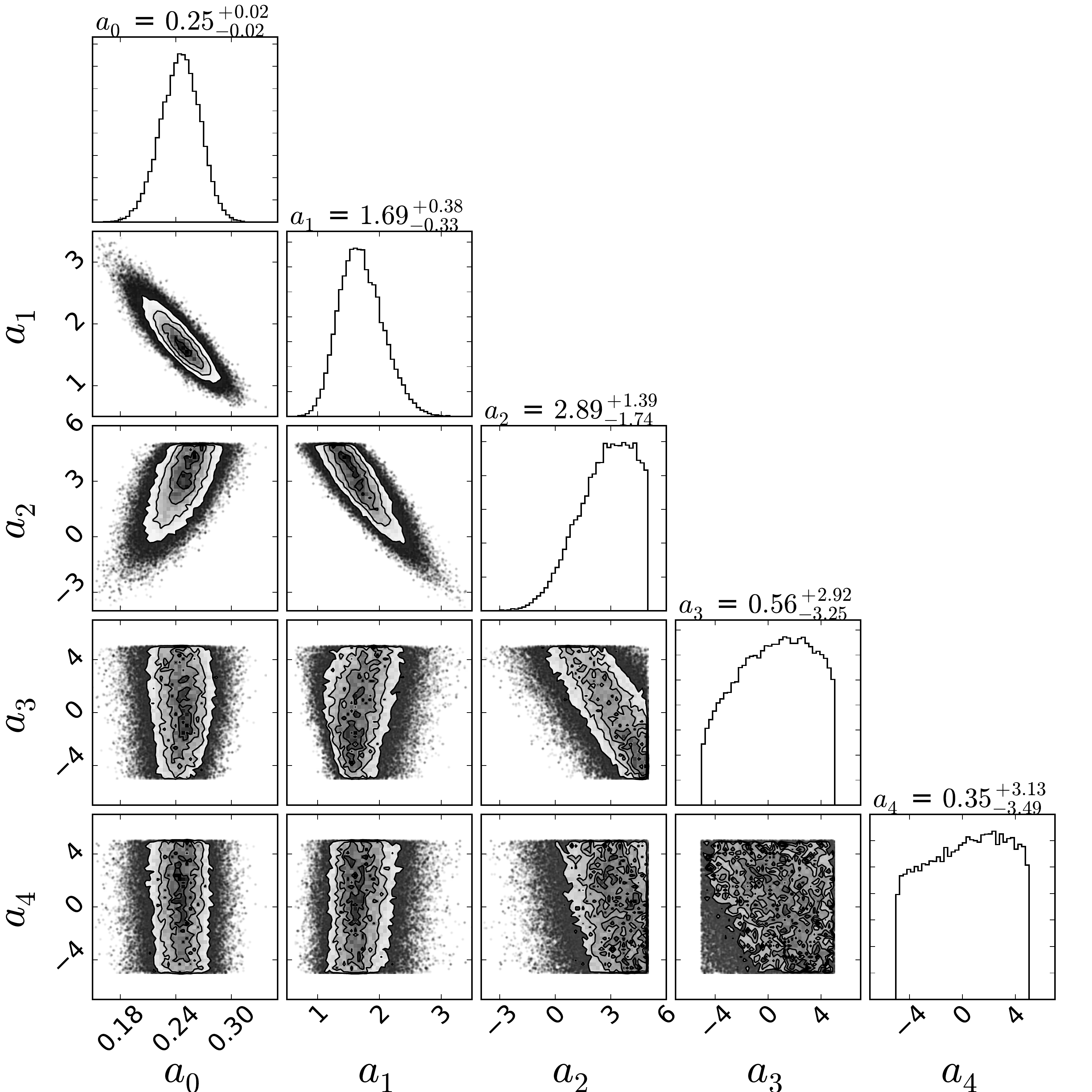}
  \caption{Projected posterior plot (see Fig.~\ref{fig:triangle_D1}) calculated at order $\kord=4$, $\kmax=4$ given 
  data set \dataset{D}{1}{5} using Prior A$'$ with $\abarzero=5$. 
    \label{fig:D15-post-k4-PriorA-abar-5}}
\end{figure}

In Fig.~\ref{fig:D15-post-k4-PriorA-abar-5}, we show a corresponding projected posterior
plot to Fig.~\subref*{fig:triangle_D1_Gaussian} but using prior Set~A$'$ 
(see Table~\ref{tab:priors}) with $\abarzero=5$.  
The marginalized posteriors on the diagonal
for the higher-order coefficients become quite skewed and highly non-Gaussian; this would significantly
complicate the determination of DoB intervals.  However, the posteriors for these
coefficients are primarily returning the prior---which in this case is flat. For the lowest-order coefficients,
the posteriors are Gaussian-like and their modes and widths agree with those from 
Fig.~\subref*{fig:triangle_D1_Gaussian} and the corresponding 
values listed in Table~\ref{tab:D1-results} up to $\kord=1$ when $\kmax=4$. We recover about the same
two-parameter correlation plot for $a_0$--$a_1$ as before, 
while those involving higher-order coefficients reflect details of the input prior.
(Note that if we chose a distributed prior for $\pr(\abar)$ and marginalized, these distributions
would be significantly smoothed; using a delta function represents an
extreme example.) Thus the meaningful results for this model are insensitive to the choice of
prior; it is only necessary that the prior restrict the range of higher-order
coefficients consistent with the expectations of naturalness.

\subsection{Prior diagnostics}

Our test cases make use of a fixed value of $\abar$
rather than marginalizing over a finite-width distribution for this naturalness parameter.  
This simplification is justified by using diagnostics for the prior that
explore the sensitivity to $\abar$.

\begin{figure}[tbh!]
  \includegraphics*[width=0.45\textwidth]{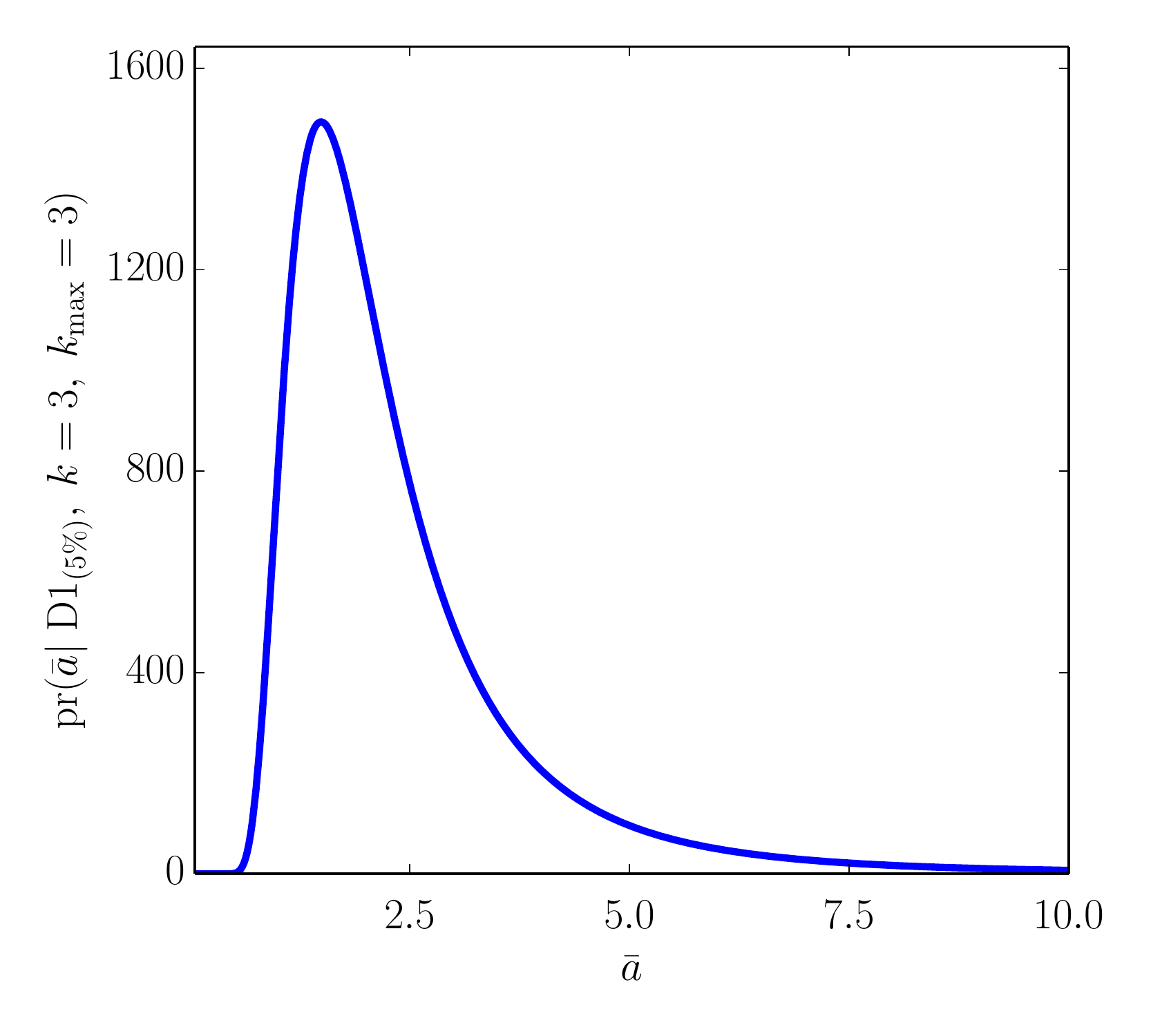}
  \caption{The posterior pdf $\pr(\abar|D,\kord,\kmax)$ calculated at $\kord=3$, $\kmax=3$ 
  using prior Set~C from Table~\ref{tab:priors} with
  $\abarmin=0.05$ and $\abarmax=20$, given
  data set \dataset{D}{1}{5}.
  \label{fig:D15-abar-post-k3-kmax3}}
\end{figure}

In general one could marginalize over a range of $\abar$ values, and 
the posterior for $\abar$, $\pr(\abar|D,\kord,\kmax)$, when used in conjunction with the $\abar$ relaxation plot discussed next, 
is then useful for identification of an appropriate marginalization range for $\abar$. The $\abar$ posterior will also
signal the presence of unnatural coefficients (see Sec.~\ref{subsec:blind}). 
\begin{widetext}
This quantity can be expressed using marginalization and Bayes' theorem as
\beq
 \pr(\abar|D,\kord,\kmax)  = \frac{1}{\pr(D|\kord,\kmax)} \int d\avec \, \int d\amarg \,
  \pr(D|\avec,\amarg,\kord,\kmax)
  \, \pr(\avec,\amarg|\abar,\kord,\kmax)\, \pr(\abar) \;.
  \label{eq:compute-abar-post}
\eeq
Note that this posterior can be obtained for different prior assumptions.
\end{widetext}
The posterior $\pr(\abar|\dataset{D}{1}{5},\kord=3,\kmax=3)$
is shown as a representative example in Fig.~\ref{fig:D15-abar-post-k3-kmax3},
in the case where the prior is Set~C.
The plot shows the region most likely for $\abar$, which in this case
implies natural coefficients. 

\begin{figure}[bth!]
  \includegraphics*[width=0.98\columnwidth]{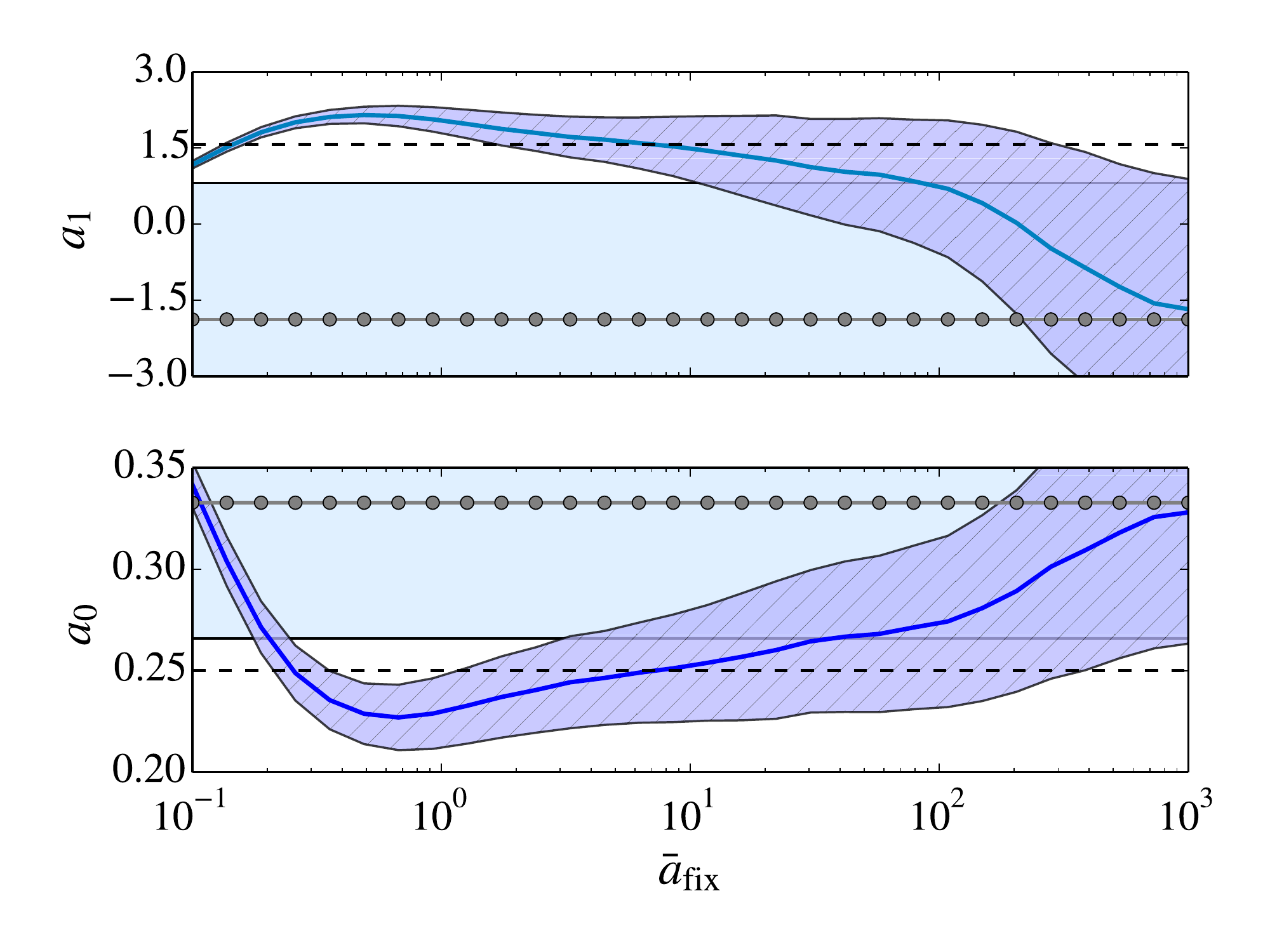}
  \caption{(color online) 
    Bayesian coefficient estimates calculated at $\kord=1$, $\kmax=4$ (solid lines with darker
    hatched error bands) as a function of $\abarzero$ using prior Set \Cprime\
    given \dataset{D}{1}{5}. The constant line with circles
    with lighter solid error bands is the least-squares estimate, which is independent of $\abarzero$. 
    The error bands represent 68\% DoBs (1-$\sigma$ errors).
  \label{fig:D1_abar_relaxation}}
\end{figure}

Another way to check the sensitivity of results to the 
naturalness parameter is to examine the parameter estimates when $\abar$ is fixed to various values $\abarzero$. 
Hence we now employ prior Set~\Cprime\ and vary $\abarzero$
from a small value to one large enough that the prior is effectively uniform.
We use such results to create
an $\abar$ relaxation plot (Fig.~\ref{fig:D1_abar_relaxation}), which shows the
projected posterior mean and width of different parameters as a
function of a fixed value $\abarzero$ for $\abar$.
$\abarzero$ sets the width of the allowed region for the 
EFT parameters:
Fig.~\ref{fig:D1_abar_relaxation} shows that too-small values
of $\abar$ (e.g., of order 0.1) will bias the posterior severely, 
while large values relax to the least-squares (uniform prior) result.  

Such ``$\abar$ relaxation" plots should be interpreted with care though, because a marginalization over $\abar$
when computing the coefficient posterior
will weight different regions of $\abar$ in the integration according to the
posterior $\pr(\abar|D,\kord,\kmax)$. This can be seen by using the rule of marginalization
to express the coefficient posterior from Eq.~\eqref{eq:post-marg-abar0} in an alternate form:
\beq
  \begin{split}
  \pr(\avec|& D,\kord,\kmax)  = \\ 
  & \int d\abar \, \pr(\avec|\abar,D,\kord,\kmax) \pr(\abar|D,\kord,\kmax)\;,
  \end{split}
\eeq
where the first term in the integrand is the posterior for the coefficients
given a specific $\abar$ and the second term is the posterior for $\abar$
that was calculated in Eq.~\eqref{eq:compute-abar-post}.
 The result of marginalizing over $\abar$ therefore
cannot directly be read from an $\abar$ relaxation plot. 
Figure~\ref{fig:D1_abar_relaxation}
simply shows how the results change for fixed values of $\abar=\abarzero$.

Ideally one finds (and marginalizes over)
a slowly varying region in $\abar$ that is consistent with naturalness expectations;
there should be little sensitivity to the endpoints of this region.
There is sometimes a plateau in the $\abarzero$ dependence, but not always. If there is a plateau, and coefficient estimates are largely independent of $\abarzero$ in the $\abar$ region
where $\pr(\abar|D,\kord,\kmax)$ is significant, then marginalization will be equivalent to 
fixing $\abar$ at an $\abarzero$ in this range. 
Taking a value for $\abarzero$ somewhat above the peak region
for $\abar$ in the $\abar$ posterior then provides a choice which is not overly restrictive. 
In the present case, we conclude that the model in question will not be
sensitive to details of how $\abar$ is marginalized in a wide region, 
and for simplicity we fix $\abarzero = 5$.

\begin{figure}[tbh!]
  \includegraphics*[width=0.98\columnwidth]{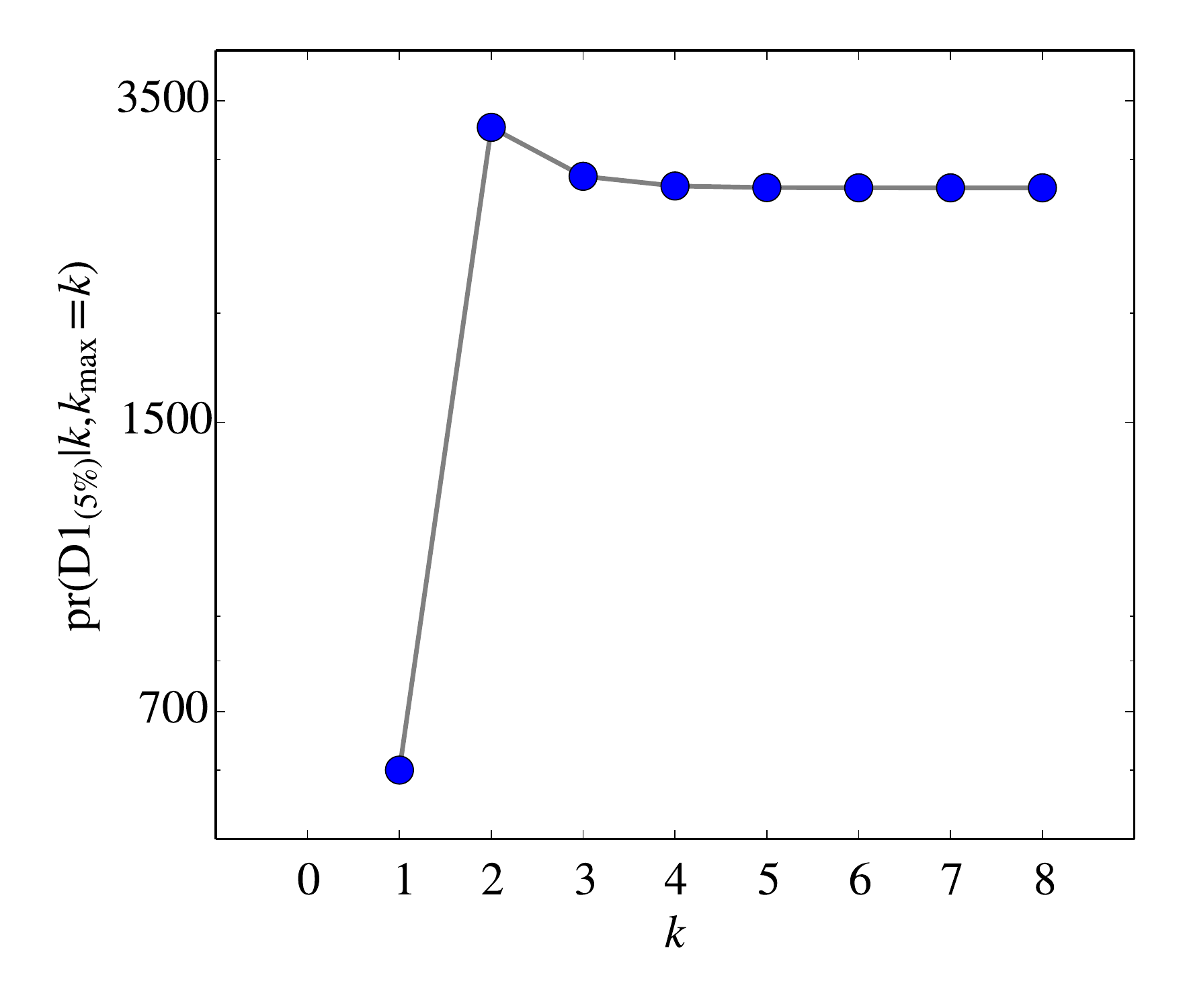}
  \caption{(color online) 
  Evidence $\pr(\dataset{D}{1}{5}|\kord,\kmax=\kord)$ for several
    values of $\kord$ using prior Set~\Cprime\ with $\abarzero=5$.
    (The evidence is not shown for $\kord=0$ since it is nearly zero).
  \label{fig:D15_evidence}}
\end{figure}

\begin{figure*}[pt]

  \subfloat{%
    \label{fig:xmax_D1_M1}%
  \includegraphics[width=0.45\textwidth]{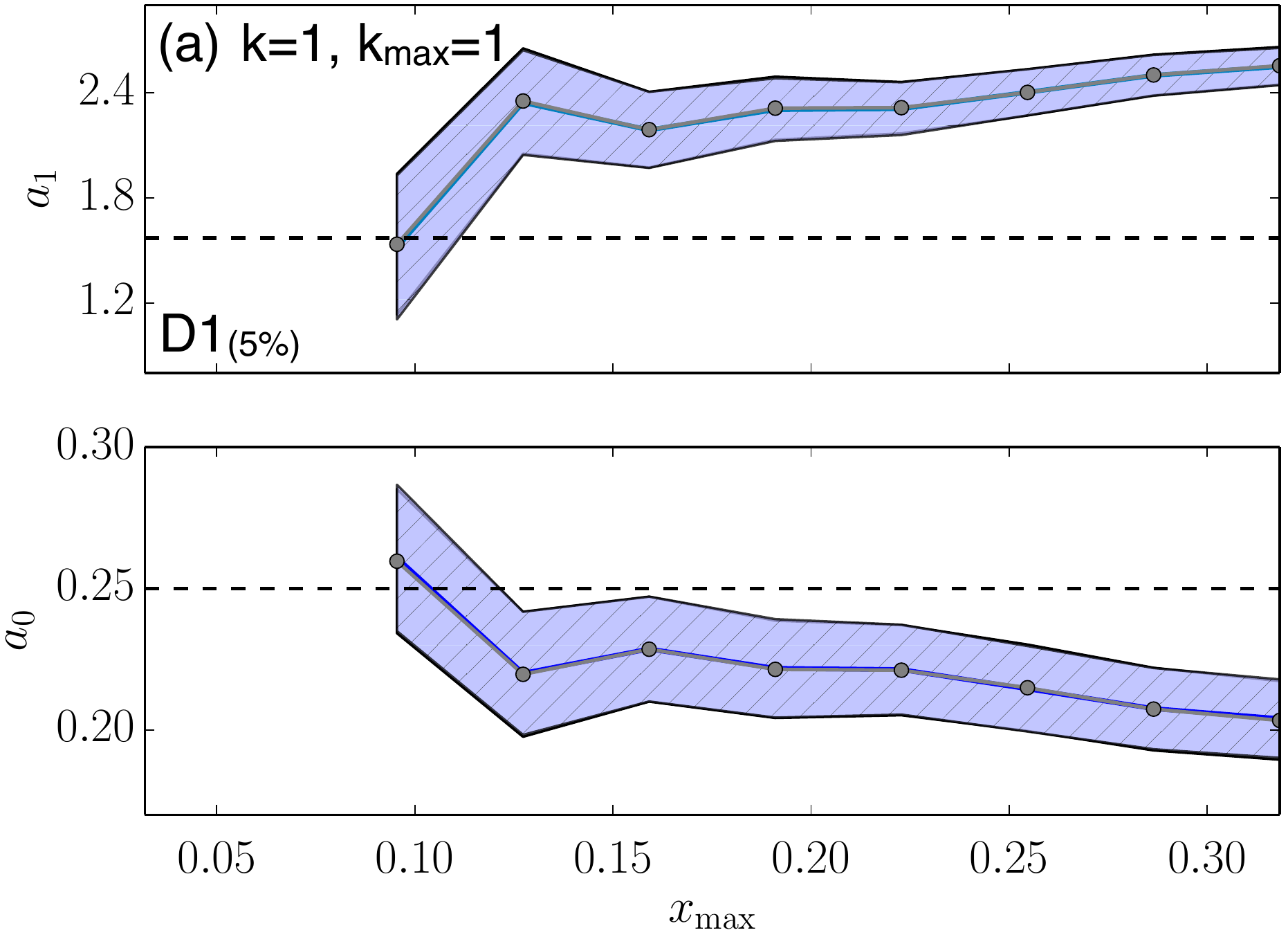}%
  }
  \hspace*{0.05\textwidth}  
  \subfloat{%
    \label{fig:xmax_D1_M2}%
  \includegraphics[width=0.45\textwidth]{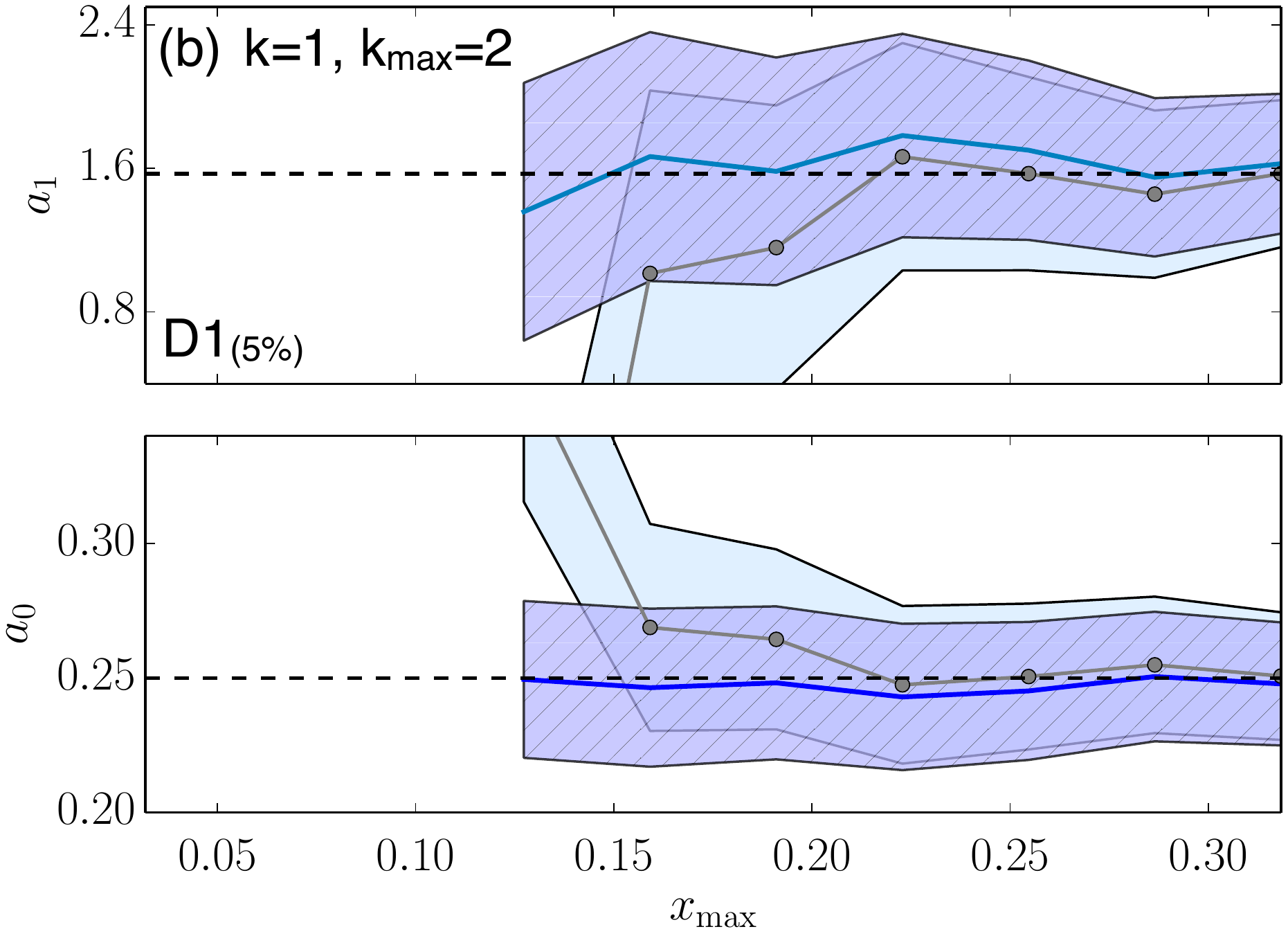}%
  } 
  \\
  \subfloat{%
    \label{fig:xmax_D1_M3}%
  \includegraphics[width=0.45\textwidth]{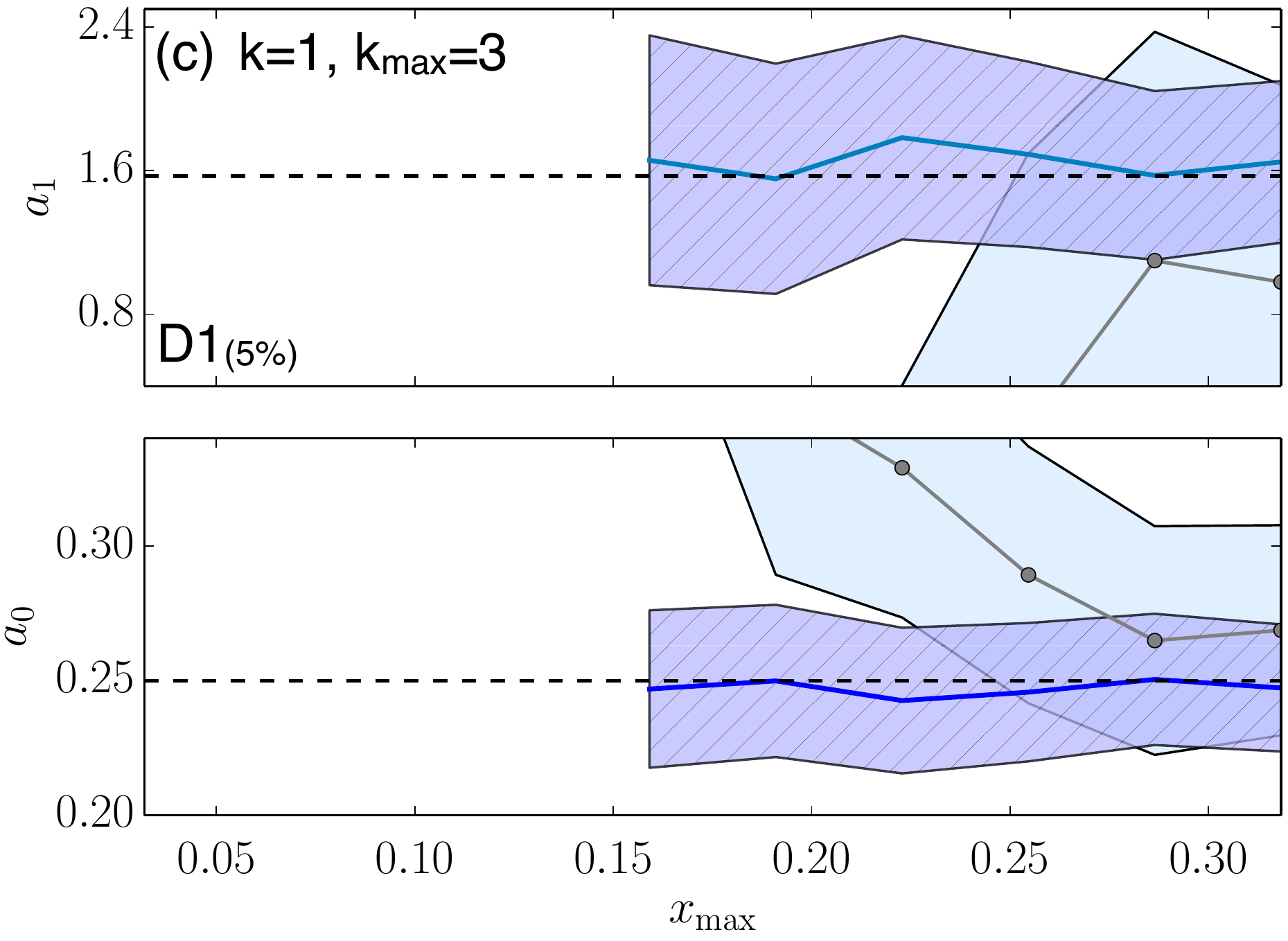}%
  }
  \hspace*{0.05\textwidth}  
  \subfloat{%
    \label{fig:xmax_D1_M4}%
  \includegraphics[width=0.45\textwidth]{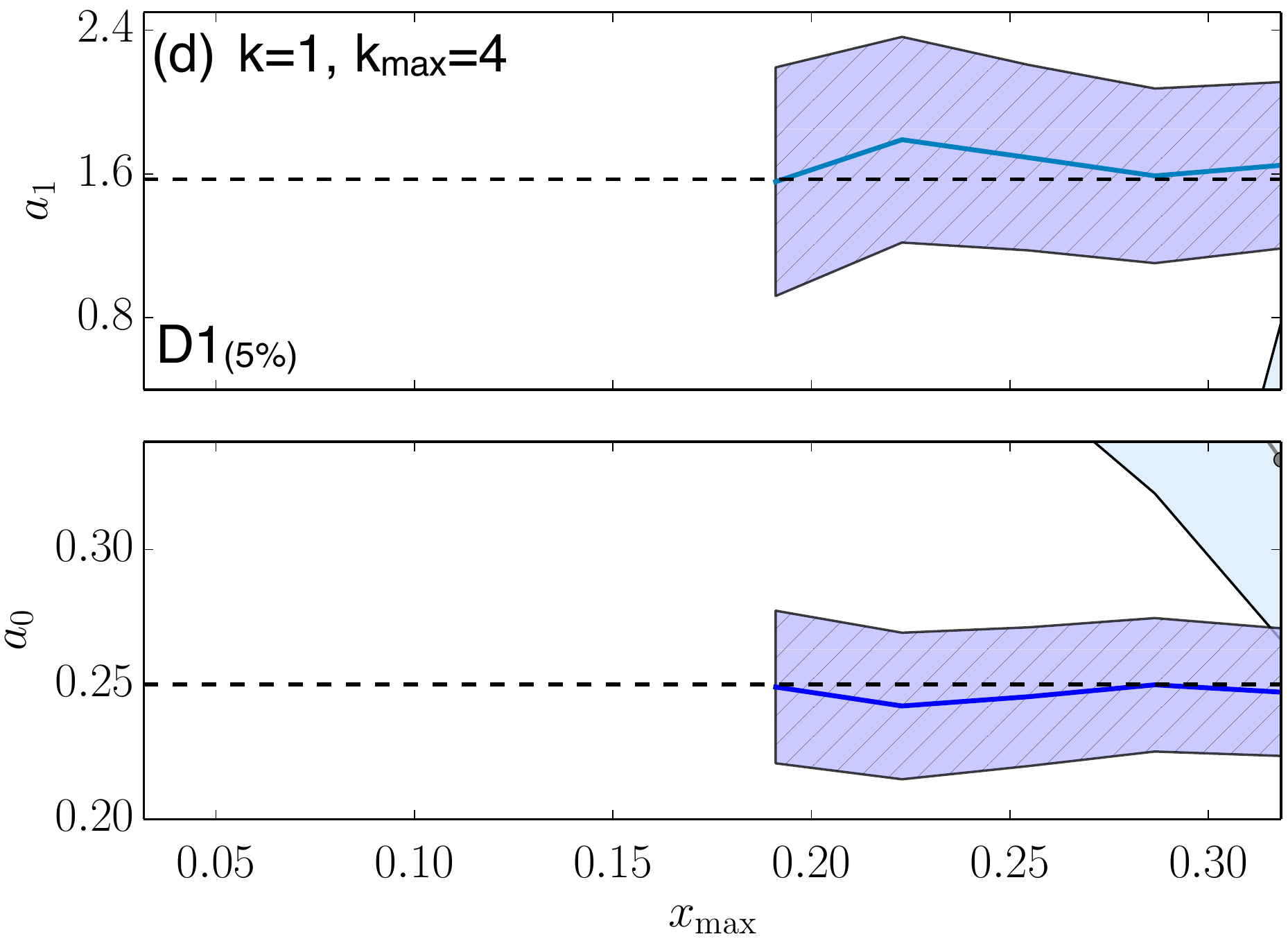}%
  } 

  \caption{(color online) 
    Bayesian coefficient estimates
    as data from data set \dataset{D}{1}{5} are sequentially added at the
    high-$x$ end. The largest $x$-value in the set is denoted as $\xmax$.
    The lines with darker hatched error bands represent estimates
    using prior Set \Cprime\ with $\abarzero=5$, and the line with circles
    with lighter solid error bands represents the least-squares estimates. 
    The error bands represent 68\% DoBs (1-$\sigma$ errors), which coincide in (a).
  \label{fig:xmax_D15}}
\end{figure*}

\subsection{Model quality: Evidence} \label{subsec:quality}

We discussed the evidence, which we define as $\pr(D|\kord,\kmax)$, in general 
terms in Sec.~\ref{subsec:bayes_factor}. In cases where the
high-order coefficients are constrained by the naturalness prior, as in the
situation reflected in Eq.~\eqref{eq:bayes_factor2}, we expect saturation behavior 
of the evidence as
parameters that are not well-constrained by the likelihood are added to the
model. The evidence for a model at order $\kord$ marginalized over
coefficients up to order $\kmax$ is
\beq
  \begin{split}
   \pr(D|\kord,\kmax)  = & \int d\abar \int d\avec \int d\amarg \\
  & \null\times \pr(D|\avec,\amarg,\kord,\kmax) \\[0.05in] 
  & \null\times \pr(\avec,\amarg|\abar,\kord,\kmax)\pr(\abar) \;,
  \end{split}
  \label{eq:evidence_linear_marg}
\eeq
where we have used the rule of marginalization from \ref{subsec:marginalization}, 
the fact that the prior for the naturalness parameter is independent of the 
truncation and marginalization orders. 

Just as in the discussion of the impact of $\kord$ and $\kmax$ on the posterior 
of $\avec$ in Sec.~\ref{subsec:setup},
our evidence calculations are also controlled solely by $\kmax$.
For any of our models calculated at $\kord$ with $\kord \leq \kmax$,
Eq.~\eqref{eq:evidence_linear_marg} collapses to an integral over all the 
coefficients up to $\kmax$:
\beq
  \begin{split}
   \pr(D| \kord\leq\kmax,&\kmax) = \int d\abar \int d\avec  \\
  & \times \pr(D|\avec,\kord=\kmax,\kmax)  \\
  & \times \pr(\avec|\abar,\kord=\kmax,\kmax)\,\pr(\abar) \;,
  \end{split}
  \label{eq:evidence_linear}
\eeq
i.e., the value of $\kord$ is irrelevant so long as it is less than or equal to $\kmax$
since all the higher-order coefficients up to $\kmax$ must also be integrated.
This result also applies for models which are nonlinear in the coefficients. For this reason we
compute only the evidence calculated at $\kord$, $\kmax = \kord$, e.g., as in Fig.~\ref{fig:D15_evidence}.
When the likelihood in Eq.~\eqref{eq:evidence_linear_marg} as a 
function of $\amarg$ is more complicated, these integrals
will not necessarily collapse so simply and the evidence may be affected separately by
$\kord$ and $\kmax$, as discussed in Sec.~\ref{subsec:setup}.

Comparing the evidence of models as parameters are added (or as we marginalize
over more high-order coefficients) provides an ideal order-by-order
comparison for the suitability of a model to describe the data~\cite{Gregory:2005,Sivia:2006}. 
Therefore when we perform parameter estimation in our Bayesian framework, we also
compute the evidence for each model to quantitatively decide how many
terms can be extracted from the data, and how many higher-order terms to
marginalize.

Figure~\ref{fig:D15_evidence} shows the evidence $\pr(\dataset{D}{1}{5}|\kord, \kmax=\kord)$
for $\kord=1$ to $\kord=8$ using prior Set~\Cprime\
with $\abarzero = 5$. The evidence is also tabulated for each of these orders
in Table~\ref{tab:D1-results}. Comparing different orders, we
see that the Bayes ratio 
$\pr(\dataset{D}{1}{5}|\kord=2)/\pr(\dataset{D}{1}{5}|\kord=1) \approx 5 $,
implying that the quadratic model is somewhat more favorable than the 
linear one.%
\footnote{We follow the empirical scale in Table~1 of Ref.~\cite{Trotta:2008qt},
for which the thresholds for weak, moderate, and strong evidence favoring one
model over another are ratios of 3, 12, and 150, respectively.} 
Comparing higher orders, the Bayes ratio for the $\kord=3$ to $\kord=2$
case is about 1, indicating that they are equally favorable. This we define as evidence
saturation, based on the discussion in Sec.~\ref{subsec:bayes_factor},
where going to higher order in the model does not improve the description of the data.
When doing the parameter estimation for our examples, if $k+1$ coefficients are expected to be
determined by the data, the order $\kmax > k$ should not affect these $k+1$
estimates; any coefficients above order $k$ will essentially return the prior. 
However, verifying that this is true is an important part of the parameter 
estimation procedure. 
As emphasized in Ref.~\cite{Lepage:2001ym}, simply truncating the fit at order $k$ amounts
to taking a delta-function prior for all coefficients of higher-order terms, since
it assumes they are precisely zero. 

\begin{figure*}[tbh]
  \subfloat{%
    \label{fig:D15_datasets}%
  \includegraphics[width=0.45\textwidth]{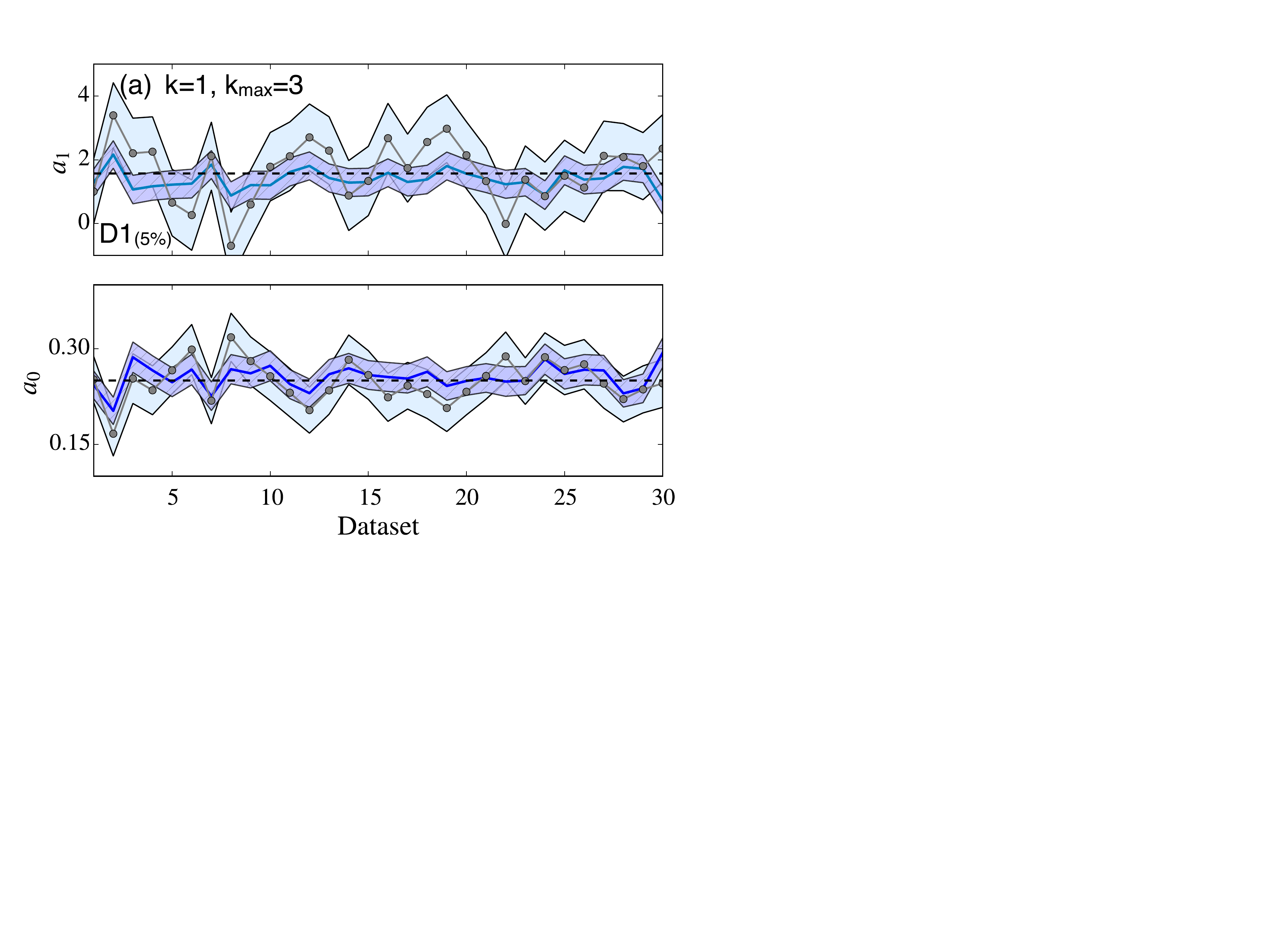}%
  }
  ~~~%
   \subfloat{%
     \label{fig:D15_accumulated}%
     \includegraphics[width=0.45\textwidth]{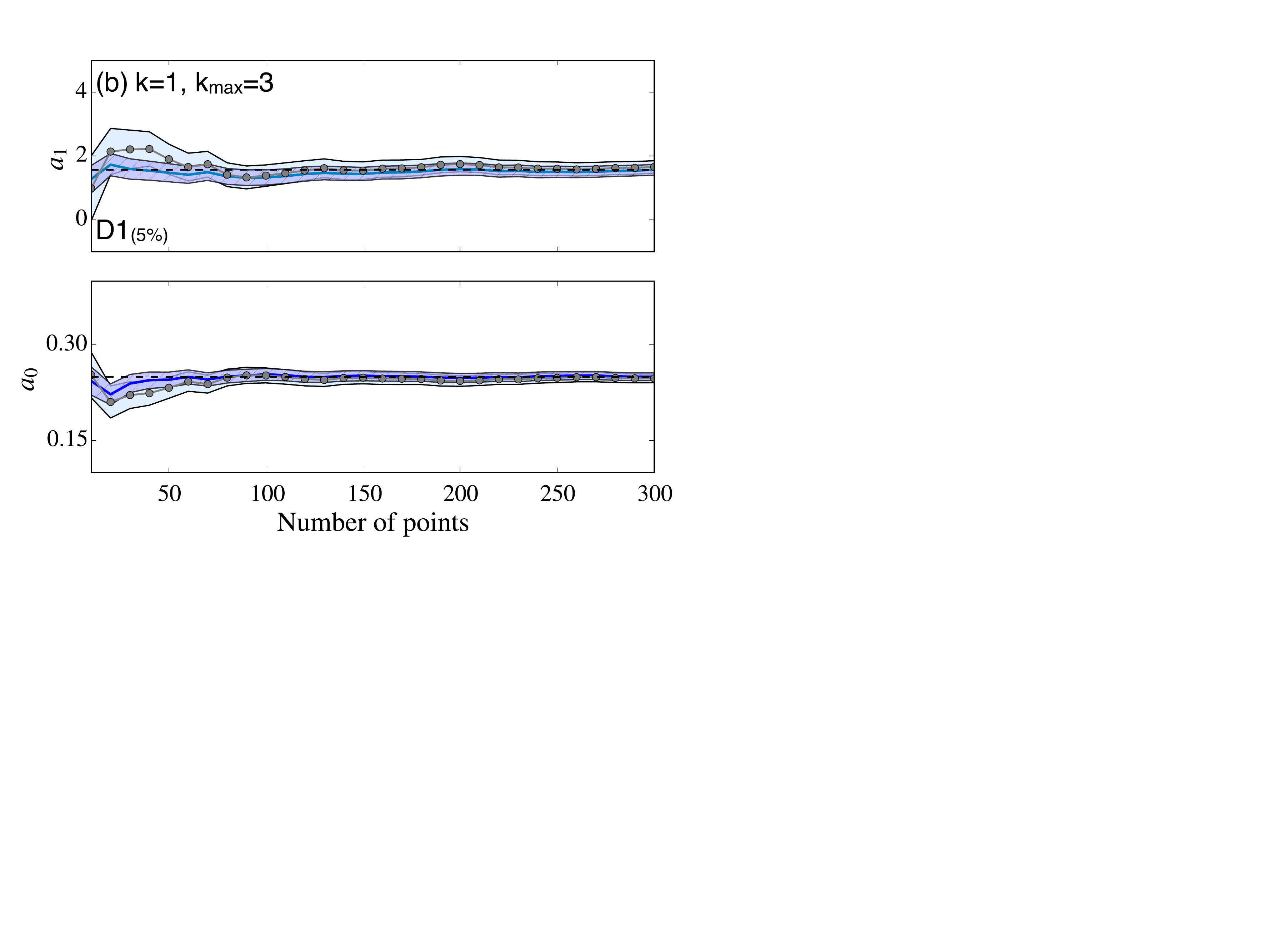}
   }
   \caption{Multi-set (a) and accumulation plots (b)
   calculated at $\kord=1$, $\kmax=3$. 
   The shaded regions denote 68\% error bands for the uniform (line with circles with lighter solid band)
   and naturalness prior (line with darker hatched band).
   The data sets used in (a) are 30 samples on the \dataset{D}{1}{5} mesh of 10 points. 
   The same data are accumulated set by set to generate (b).
   In each case the prior was Set~\Cprime\ with $\abarzero = 5$. \label{fig:D15-dataset-accum}}
\end{figure*}

\subsection{$\xmax$ plots}

Variable $\xmax$ plots show the projected maxima and widths of the parameters
(which, for this problem, are the means and 1-$\sigma$ errors)
as a function of the maximum parameter value to which data are fit. 
In this way we examine the
coefficients as we sequentially add data sampled at larger values of $x$,
where higher-order contributions become increasingly important. The effect on the
computed posterior can be seen from Eq.~\eqref{eq:objective-func}, where the data
at different values of $x$ form the residuals of the likelihood
in Eq.~\eqref{eq:least-squares-likelihood}, which goes into the coefficient
posterior calculation. Plotting the evolution
of the parameters as a function of $\xmax$ shows the influence of each datum on
the stability of the posterior in each dimension. Projected posterior plots such as those
in Fig.~\ref{fig:triangle_D1} would be necessary
to further analyze the correlations between orders at each value of $\xmax$.

A series of such plots are shown in Fig.~\ref{fig:xmax_D15} for data set
\dataset{D}{1}{5} at four different fixed orders. The data set can be seen
in Figs.~\ref{fig:D1-ls-k4-fit} and \ref{fig:D1-bayes-abar-5-k4-fit} where we
plot predictions from parameter estimates using different priors on the
coefficients. We plot the $k=1$ parameter results
as the order of the marginalization $\kmax$ is increased, 
starting in Fig.~\subref*{fig:xmax_D1_M1}
with $k=1$ (no higher-order coefficients) and going up to $\kmax=4$ in
Fig.~\subref*{fig:xmax_D1_M4}. We plot the estimates with the naturalness
prior as solid lines with error bands and compare these results with
the least-squares results, plotted as lines with circles.

Some observations:
\bi
 \I The $\kmax = 1$ plot illustrates underfitting: the linear model
 only works well for the most infrared data (here the smallest $\xmax$ point only)
 and significantly deviates from the true result elsewhere.
 Consequently, the result for larger $\xmax$ shows no stability.
 The prior for $a_0$ and $a_1$ is seen to be irrelevant here, and the least-squares
 results are the same as the estimates with the naturalness prior.
 \I The $\kmax = 2$ plot with no prior shows \emph{overfitting} for the
 lowest values of $\xmax$, as there are too many terms available for
 the fit data, given the size of the (simulated) experimental error.
 As $\xmax$ increases, the least-squares result becomes reliable.
 With the naturalness prior included, there is stability for $a_0$ and $a_1$
 over the entire range.  
 This marginalization over $a_2$ is the key feature for stability with $\xmax$.
 \I The $\kmax = 3$ plot shows that the uniform-prior results are off
 the scale for much of the displayed $\xmax$ range, in strong contrast to the result with the
 naturalness prior, which gives
 stable central values and 1-$\sigma$ errors. These naturalness-prior results
 are similar to the corresponding results when $\kmax = 2$.  These patterns
 continue---for both the uniform-prior and naturalness-prior results---when $\kmax=4$ [shown
 in Fig.~\subref*{fig:xmax_D1_M4}] and beyond (not shown).
\ei
This example illustrates the utility of using $\xmax$ plots to check for
instability with respect to the data range, which can signal
underfitting and/or overfitting.

The stability with $\xmax$ indicates that all high-$x$ (UV) effects have been sufficiently
accounted for, and that adding more high-$x$ data does not improve the estimates.
The $k$ value at which these particular extractions become stable
with respect to $\xmax$ corresponds to that at which the evidence saturates. In this example,
stability with $\xmax$ begins when $\kmax=2$, which is where the evidence saturates in
Fig.~\ref{fig:D15_evidence}. In fact, we will see in  Sec.~\ref{subsec:pastbreakdown} that
$\xmax$ plots can show stability when the evidence is past the saturation peak.
This emphasizes the need to examine all the diagnostics before reaching a conclusion
on the validity of the parameter estimation.

\subsection{Multi-set and accumulation} \label{sec:multiset-accum}

\begin{figure}
  \includegraphics[width=0.48\textwidth]{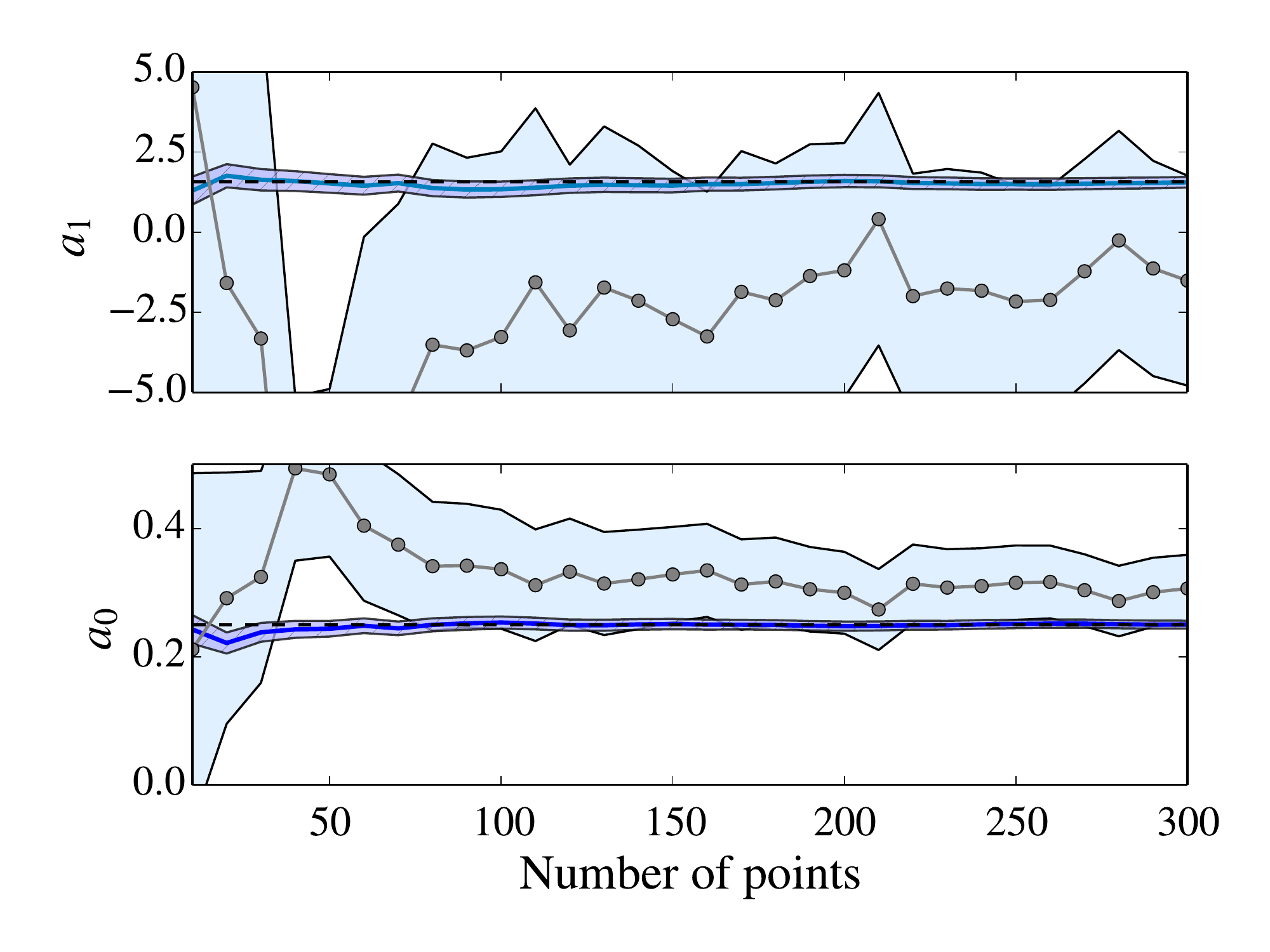}
  \caption{Accumulation plot for \dataset{D}{1}{5} at $\kord=1$, $\kmax=6$. Same implementation
  as Fig.~\protect\subref*{fig:D15_accumulated} but with
  $\kmax=6$. \label{fig:D15-accum-k1-kmax6}}
\end{figure}

Multi-set [Fig.~\subref*{fig:D15_datasets}] and accumulation
[Fig.~\subref*{fig:D15_accumulated}] plots are useful when there is enough data to
subdivide it into a collection of smaller but (roughly) equivalent data sets.
Multi-set plots provide a visualization of how fluctuations in the data affect
the parameters. Figure~\subref*{fig:D15_datasets} illustrates the fluctuations of
the parameters $a_0$ and $a_1$, estimated with the naturalness prior, as compared to the least-squares
estimates, for the case $\kmax = 3$. For 30 data sets sampled on the same
grid with the same random error as \dataset{D}{1}{5}, we compare the maxima and
width of the projected posteriors for each coefficient. This plot illustrates
how large the fluctuations in the coefficients are as the data fluctuates. The added restriction from the naturalness prior reduces
both the spread in the calculated central values and the size of the coefficients' error,
compared to the result using the uniform prior.

Accumulation plots illustrate the utility of the prior when few data are available. Reading
Fig.~\subref*{fig:D15_accumulated} from left-to-right shows that when there is less data available,
the naturalness prior increases the precision of the estimates. Once there is enough data the
uniform prior and naturalness prior results do not differ since the likelihood is the dominant
component of the posterior. However, the danger with a uniform prior is that overfitting can
degrade the parameter estimation, even when there is a large amount of data, see Fig.~\ref{fig:D15-accum-k1-kmax6}.
Even with a large data set the uniform prior on included coefficients---and $\delta$-function
prior on omitted coefficients---means care must be taken in choosing $\kmax$, so that neither overfitting or underfitting takes place. 
Bayesian parameter estimation avoids this delicate selection,
because the estimates of the low-order coefficients are stable with respect to $\kmax$.

\subsection{Residual plots}

\begin{figure}[tb!]
  \includegraphics[width=0.48\textwidth]{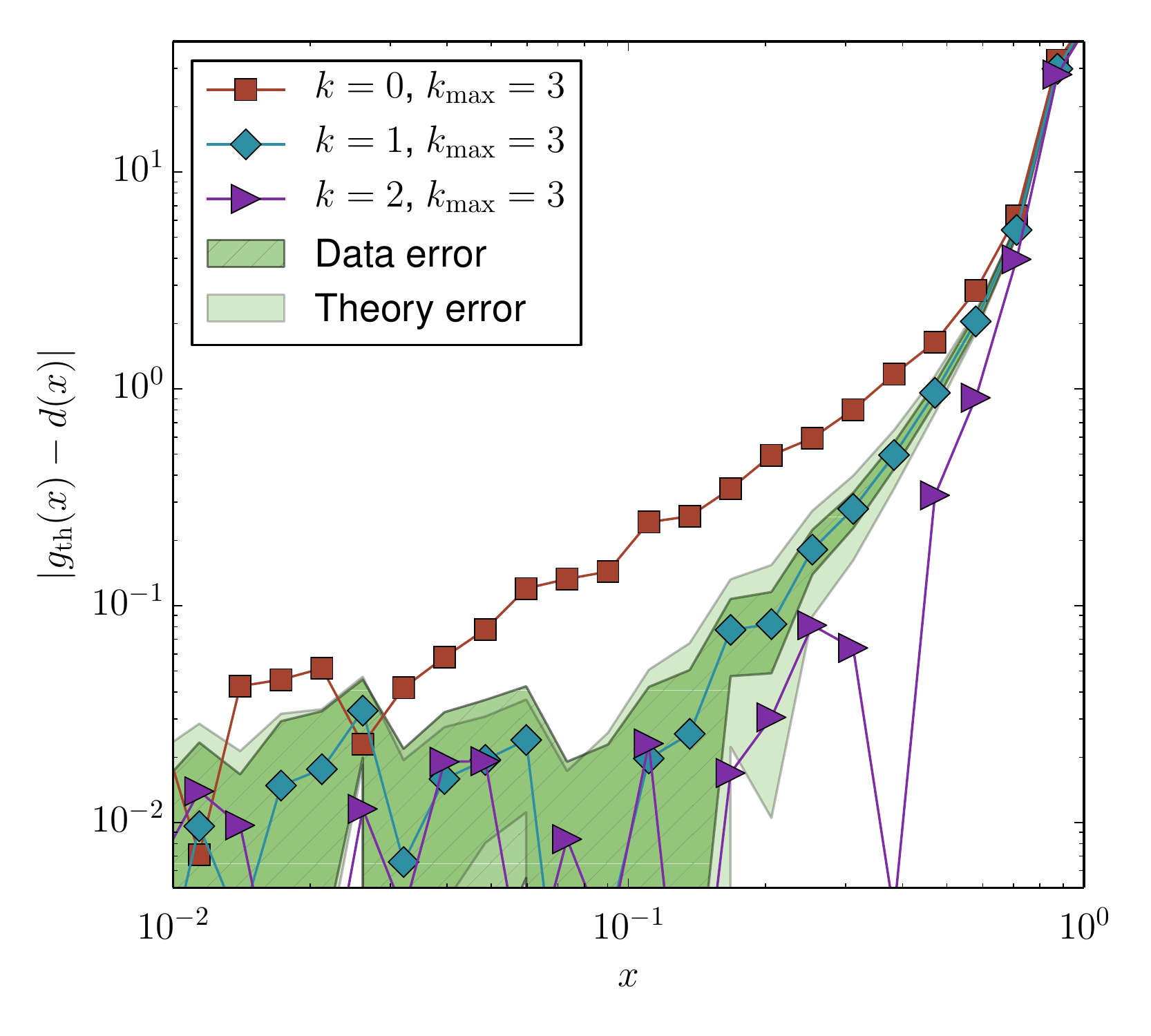}
  \caption{Plot of residuals for predictions of Model~D 
  at leading order ($\kord=0$), next-to-leading order ($\kord=1$), 
  and next-to-next-to-leading order ($\kord=2$),
  all with $\kmax=3$, given
  \dataset{D}{1}{5}. The 68\% error bands from both theory and data are shown on
  the $\kord=1$ prediction as a representative example. 
  \label{fig:D15-Lepage-kmax3-k0-to-2}}
\end{figure}

For sufficiently small values of the expansion parameter $x$, the truncation
errors for an EFT calculated to order $k$ should scale like the first
omitted term (e.g., like $x^{k+1}$ if all orders are present).
This can be tested by a log-log plot of the absolute values of the residuals
(difference of prediction and data) as a function of $x$.
Plots of these type were introduced to analyze EFT behavior  in
Ref.~\cite{Lepage:1997cs} and are commonly called ``Lepage plots''. 
A successful realization of the EFT should ideally
reveal a clear signal of power-law behavior,
with the slope given by the order in the expansion, in an extended region of $x$.
This signal can be masked by data errors, particularly at low $x$,
and by still higher-order truncation errors as the breakdown scale is neared.
A Lepage plot can manifest the different regions, help to disentangle
regulator artifacts from errors due to truncating the EFT Hamiltonian,
and, in principle, approximately determine the breakdown scale~\cite{Furnstahl:2014xsa}.

\begin{figure*}[p!]
  \includegraphics*[width=0.85\textwidth]{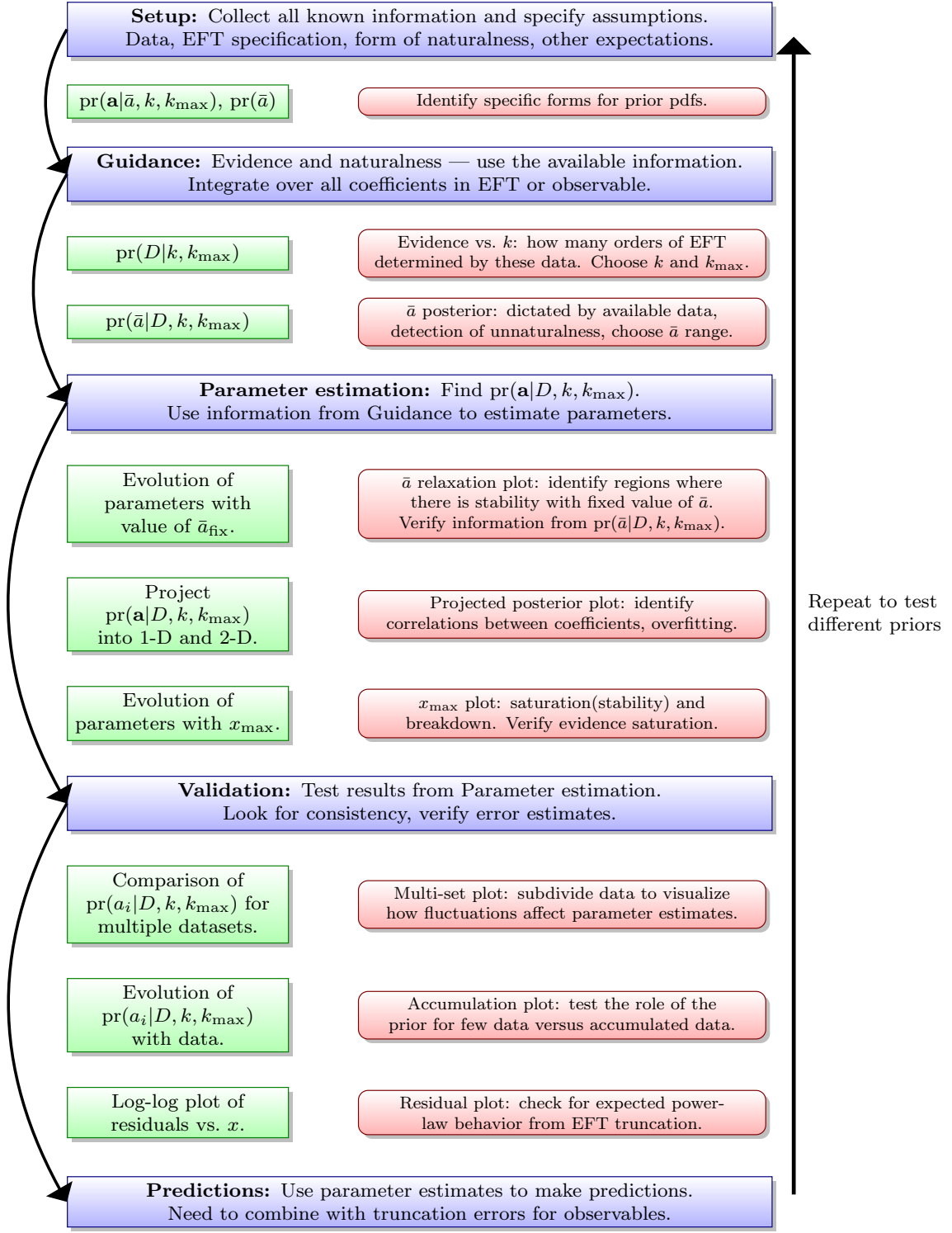}
  \caption{(color online) Flowchart for parameter estimation applying 
  diagnostic tools from Table~\ref{tab:diagnostic_list}. 
  }
  \label{fig:flowchart}
\end{figure*}

We show an example of such a plot 
for Model~D using \dataset{D}{1}{5} in Fig.~\ref{fig:D15-Lepage-kmax3-k0-to-2}.
The three sets of points are residuals at orders $\kord = 0$, 1, and 2,
from a parameter estimation at $\kmax=3$.
Data and theoretical error bands have been added for the $\kord=1$ residuals
to help identify the crossover between the region where data errors
mask the signal of the first omitted term and where the anticipated
power-law behavior should be seen.
In this example, the crossover occurs at successively higher values of $x$,
with at best a very narrow region of power-law behavior for $\kord=2$.

In practice data are frequently sparse and too noisy to robustly discern 
from Lepage plots whether power-law exponents are in accordance with
EFT expectations. 
A different type of error plot that avoids this problem
compares theory residuals from different values of an
EFT regulator scale (i.e., cutoff) rather than between 
theory and data~\cite{Griesshammer:2015osb,Griesshammer:2004pe,Bedaque:2002yg}.  
Examples and further discussion of residual plots are given in 
Ref.~\cite{Furnstahl:2014xsa} and we do not consider them further in the
present work.


\section{The Parameter Estimation Process}
\label{sec:model-problems}

\subsection{Overview}  \label{sec:model-H}

In Fig.~\ref{fig:flowchart}, we present a possible flowchart for the full
parameter estimation process, which orders and builds on the diagnostic tools
described in the last section.
The process starts with the Setup of the model and specification
of all available information and theoretical expectations (including
the form of the priors but not their widths), 
and ends with Predictions of observables from the fit parameters
(or, more precisely, from the posteriors for the parameters).
The intermediate steps are divided into Guidance, Parameter estimation,
and Validation.
In this section we consider each of these in turn, not exhaustively, but 
to highlight how individual diagnostics can offer different
insights into parameter estimation for EFTs.
We emphasize that while we have for clarity described the process as a forward flow,
in practice one would backtrack if later diagnostics do not support
earlier conclusions.

\begin{figure}[tbh!]
    \includegraphics[width=0.4\textwidth]{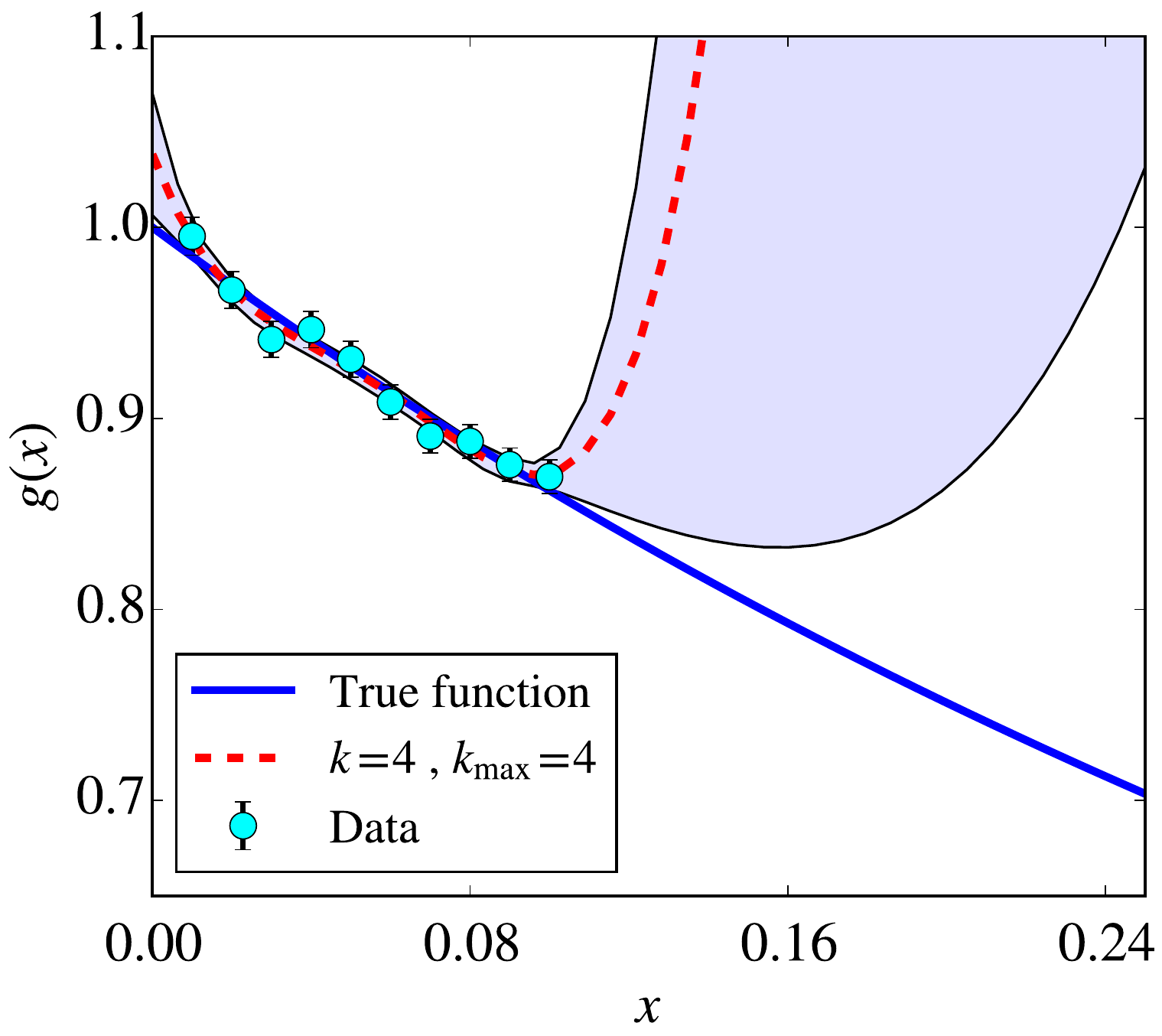}
    \caption{(color online)
    Comparison of data set \dataset{H}{0}{1} (corresponding to the first row of
    Table~\ref{tab:model-H-labels}), the underlying function for Model~H from
    Eq.~\eqref{eq:model-H}, and a least-squares prediction calculated at order
    $\kord=4$, $\kmax=4$ from that data set. The error bands represent 1-$\sigma$ (68\% DoBs).
    \label{fig:H0_pred_uniform_k4}}
\end{figure}

\begin{figure}[tbh!]
    \includegraphics[width=0.4\textwidth]{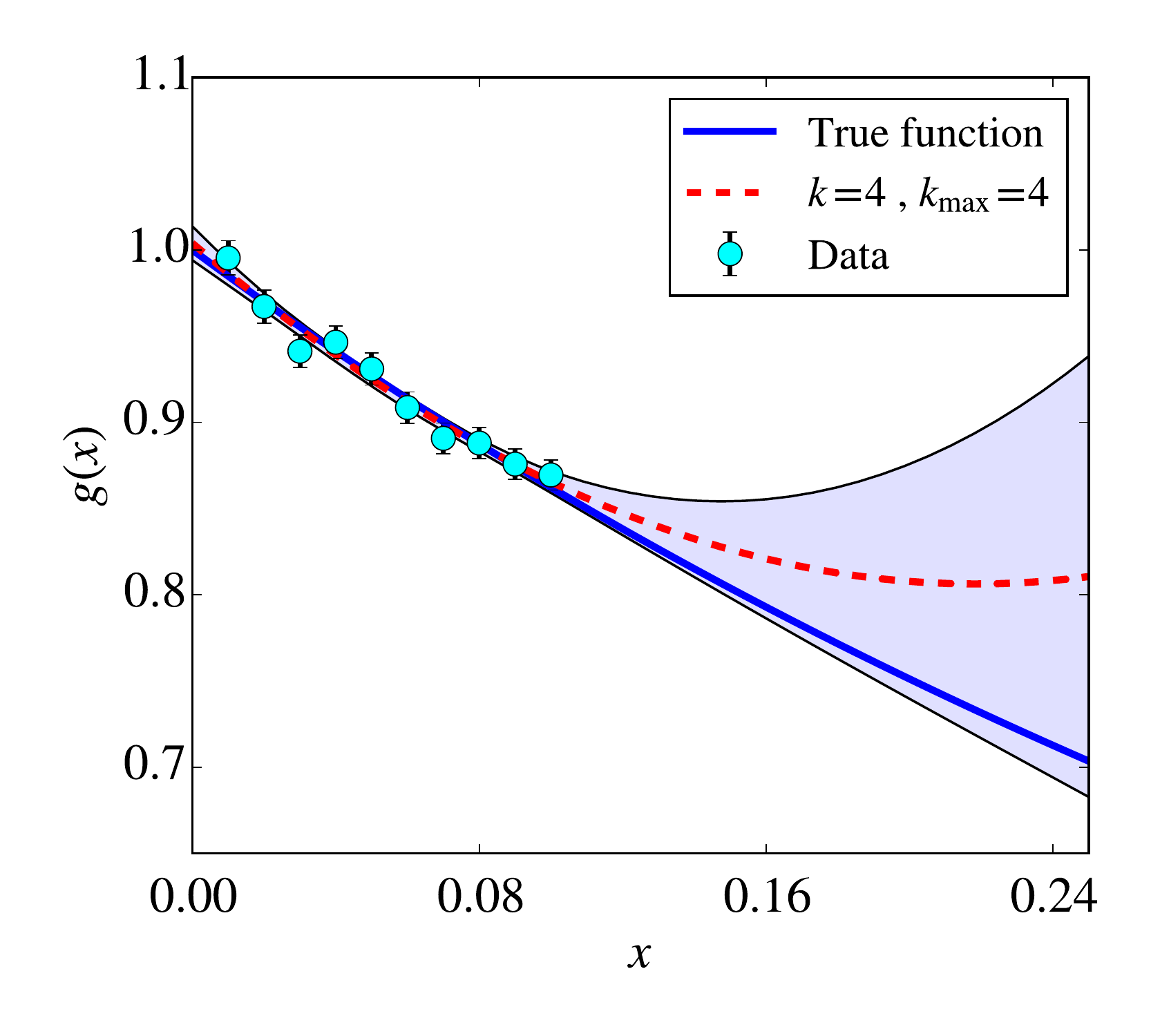}
    \caption{(color online)
    Comparison of data set \dataset{H}{0}{1} (corresponding to the first row of
    Table~\ref{tab:model-H-labels}), the underlying function for Model~H from
    Eq.~\eqref{eq:model-H}, and a Bayesian prediction calculated at order
    $\kord=4$, $\kmax=4$ using prior Set \Cprime\ with
    $\abarzero = 5$, from that data set. The error bands represent 1-$\sigma$ (68\% DoBs).
    \label{fig:H0_pred_abar_5_k4}}
\end{figure}

We choose another convenient model for these explorations:
\beq
    g(x) = \frac{\beta^2}{(\beta + x)^2}
    \;,
    \label{eq:model-H}
\eeq
with fixed $\beta = 1.3$.
The Taylor expansion, which identifies the coefficients we seek to estimate,
is
\beq
     g(x) = 1 - 1.54 x + 1.78 x^2 - 1.82 x^3 + 1.75 x^4 + \cdots \;.
        \label{eq:model-H-th}
\eeq
Note that the pole is at negative $x$ so that the
radius of convergence is ``hidden" in the data. This also
means that the coefficients in the Taylor expansion have alternating sign. 
The magnitudes of the coefficients are
$\sim 1$ up until $\Order(x^{10})$, where they begin decreasing.
We consider
a variety of data sets, of varying precision, and with data
over different ranges in $x$ and with different numbers of points. 
These sets are
enumerated in Table~\ref{tab:model-H-labels} (and available in files in
the supplementary material). We add a subscript to 
the label to indicate the percent relative error.
Figures~\ref{fig:H0_pred_uniform_k4} and \ref{fig:H0_pred_abar_5_k4} 
show the underlying model function from Eq.~\eqref{eq:model-H}, along
with one data set \dataset{H}{0}{1}. Both
fits agree quite well in the region where there is data. But, outside
that interval, the uniform-prior predictions in Fig.~\ref{fig:H0_pred_uniform_k4}
show the consequences of overfitting. The result is not in agreement with the
underlying function already a disappointingly small distance above the fit region. 
And the problems are not just at high $x$: the overfitting reduces precision at low 
$x$ too. The Bayesian result in Fig.~\ref{fig:H0_pred_abar_5_k4} does not
suffer from these problems.

\begin{table}[th]
    \caption{Model~H data set labels for sampling
    grid ranges and number of points. The breakdown scale
    is $x=1.3$.\label{tab:model-H-labels}}
    \begin{tabular}{|c|c|l|c|}
        \hline
        Label & \# of pts. & \multicolumn{1}{c|}{Grid} & Spacing \\
        \hline
        H0    & 10 &$0.01 \leq x \leq 0.1$ & linear \\
        H1    & 10 &$0.05 \leq x \leq 0.5$ & linear \\
        H2    & 15 &$0.05 \leq x \leq 0.75$ & linear \\
        H3    & 10 &$0.05 \leq x \leq 0.5$ & quadratic \\
        \hline
        H4    & 15 &$0.10 \leq x \leq 1.5$ & linear \\
        \hline
    \end{tabular}
\end{table}

The data sets in Table~\ref{tab:model-H-labels} are representative of different
situations that might be encountered in EFT parameter estimation.
\dataset{H}{0}{1} has very small errors and is sampled on a mesh where the expansion parameter is very
small. \dataset{H}{1}{5} is sampled over a small range
of $x$ with a fairly large data error. \dataset{H}{2}{5} expands upon the \dataset{H}{1}{5}
mesh by adding an additional 5 ultraviolet (UV) 
measurements to the mesh to probe the improvement of
leading-order extractions when UV data are added. Finally, \dataset{H}{3}{1} is an
accurate data set measured on a quadratically spaced mesh (in the same
range as \dataset{H}{1}{5}). This both simulates an application
to EFT expansions with even powers only and results in a data set with
several high-precision measurements at small $x$. 
\dataset{H}{4}{5} is used in Sec.~\ref{subsec:pastbreakdown} as
a case study of an EFT fit beyond its breakdown scale. 

\begin{table*}[p]
   \caption{
    Coefficient estimates from sampling of $\pr(\avec|\dataset{H}{0}{1},\kord,\kmax)$
   given the expansion from Eq.~\eqref{eq:model-th-expansion} 
   (these results are controlled by $\kmax$ only, see Sec.~\ref{subsec:setup}).
   The left side of the table is for a uniform prior, 
   which is equivalent to a least-squares fit, and includes the $\chi^2$/dof values.
   The right side of the table is using prior Set~\Cprime\ 
   from Table~\ref{tab:priors} with $\abarzero = 5$, and includes the evidence.
   For both priors the posterior pdf is a multi-dimensional
   Gaussian.
   \label{tab:H0-results-both}
    }
  \begin{tabular}{|c|c||c|c|c|c||c|c|c|c|}
    \hline
     \multicolumn{2}{|c||}{} & \multicolumn{4}{|c||}{Uniform prior} &  \multicolumn{4}{|c|}{Gaussian prior} \\
    \hline
      $\kord$ & $\kmax$ & $\chi^2/$dof & $a_0$ & $a_1$ & $a_2$ 
      & Evidence & $a_0$ & $a_1$ & $a_2$ \\ 
    \hline
       0 & 0  &  20 & 0.92$\pm$0.00   &                &              & $\sim 0$ & 0.92$\pm$0.00 &  &  \\
       1 & 1  &  0.90 & 0.99$\pm$0.01 & $-1.3\pm$0.1 &              & $7.2\times 10^{9}$ & 0.99$\pm$0.01 & $-1.3\pm$0.1 &   \\
       2 & 2  &  0.64 & 1.0$\pm$0.01 & $-2.1\pm$0.5 & 6.7$\pm$4 & $1.0\times 10^{10}$ & 1.0$\pm$0.01 & $-1.8\pm$0.4 & 4.0$\pm$3 \\
       2 & 3  &  0.74 & 1.0$\pm$0.02 & $-2.2\pm$1 & 9.8$\pm$30   & $1.0\times 10^{10}$ & 1.0$\pm$0.01 & $-1.8\pm$0.4 & 4.0$\pm$3\\
       2 & 4  &  0.67 & 1.0$\pm$0.03 & $-5.6\pm$4 & 130$\pm$100   & $1.0\times 10^{10}$ & 1.0$\pm$0.01 & $-1.8\pm$0.4 & 4.0$\pm$3\\
       2 & 5  &  0.54 & 1.1$\pm$0.1 & $-14\pm$9 & 580$\pm$400  & $1.0\times 10^{10}$ & 1.0$\pm$0.01 & $-1.8\pm$0.4 & 4.0$\pm$3\\
       2 & 6  &  0.69 & 1.1$\pm$0.1 & $-8.6\pm$20 & 190$\pm$2000   & $1.0\times 10^{10}$ & 1.0$\pm$0.01 & $-1.8\pm$0.4 & 4.0$\pm$3 \\
    \hline
        \multicolumn{3}{|c|}{True values}  & 1.0 & $-1.54$ & 1.78 & & 1.0 & $-1.54$ & 1.78 \\
    \hline
    \end{tabular}

   \caption{
   Same as Table~\ref{tab:H0-results-both} except sampling from 
   $\pr(\avec|\dataset{H}{1}{5},\kord,\kmax)$.
    \label{tab:H1-results-both}
    }
  \begin{tabular}{|c|c||c|c|c|c||c|c|c|c|}
    \hline
     \multicolumn{2}{|c||}{} & \multicolumn{4}{|c||}{Uniform prior} &  \multicolumn{4}{|c|}{Gaussian prior} \\
    \hline
      $\kord$ & $\kmax$ & $\chi^2/$dof & $a_0$ & $a_1$ & $a_2$ 
      & Evidence & $a_0$ & $a_1$ & $a_2$ \\ 
    \hline
        0 & 0  &  13 & 0.65$\pm$0.01 &             &                   & $\sim 0$  & 0.65$\pm$0.01 &  &  \\
        1 & 1  &  2.3 & 0.89$\pm$0.03 & $-0.74\pm$0.1 &               & $1.4\times 10^{2}$ & 0.89$\pm$0.03 & $-0.74\pm$0.08 &   \\
        2 & 2  &  2.3 & 0.95$\pm$0.05 & $-1.3\pm$0.4 & $0.97\pm$0.6 & $6.2\times 10^{1}$ & 0.95$\pm$0.05 & $-1.3\pm$0.4 & 0.95$\pm$1 \\
        2 & 3  &  2.5 & 1.0$\pm$0.08 & $-2.1\pm$1 & $4.4\pm$4   & $3.9\times 10^{1}$ & 0.96$\pm$0.06 & $-1.5\pm$0.7 & 1.8$\pm$3 \\
        2 & 4  &  2.9 & 0.92$\pm$0.1 & $-0.04\pm$3 & $-10\pm$20   & $3.3\times 10^{1}$ & 0.97$\pm$0.06 & $-1.6\pm$0.7 & 1.8$\pm$3 \\
        2 & 5  &  3.5 & 0.75$\pm$0.2 & $5.5\pm$7 & $-66\pm$70    & $3.0\times 10^{1}$ & 0.97$\pm$0.06 & $-1.6\pm$0.7 & 1.9$\pm$3 \\
        2 & 6  &  4.0 & 0.13$\pm$0.5 & $29\pm$20 & $-380\pm$200     & $3.0\times 10^{1}$ & 0.97$\pm$0.06 & $-1.6\pm$0.7 & 1.7$\pm$3 \\
    \hline
        \multicolumn{3}{|c|}{True values}  & 1.0 & $-1.54$ & 1.78 & & 1.0 & $-1.54$ & 1.78 \\
    \hline
    \end{tabular}

   \caption{
    Same as Table~\ref{tab:H0-results-both} except sampling from 
   $\pr(\avec|\dataset{H}{2}{5},\kord,\kmax)$.
    \label{tab:H2-results-both}
    }
  \begin{tabular}{|c|c||c|c|c|c||c|c|c|c|}
    \hline
     \multicolumn{2}{|c||}{} & \multicolumn{4}{|c||}{Uniform prior} &  \multicolumn{4}{|c|}{Gaussian prior} \\
    \hline
      $\kord$ & $\kmax$ & $\chi^2/$dof & $a_0$ & $a_1$ & $a_2$ 
      & Evidence & $a_0$ & $a_1$ & $a_2$ \\ 
    \hline
        0 & 0  &  29 & 0.53$\pm$0.01 &          &                      &  $\sim 0$          & 0.53$\pm$0.01 &  &   \\
        1 & 1  &  2.4 & 0.89$\pm$0.02 & $-0.70\pm$0.04 &               & $1.7\times 10^{5}$ & 0.89$\pm$0.02 & $-0.70\pm$0.04 &   \\
        2 & 2  &  1.2 & 1.0$\pm$0.04 & $-1.4\pm$0.2 & 0.77$\pm$0.2     & $2.7\times 10^{7}$ & 1.0$\pm$0.04 & $-1.4\pm$0.2 & 0.77$\pm$0.2 \\
        2 & 3  &  1.3 & 1.0$\pm$0.1 & $-1.7\pm$0.5 & 1.6$\pm$1         & $6.0\times 10^{6}$ & 1.0$\pm$0.05 & $-1.6\pm$0.5 & 1.4$\pm$1 \\
        2 & 4  &  1.3 & 1.1$\pm$0.1 & $-2.6\pm$1 & 5.9$\pm$5           & $2.6\times 10^{6}$ & 1.0$\pm$0.06 & $-1.7\pm$0.6 & 1.6$\pm$2 \\
        2 & 5  &  1.3 & 1.0$\pm$0.1 & $0.63\pm$3 & $-17\pm$20          & $1.6\times 10^{6}$ & 1.0$\pm$0.06 & $-1.8\pm$0.7 & 2.0$\pm$2 \\
        2 & 6  &  1.3 & 1.1$\pm$0.2 & $-3.5\pm$5 & 23$\pm$50           & $1.2\times 10^{6}$ & 1.1$\pm$0.06 & $-1.9\pm$0.7 & 2.2$\pm$2 \\
    \hline
        \multicolumn{3}{|c|}{True values}  & 1.0 & $-1.54$ & 1.78 & & 1.0 & $-1.54$ & 1.78 \\
    \hline
    \end{tabular}

   \caption{
   Same as Table~\ref{tab:H0-results-both} except sampling from 
   $\pr(\avec|\dataset{H}{3}{1},\kord,\kmax)$.
    \label{tab:H3-results-both}
    }
  \begin{tabular}{|c|c||c|c|c|c||c|c|c|c|}
    \hline
     \multicolumn{2}{|c||}{} & \multicolumn{4}{|c||}{Uniform prior} &  \multicolumn{4}{|c|}{Gaussian prior} \\
    \hline
      $\kord$ & $\kmax$ & $\chi^2/$dof & $a_0$ & $a_1$ & $a_2$ 
      & Evidence & $a_0$ & $a_1$ & $a_2$ \\ 
    \hline
        0 & 0  & 380 & 0.68$\pm$0.00 &             &                   &   $\sim 0$          & 0.68$\pm$0.00 &  &  \\
        1 & 1  &  4.4 & 0.94$\pm$0.00 & $-0.87\pm$0.01 &               &  $8.8\times 10^{3}$ & 0.94$\pm$0.00 & $-0.87\pm$0.01 &  \\
        2 & 2  & 0.88 & 0.98$\pm$0.01 & $-1.23\pm$0.1 & $0.64\pm$0.1  &  $3.6\times 10^{8}$ & 0.98$\pm$0.01 & $-1.2\pm$0.07 & $0.64\pm$0.12 \\
        2 & 3  & 0.73 & 0.96$\pm$0.02 & $-0.96\pm$0.2 & $-0.50\pm$0.9 &  $1.7\times 10^{8}$ & 0.96$\pm$0.01 & $-0.96\pm$0.2 & $-0.48\pm$0.8 \\
        2 & 4  & 0.62 & 0.94$\pm$0.03 & $-0.35\pm$0.6 & $-4.8\pm$4  &  $1.1\times 10^{8}$ & 0.96$\pm$0.02 & $-0.94\pm$0.3 & $-0.66\pm$1 \\
        2 & 5  & 0.58 & 0.90$\pm$0.05 & $0.80\pm$2 & $-17\pm$10      &  $8.7\times 10^{7}$ & 0.96$\pm$0.02 & $-0.91\pm$0.3 & $-0.87\pm$2 \\
        2 & 6  & 0.78 & 0.90$\pm$0.1 & $0.70\pm$4 & $-15\pm$50       &  $8.0\times 10^{7}$ & 0.96$\pm$0.02 & $-0.88\pm$0.3 & $-0.99\pm$2 \\
    \hline
        \multicolumn{3}{|c|}{True values}  & 1.0 & $-1.54$ & 1.78 & & 1.0 & $-1.54$ & 1.78 \\
    \hline
    \end{tabular}
\end{table*}

Tables~\ref{tab:H0-results-both}, \ref{tab:H1-results-both}, \ref{tab:H2-results-both},
and \ref{tab:H3-results-both} each show the results from the Parameter estimation
stage using the data sets enumerated above, which are
generated at random and are not selected to be ``typical'' in any way.
As a consequence, the impact of fluctuations is manifested.
Without accounting for the errors, one might jump to false conclusions by
naive comparisons to the true values.
We see two examples immediately. First, the estimates for $a_1$ in Table~\ref{tab:H1-results-both}
with the Gaussian prior are much closer to the true values than those
in Table~\ref{tab:H0-results-both}.  But, the latter's tighter 68\% DoB limits say it
is the more precise prediction; it is just chance that the extraction from the H1 data set results
in a central value that is close as the DoB limits are quite wide.
Second, in Table~\ref{tab:H3-results-both}, the estimates with a Gaussian prior differ
from the true values well
beyond the quoted error bars.  But, as documented below, this particular data
set happened to have $2\sigma$ fluctuations in the data points, which leads to estimates
for which the true value is outside the 68\% DoB interval.  This should not be
a surprise; indeed, it is expected one-third of the time! The occurrence of a large
fluctuation in \dataset{H}{3}{1} 
provides a testing ground for our multi-set analysis (see Sec.~\ref{sec:multiset-accum}); this will reveal the impact of fluctuations on parameter estimates.

\subsection{Guidance: Exploring the prior, $\kord$, and $\kmax$}

The goal of the Guidance stage of our parameter estimation process is to take a 
particular data set and determine two aspects of the associated fit: the number of EFT parameters that can be reliably extracted and the range of naturalness parameters
that should be used in that extraction.
(Note: these procedures can be adapted to priors for other information.)
The evidence  $\pr(D|\kord,\kmax)$ illuminates the information content of the data
with respect to the order of our fit. 
It is important to do the parameter estimation in the $k$-region where the
evidence has saturated, or else the estimates of those coefficients 
that can be determined will not be stable with respect to
the order of marginalization $\kmax$. In Sec.~\ref{subsec:quality} it was discussed
that for these model problems when $\kord \leq \kmax$, the evidence is controlled 
exclusively by $\kmax$. Therefore when we calculate the evidence, we do so for
$\kord=\kmax$ as we did in Fig.~\ref{fig:D15_evidence}.  
For $\kmax$s not in the saturation region the coefficients are underfit---even in the
presence of a naturalness prior. Meanwhile, the pertinent values of $\abar$ are 
indicated by the posterior $\pr(\abar|D,\kord,\kmax)$, which
shows the data's content with respect to the naturalness of the coefficients.

The Guidance diagnostics are interrelated and should be compared for different
prior assumptions---in this case we will compare Set~C (or \Cprime) and Set~A
as an example. As confirmed below,
for Model~H we can fix $\abar$ to a reasonable value of $\abarzero$ and the parameter estimates are largely insensitive to 
this choice. 
More detailed comparisons would be needed if results were more sensitive to the
choice of prior.
In the following we use the results in Tables~\ref{tab:H0-results-both}, 
\ref{tab:H1-results-both}, \ref{tab:H2-results-both}, and \ref{tab:H3-results-both}
to evaluate what we learn from the Guidance stage.

\begin{figure}[tbh]
    \includegraphics[width=0.46\textwidth]{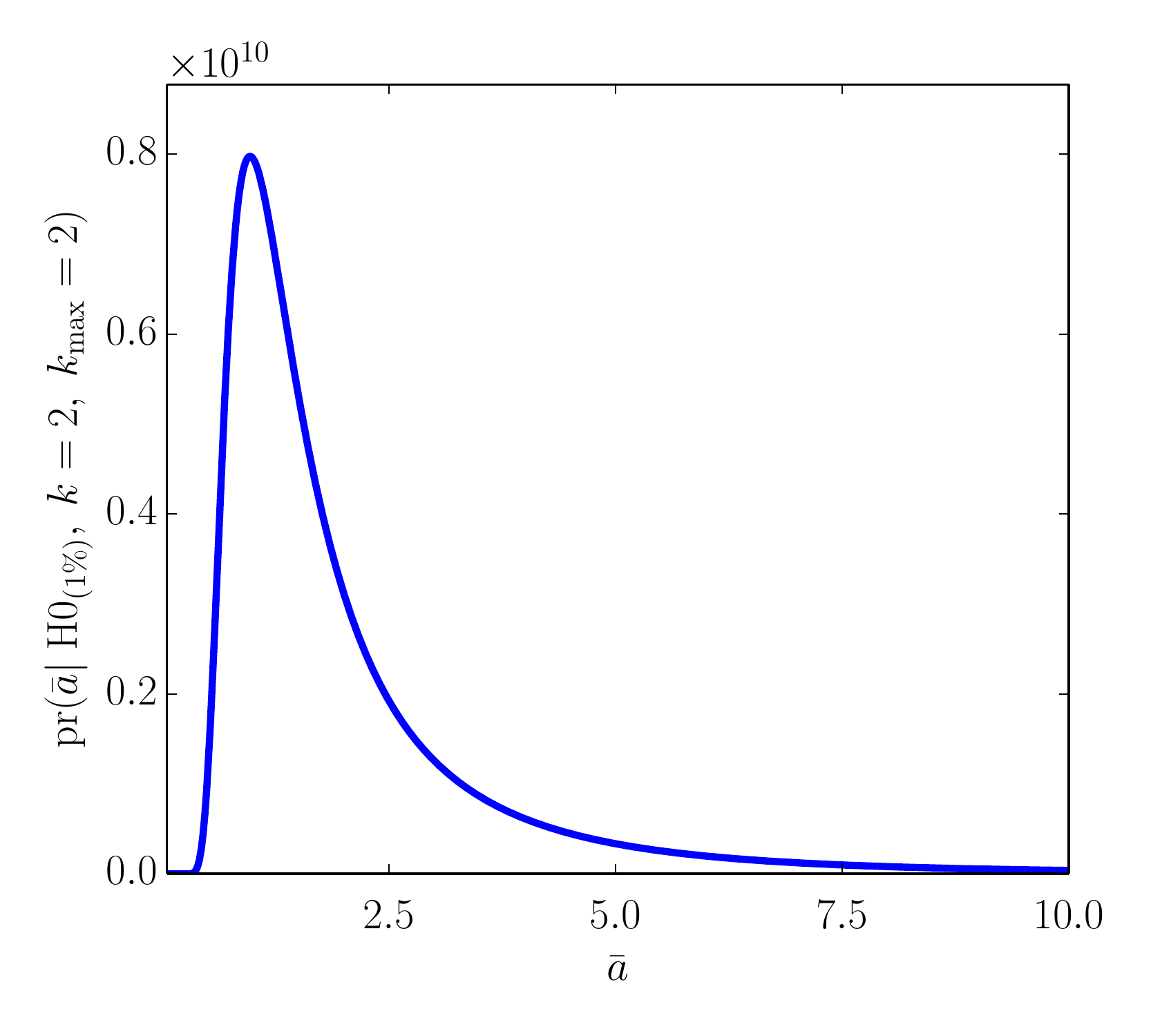}
    \caption{The posterior pdf $\pr(\abar|D,\kord,\kmax)$  calculated at $\kord=2$, $\kmax=2$
    assuming prior Set~C from Table~\ref{tab:priors} with
  $\abarmin=0.05$ and $\abarmax=20$, given
  data set \dataset{H}{0}{1}.
    \label{fig:H01_abar_post_C0p05to20} }
\end{figure}

\begin{figure}[tbh]
    \includegraphics[width=0.46\textwidth]{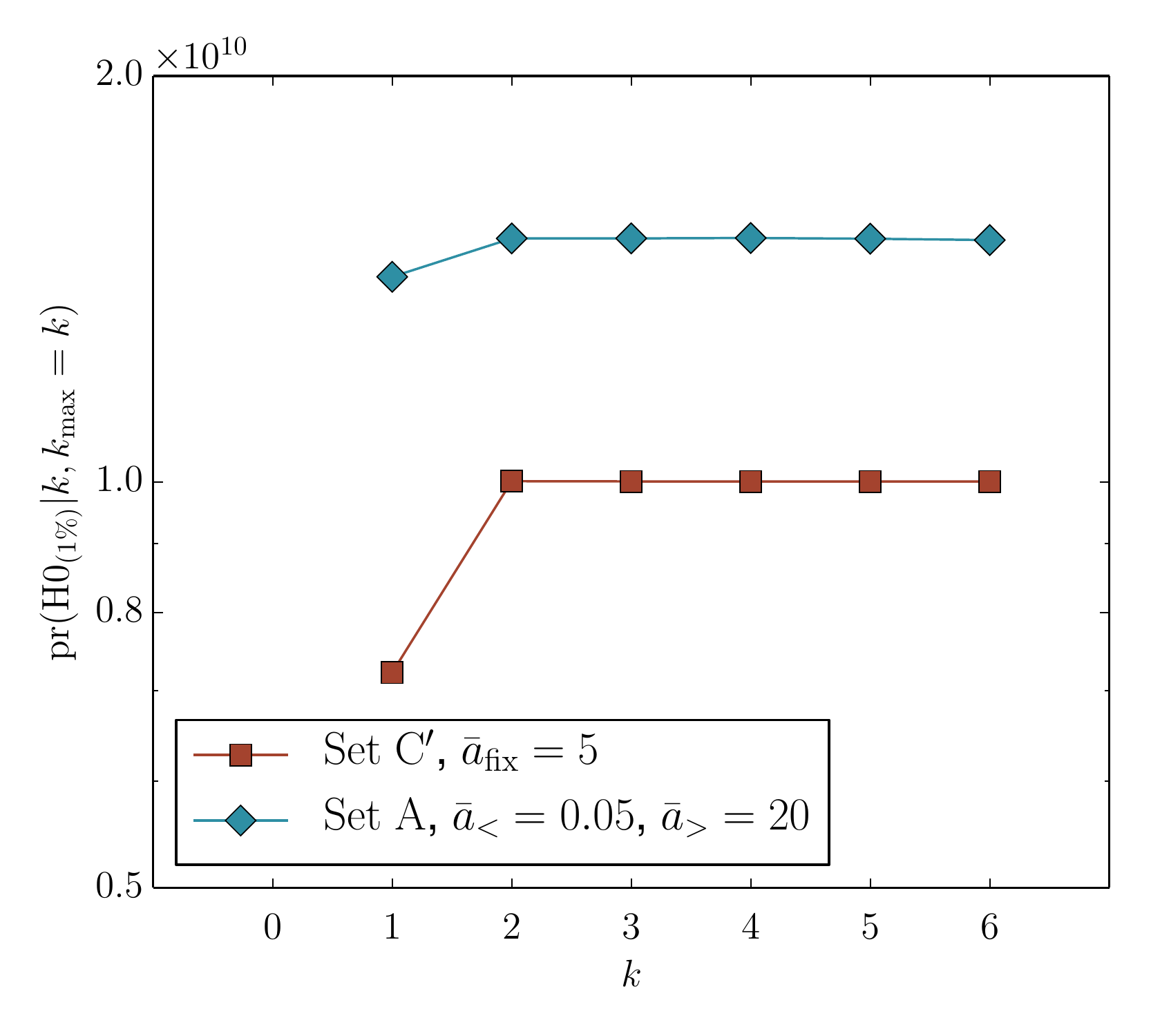}
    \caption{Evidence $\pr(\dataset{H}{0}{1}|\kord,\kmax=\kord)$ using different prior
    assumptions for several values of $\kord$ with $\kmax = \kord$.
    (The evidence for $\kord = 0$ is not shown for either prior because it is nearly zero). 
        Note that only ratios of the evidence at different $k$ for the same data and priors 
    are significant.
    \label{fig:H01_evi_compare_A_Cp} }
\end{figure}

\textbf{Small error, very small range.}
We first consider \dataset{H}{0}{1}, a very accurate data set sampled at small 
$x$ values, and examine what the evidence and $\abar$ posterior plots
tell us in advance of parameter estimation.
The $\abar$ posterior in Fig.~\ref{fig:H01_abar_post_C0p05to20} uses prior Set~C.  It
shows that the most likely value for $\abar$ given this information is about~1,
and that choosing $\abarzero = 5$ will not be overly restrictive.
The evidence plot for Set~\Cprime\ in Fig.~\ref{fig:H01_evi_compare_A_Cp}
shows saturation at $\kord =2$, with no real improvement
from $\kord=1$.
This suggests that the limited range of this data set in $x$ means that we cannot extract 
information past the quadratic order (and that order will have limited information), 
and that we should marginalize over parameters $a_{i>2}$. This conclusion
is the same for different priors, and the evidence saturation behavior
is compared for Set~A and Set~\Cprime\ in Fig.~\ref{fig:H01_evi_compare_A_Cp}.

\begin{figure}[tbh]
    \includegraphics[width=0.46\textwidth]{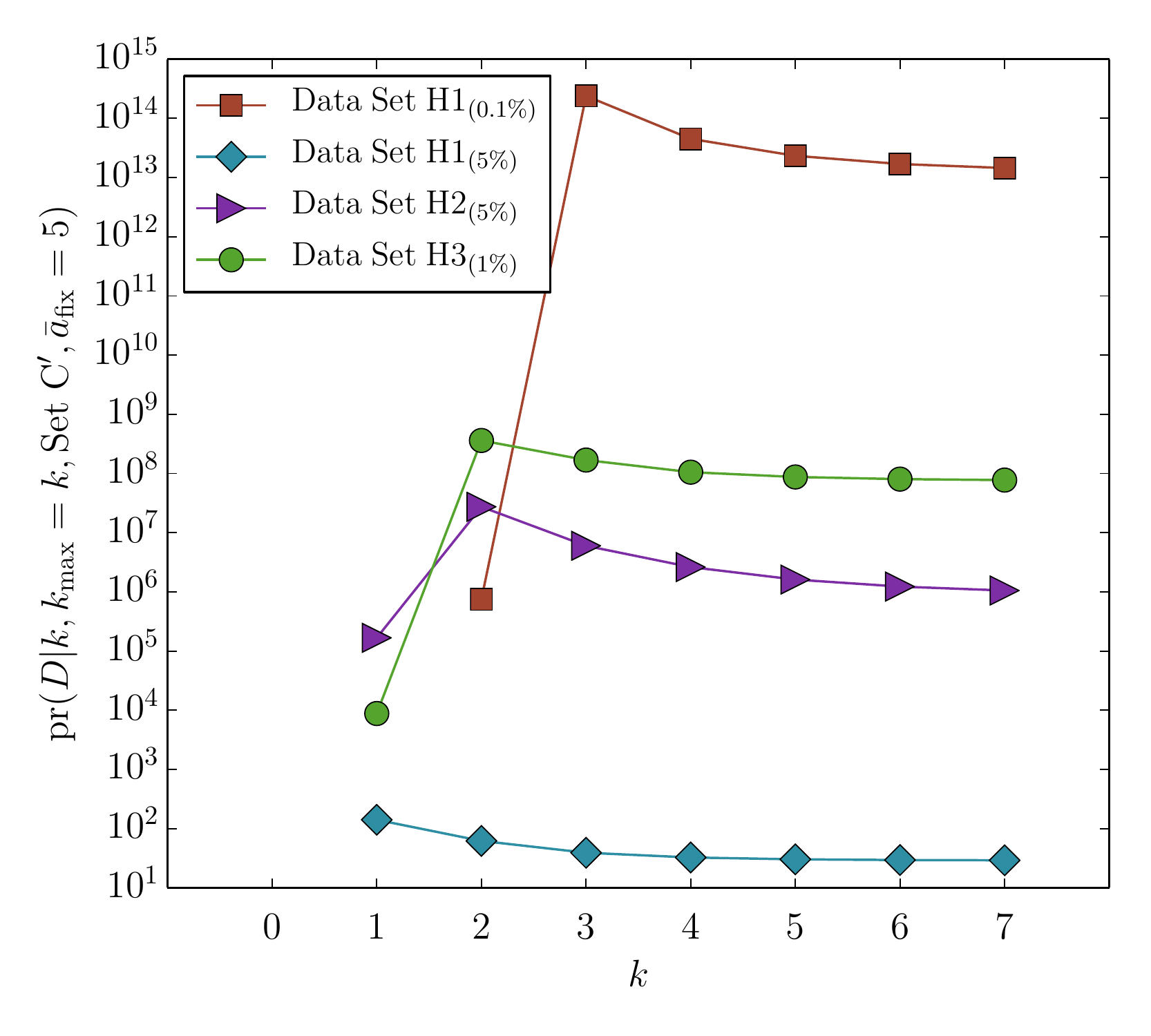}
    \caption{Data set evidence $\pr(D|\kord,\kmax=\kord)$ for several
    values of $\kord$ using prior Set~\Cprime\ with $\abarzero=5$ in each case.
    (The evidence is not shown for all calculations at $\kord=0$ because they are nearly zero, and also 
    the evidence for data set \dataset{H}{1}{0.1} at $\kord=1$ for the same reason). 
    \label{fig:H_evi_sets_compare_Cp} }
\end{figure}

\begin{figure*}[tbh!]
  \subfloat{%
    \label{fig:H01_xmax_k1_kmax1}%
  \includegraphics[width=0.45\textwidth]{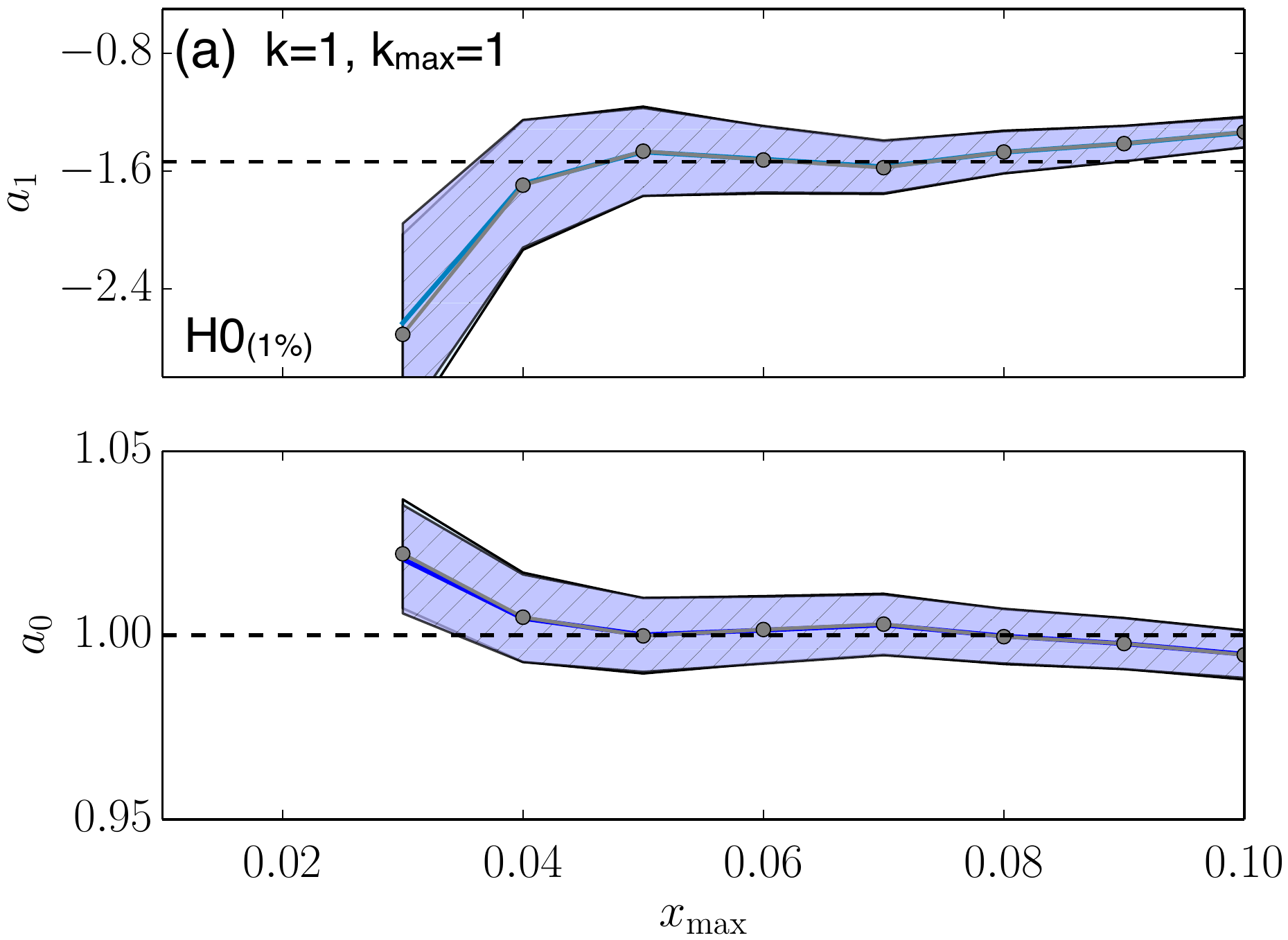}%
  }
  ~~~%
   \subfloat{%
     \label{fig:H01_xmax_k1_kmax2}%
     \includegraphics[width=0.45\textwidth]{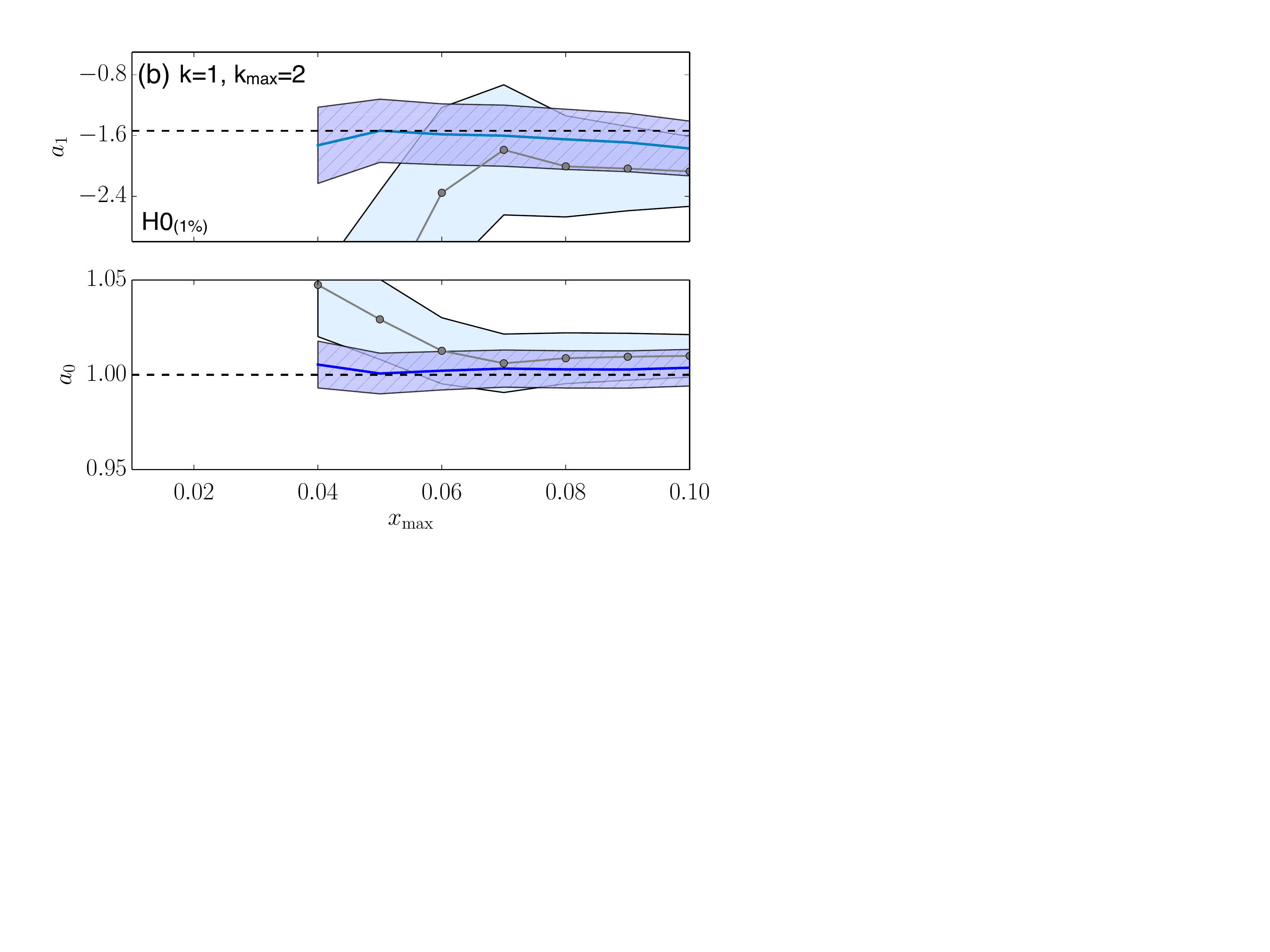}
   }
   \caption{(color online)
    Bayesian coefficient estimates
    as data from data set \dataset{H}{0}{1} are sequentially added at the
    high-$x$ end. The largest $x$-value in the set is denoted as $\xmax$.
    The lines with darker hatched error bands represent estimates
    using prior Set \Cprime\ with $\abarzero=5$, and the line with circles
    with lighter solid error bands represents the least-squares estimates. 
    The error bands represent 68\% DoBs (1-$\sigma$ errors), which coincide
    in (a).
    \label{fig:H01_xmax}}
\end{figure*}

The expectations developed in the Guidance stage for \dataset{H}{0}{1} 
are verified by the parameter estimates in 
Table~\ref{tab:H0-results-both}.
In particular, the results on the right side of the table using a Gaussian prior 
are consistent with determining successive orders with decreasing accuracy. The last coefficient
determined by the data is $a_2$, which has a large error, consistent with the evidence
saturation. Higher-order coefficients simply return the prior.

It is instructive to compare to the results on the 
left side of the table, which are extracted using a uniform prior, i.e.,
assuming no prior knowledge of coefficient size. One might think that a high-quality
data set with many points near $x=0$ should be dominated by the lowest-order 
terms, but these results show that the effects of fine-tuning are severe. 
In this case Table~\ref{tab:H0-results-both} shows that we have underfitting 
at $\kmax=1$, 
but, by the time we use $\kmax=2$ for our fit, $a_2$ and $a_1$ have both acquired large 
(in absolute terms) central values. This overfitting only gets more marked as $\kmax$ 
increases. 
In general the uncertainties allow $a_0$ and $a_1$ to be within the 68\% 
(or, at worst 95\%) DoB interval (once $\kmax>1$) but the 68\% interval is so wide in the case 
of $a_1$ that this statement is of little practical use. This is disappointing: 
one might have expected better from this low-$x$ (infrared) data. 
A projected posterior plot verifies that it occurs because $a_1$ and 
$a_2$ are highly anti-correlated, and there is nothing in this data set to pin down
the value of $a_2$.

At higher orders, plotting the uniform prior prediction,
e.g., $\kord=\kmax=4$ in Fig.~\ref{fig:H0_pred_uniform_k4}, the fine-tuning 
(overfitting) problem is manifest.
Similar results using the uniform prior for our
other data sets are shown in Tables~\ref{tab:H1-results-both}, \ref{tab:H2-results-both},
and \ref{tab:H3-results-both}, where generic underfitting/overfitting 
is also present. These issues make it clear that not including known information
results in worse predictions, which can manifest in subtle ways. For example, the $a_0$
results for the uniform prior in Table~\ref{tab:H0-results-both} are not too far off from the
true value but are in fact influenced by fine-tuning. The remaining significant
error bar on $a_1$ shows that data in this narrow $x$ window does not have
a good lever arm to accurately determine the slope of the underlying function at $x=0$,
a feature that is quantified by the evidence.

\textbf{Large error, small range.}
In the case of \dataset{H}{1}{5}, we might expect better constraints on
$a_2$ from data that covers a wider but still small range of $x$ than \dataset{H}{0}{1}. 
Fig.~\ref{fig:H_evi_sets_compare_Cp} shows a slight peak at
$\kord=1$ and saturation at higher $\kord$ for \dataset{H}{1}{5}.
The interaction of the prior and likelihood is not as 
clean as in Eq.~\eqref{eq:bayes_factor2} when $\kord=1$. By using only
the $\kord=1$ order EFT where the evidence peaks rather than the saturation region,
we are not including prior information that the EFT
has higher-order coefficients---this results in a poor extraction, as is
evident on the right of
Table~\ref{tab:H1-results-both}. Estimates should be made in the
saturation region, and the coefficients extracted at lower orders become stable
with respect to $\kmax$ because we appropriately account for higher-order effects. 
The uniform-prior results for \dataset{H}{1}{5} in Table~\ref{tab:H1-results-both}
show that the $\kmax=1$ result is not very good, but that in fact $\kmax=1$ and $\kmax=2$ have
the same values for the $\chi^2$/dof. Hence the lowest $\chi^2$ or the evidence peak
are not reliable diagnostics of good parameter estimates---only when the evidence has saturated
 do the values of coefficients determined by the data cease to change with increasing 
$\kmax$.

For comparison, we also calculate the evidence for data set \dataset{H}{1}{0.1},
which has 50 times smaller relative error than \dataset{H}{1}{5}. 
For this highly precise data sampled on
the H1 mesh from Table~\ref{tab:model-H-labels}, the evidence saturates
at $\kord=3$ in Fig.~\ref{fig:H_evi_sets_compare_Cp},
indicating that there are limitations from where the data are sampled even if
it is very accurate and precise. There is still a small peak in the evidence as there was in the
case of 5\% error.

\textbf{Large error, larger range.}
\dataset{H}{2}{5} has the same density of points as \dataset{H}{1}{5} but
has 5 extra points up to $x=0.75$. Does the addition of 5 more UV points
improve the lower-order estimates by constraining the higher-order ones
better? Yes: Fig.~\ref{fig:H_evi_sets_compare_Cp} shows that the evidence with $\kord=1$
is now several orders of magnitude lower than the evidence in the saturation
region near $\kord=3$.  The evidence is highest at $\kord=2$ but
the estimates in Table~\ref{tab:H2-results-both}
using a naturalness prior are not stable with respect to 
changing $\kmax$ until the evidence saturates.

\begin{figure*}[tbh]
  \subfloat{%
    \label{fig:H15_xmax_k1_kmax2}%
  \includegraphics[width=0.45\textwidth]{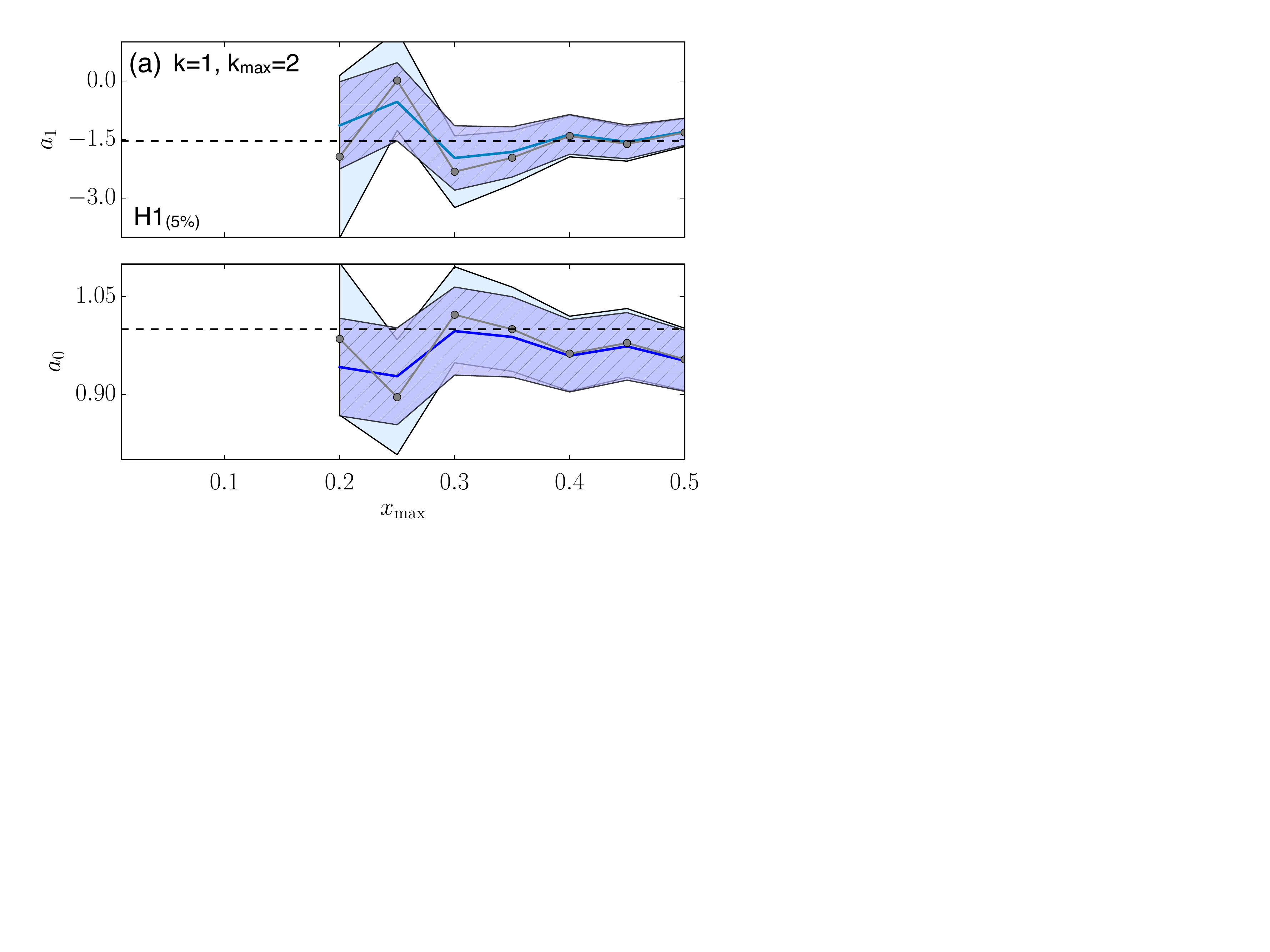}%
  }
  ~~~%
   \subfloat{%
     \label{fig:H15_xmax_k1_kmax3}%
     \includegraphics[width=0.45\textwidth]{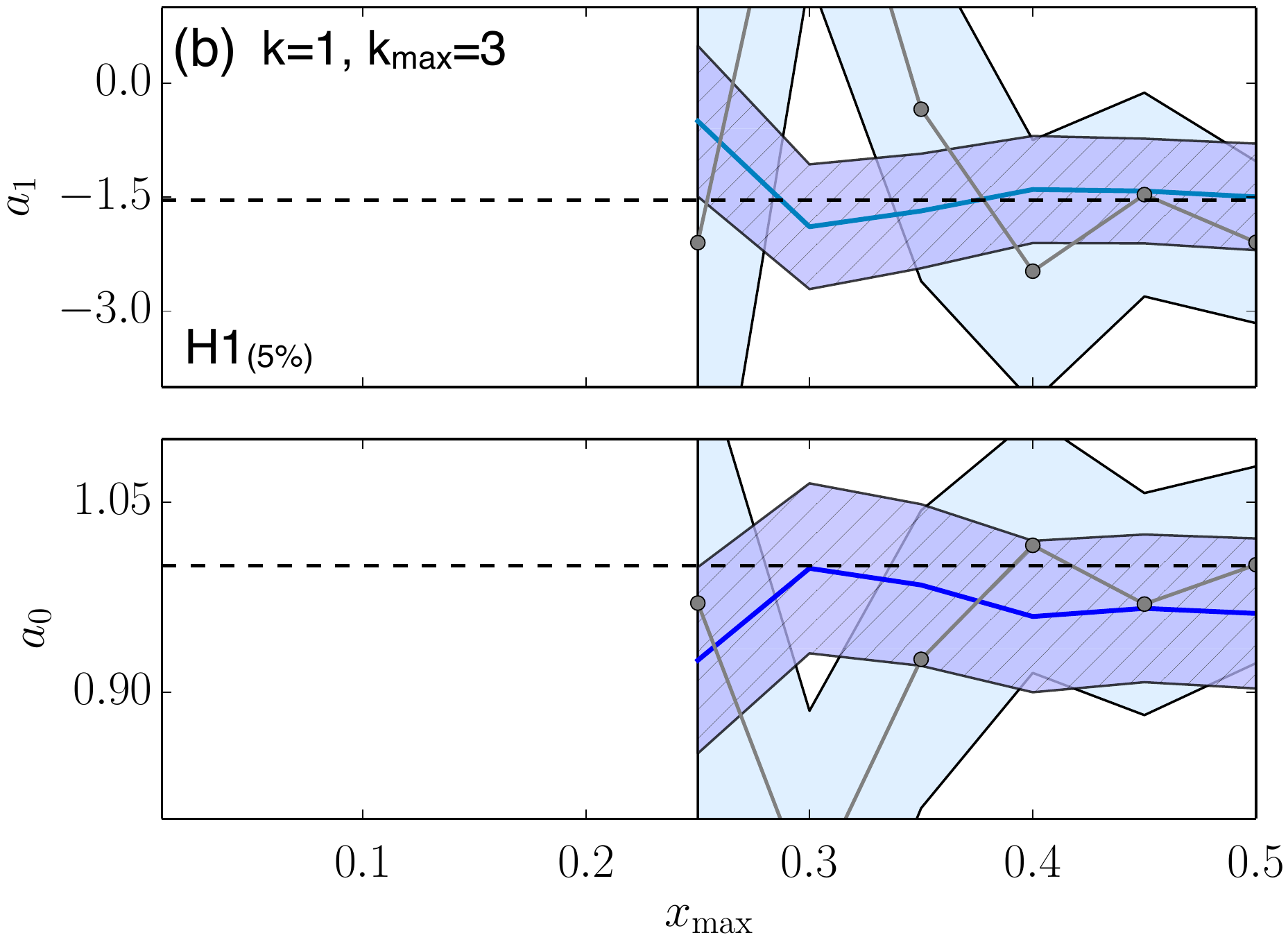}
   }
   \caption{(color online) $\xmax$ plots for data set \dataset{H}{1}{5} with same description
   as Fig.~\ref{fig:H01_xmax}.}
\end{figure*}

\begin{figure*}[tbh]
  \subfloat{%
    \label{fig:H25_xmax_k1_kmax2}%
  \includegraphics[width=0.45\textwidth]{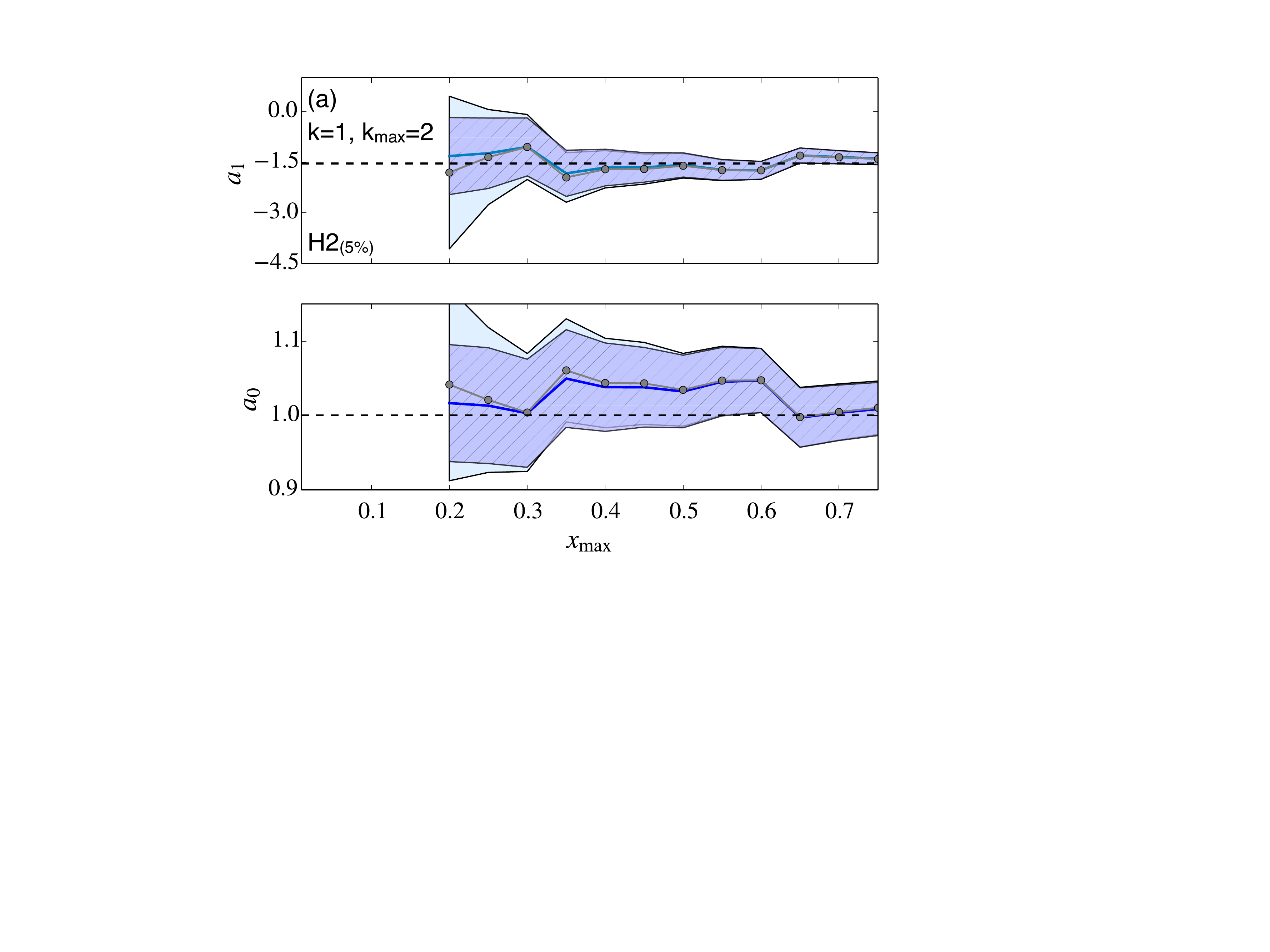}%
  }
  ~~~%
   \subfloat{%
     \label{fig:H25_xmax_k1_kmax3}%
     \includegraphics[width=0.45\textwidth]{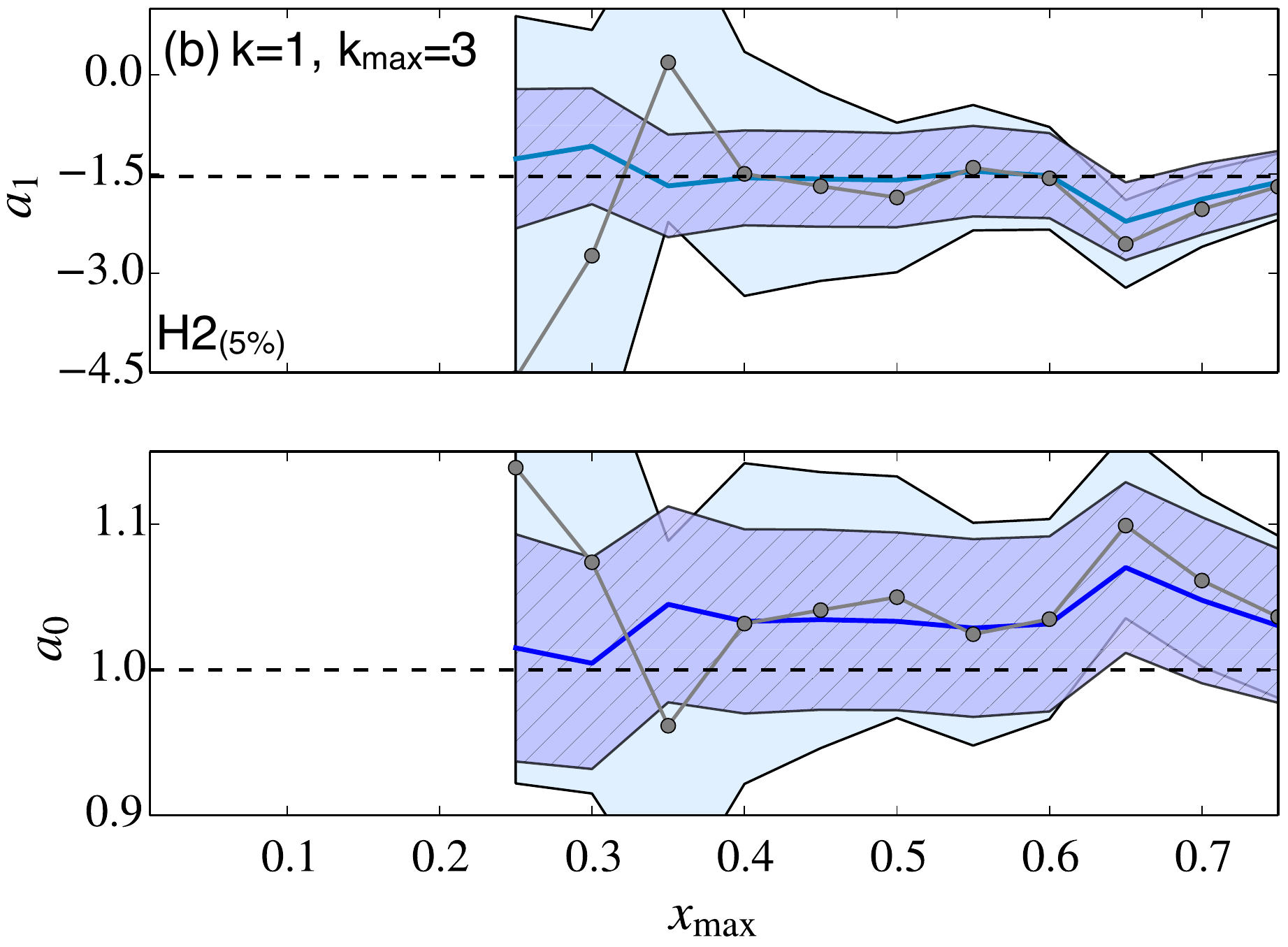}
   }
   \caption{(color online) 
   $\xmax$ plots for data set \dataset{H}{2}{5} with same description
   as Fig.~\ref{fig:H01_xmax}.}
\end{figure*}

\begin{figure*}[tbh]
  \subfloat{%
    \label{fig:H31_xmax_k1_kmax2}%
  \includegraphics[width=0.45\textwidth]{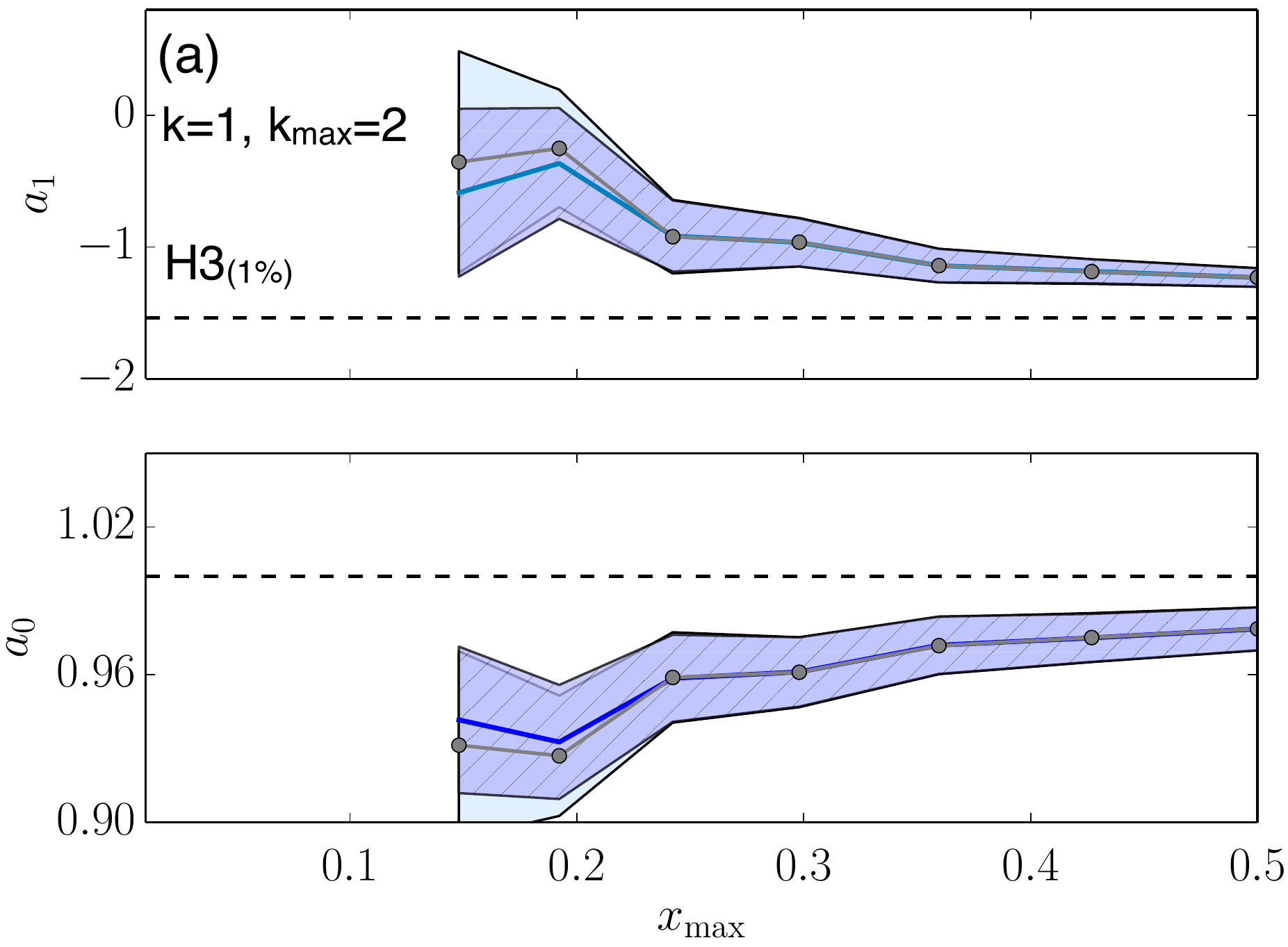}%
  }
  ~~~%
   \subfloat{%
     \label{fig:H31_xmax_k1_kmax3}%
     \includegraphics[width=0.45\textwidth]{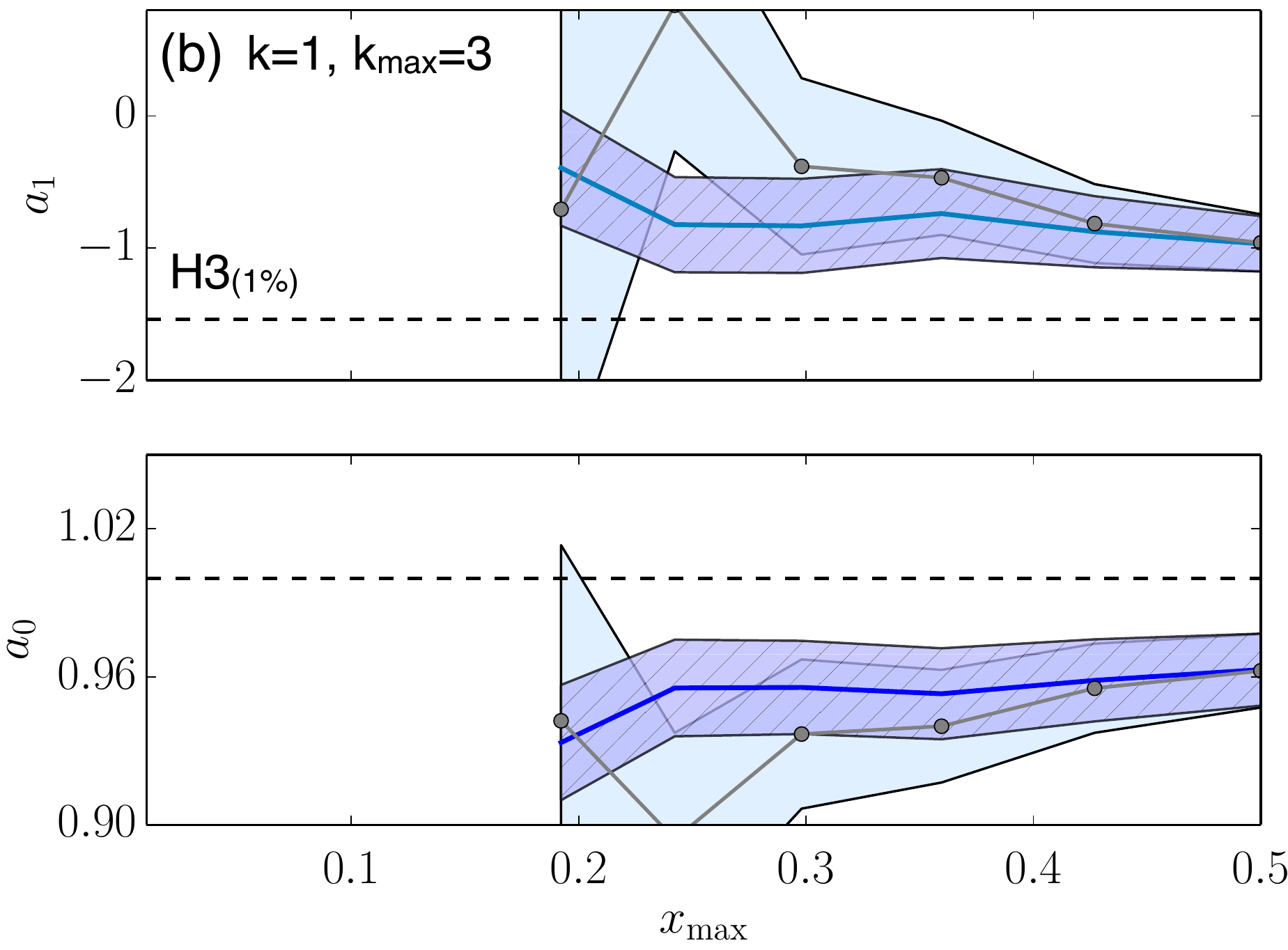}
   }
   \caption{(color online)
   $\xmax$ plots for data set \dataset{H}{3}{1} with same description
   as Fig.~\ref{fig:H01_xmax}.}
\end{figure*}

\textbf{Small error, small range.}
Lastly, we examine \dataset{H}{3}{1}, which samples the same region of $x$
as \dataset{H}{1}{5}, but has more data points at small $x$ and has higher
precision. The uniform-prior results are shown on the left
side of Table~\ref{tab:H3-results-both},
and we see the generic overfitting/underfitting. 
Figure~\ref{fig:H_evi_sets_compare_Cp} shows saturation at $\kord=2$ for this set. Compared to
the evidence for \dataset{H}{1}{5}, the decisiveness of the evidence at $\kord=2$ to $\kord=1$
is better. Turning to parameter estimation results using
prior Set~\Cprime\ on the right of Table~\ref{tab:H3-results-both},
the extraction stabilizes as we enter the saturation region but the results
do not seem consistent with the true values.
This is not a failure, however; this particular \dataset{H}{3}{1} data set
is one of the 32\% of data sets that will not produce LECs in the 68\% band.
The $a_0$ extraction is driven by the lowest-$x$ data point so that it is less than the
true value, $a_0~=~1$.
We will discuss this data set further when considering multi-set plots below.

\textbf{Summary.}
A uniform prior (least-squares method) is limited to the 
information in the data sets themselves. 
Whenever lower-order coefficients are correlated with higher-order
ones, there will be underfitting of coefficients associated with the last 
included order---and overfitting of lower-order coefficients. 
While high-precision data can ameliorate the situation for a few coefficients, 
the potential of such a data set is still under-realized
in a least-squares analysis. 
More generally, such difficulties cannot be avoided by choosing a particular
$x$ domain for the fit.

The Bayesian procedure is much more effective but is still constrained 
by the limitations of the data set, and these limitations are quantifiable using the evidence.
Using a wider region, but with less precise data, decreases the accuracy
of the $a_0$ determination markedly. ($a_0$ is determined with essentially 
the same accuracy from \dataset{H}{1}{5} and \dataset{H}{2}{5}.) The decrease in the accuracy of $a_1$ is not
as bad in \dataset{H}{1}{5} and \dataset{H}{2}{5} as \dataset{H}{0}{1}, because some of the loss in data precision is made up by the increased 
lever arm. 
All three data sets suggest that only a precise data set, over a wide range
of $x$, will be able to give unambiguous information on $a_2$. Perhaps most 
notably, in each case, $\dataset{H}{0}{1}$, \dataset{H}{1}{5}, and $\dataset{H}{2}{5}$, 
both the central 
values and 68\% DoB intervals stabilize with $\kmax$ once $\kmax \geq 3$.
Those stabilized results are consistent with the underlying values 
of all of $a_0$, $a_1$, and $a_2$.

\subsection{Parameter estimation: $\xmax$ analyses}

For the Parameter estimation step in the process, the principal diagnostic
is always the projected posterior plot, which illuminates the correlations
and shows features such as which
parameters are dominated by the prior.
We have already examined examples of projected posterior plots and associated tables
of parameter means and standard deviations, so
here we focus instead on the added utility of 
$\xmax$ plots. These track the evolution of parameter
estimates as the data set is built up sequentially with more and more high-$x$
(UV) data, thus making it clear that choosing ranges of data to fit certain orders
in the EFT may omit important information. 

\textbf{Small error, very small range.}
We saw in Fig.~\ref{fig:H01_evi_compare_A_Cp} that evidence saturation for
\dataset{H}{0}{1} occurs at $\kord=2$, which is slightly favored over $\kord=1$.
But, Table~\ref{tab:H0-results-both} showed that, while the posterior 
contains some information on $a_2$, it is largely undetermined.
Therefore we explore the $\kord=1$ results (the first two coefficients), 
marginalizing over higher coefficients for $\kmax>1$. Figure~\subref*{fig:H01_xmax_k1_kmax1} shows the $\kord=1$,
$\kmax=1$\ $\xmax$ plot.
Here the uniform and prior Set~\Cprime\
parameter estimates overlap and are not stable with increasing $\xmax$, indicating
underfitting. When we marginalize to $\kmax=2$, where the evidence saturated,
in Fig.~\subref*{fig:H01_xmax_k1_kmax2}, the naturalness prior parameter estimates are stable
with $\xmax$. The uniform-prior results are not and overfitting occurs,
especially for small $\xmax$. Once $\xmax \approx 0.07$, the uniform prior
error bars encompass the true value but are significantly larger than the naturalness
results. As $\kmax$ is increased further into the saturation region, the parameter
estimates with the naturalness prior continue to be stable with $\xmax$ (not shown).

\textbf{Large error, different ranges.}
Turning to \dataset{H}{1}{5} and \dataset{H}{2}{5}, which are sampled at larger ranges of
$x$ with larger errors, we explore the $\xmax$ behavior at $\kmax$ values near
the evidence saturation.
Figures~\subref*{fig:H15_xmax_k1_kmax2} and \subref*{fig:H15_xmax_k1_kmax3} show the results
at $\kord=1$ with $\kmax=2,3$ repectively for \dataset{H}{1}{5}. In light of
the evidence in Fig.~\ref{fig:H_evi_sets_compare_Cp}, $\kmax=3$ is approximately
in the saturation region while $\kmax=2$ does not account for enough higher orders.
At $\kmax=2$ the estimates at the largest $\xmax$ overlap with the uniform-prior results,
while the corresponding estimate for $\kmax=3$ is slightly better. The results for $\kmax=3$ are also
fairly stable with $\xmax$, consistent with the evidence saturation.

For \dataset{H}{2}{5}, with extra UV data, we show the 
$\kord=1$, $\kmax=2,3$ results in Figs.~\subref*{fig:H25_xmax_k1_kmax2} and
\subref*{fig:H25_xmax_k1_kmax3}, which again explores the transition region near
evidence saturation seen in Fig.~\ref{fig:H_evi_sets_compare_Cp}.
As we saw for \dataset{H}{1}{5}, the estimates are not stable with $\xmax$ until
$\kmax=3$ when the evidence truly saturates. In particular the final estimates with the largest
$\xmax$ in Fig.~\subref*{fig:H25_xmax_k1_kmax2} significantly underestimate the error in
$a_1$.

\textbf{Small error, small range.}
The evidence for \dataset{H}{3}{1} in Fig.~\ref{fig:H_evi_sets_compare_Cp} clearly
saturates. We compare both $\kmax=2,3$, just showing the first two coefficients ($\kord=1$), but not
$\kmax=1$ because it is in the underfitting regime. Surprisingly perhaps from the perspective
of the evidence, $\kmax=2$ in Fig.~\subref*{fig:H31_xmax_k1_kmax2} still gives
underfitting and for no $\xmax$ is the estimate consistent with the true parameter
to the 1-$\sigma$ level. Even for low $\xmax$, underfitting still occurs.
But the essential feature here is the absence of stability with respect to $\xmax$, {\it not}
that the 68\% DoBs miss the true parameter values. In contrast, 
for $\kmax=3$, shown in Fig.~\subref*{fig:H31_xmax_k1_kmax3},
there is stability with $\xmax$, even as the uniform-prior results begin
overfitting. But, even here, the 68\% DoB intervals miss the true value. Such ``occasional
misses" are, though, consistent with a statistical interpretation of these error bands.

\textbf{Summary.}
Using $\xmax$ plots rather than a fixed range of data provides
valuable insight into the parameter extraction.
They establish the region of $x$ that provides
stable predictions, help to identify where in $x$ new information on
coefficients stops,
and confirm the evidence saturation from the Guidance.

\begin{figure*}[tbh]
  \subfloat{%
    \label{fig:H01_datasets_k1_kmax3}%
  \includegraphics[width=0.45\textwidth]{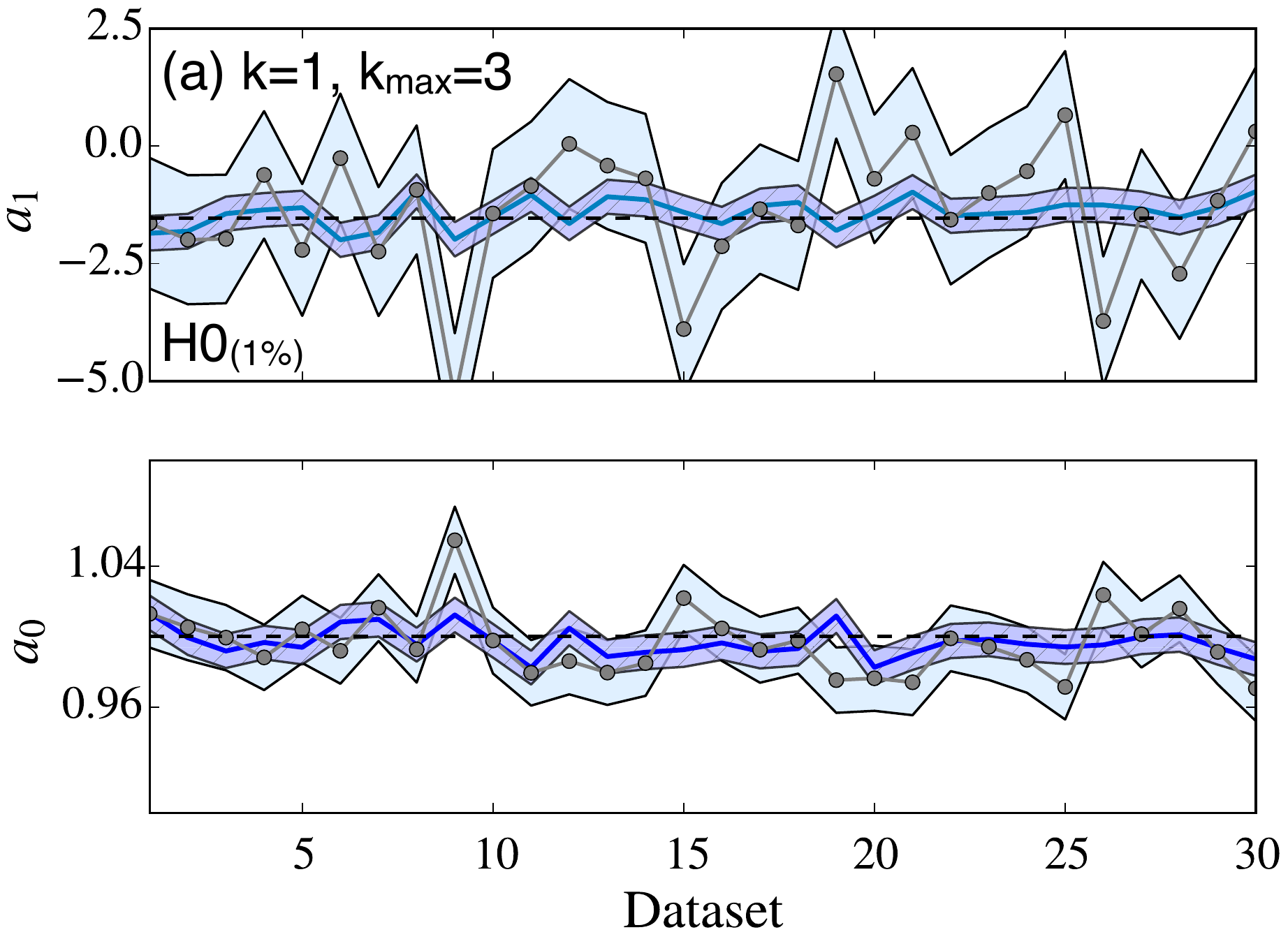}%
  }
  ~~~%
   \subfloat{%
     \label{fig:H01_accumulated_k1_kmax3}%
     \includegraphics[width=0.45\textwidth]{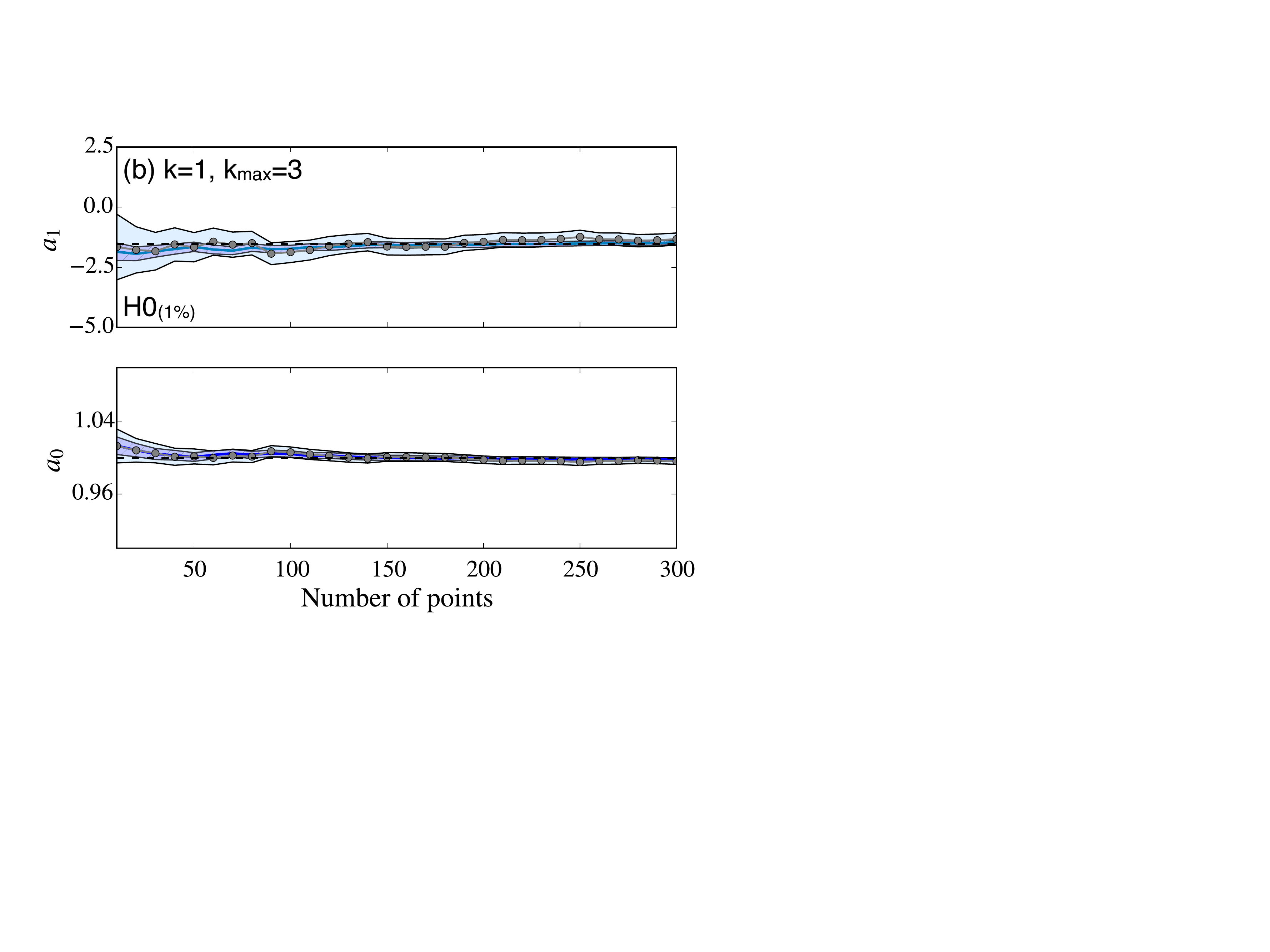}
   }
   \caption{Multi-set (a) and accumulation plots (b)
   calculated at $\kord=1$, $\kmax=3$. 
   The shaded regions denote 68\% error bands for the uniform (line with circles with lighter solid band)
   and naturalness prior (line with darker hatched band).
   The data sets used in (a) are 30 samples on the \dataset{H}{0}{1} mesh from Table~\ref{tab:model-H-labels}. 
   The same data are accumulated set by set to generate (b).
   In each case the prior was Set~\Cprime\ with $\abarzero = 5$.
   \label{fig:datasets-H01}} 
\end{figure*}

\begin{figure*}[tbh]
  \subfloat{%
    \label{fig:H15_datasets_k1_kmax4}%
  \includegraphics[width=0.45\textwidth]{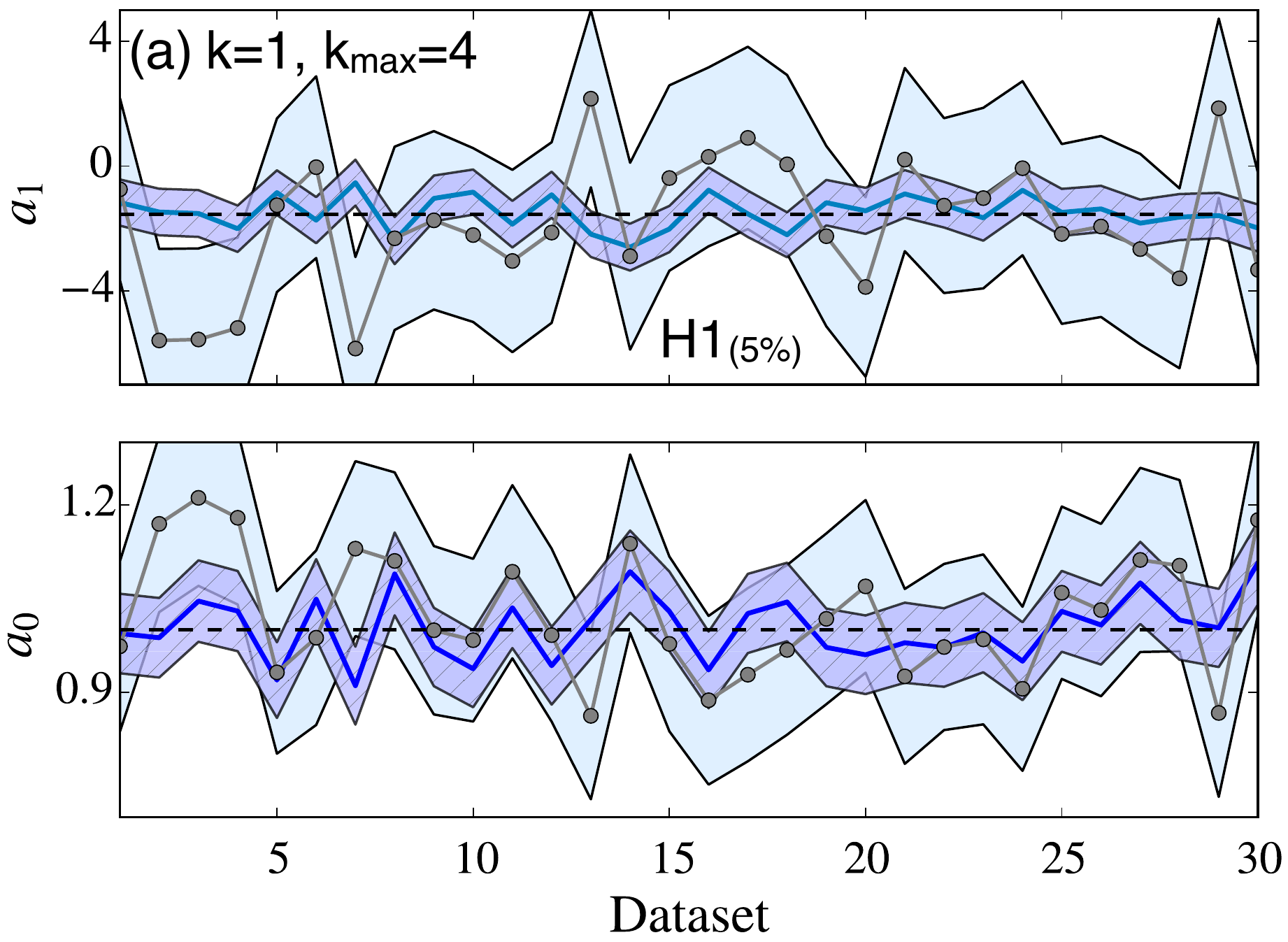}%
  }
  ~~~%
   \subfloat{%
     \label{fig:H15_accumulated_k1_kmax4}%
     \includegraphics[width=0.45\textwidth]{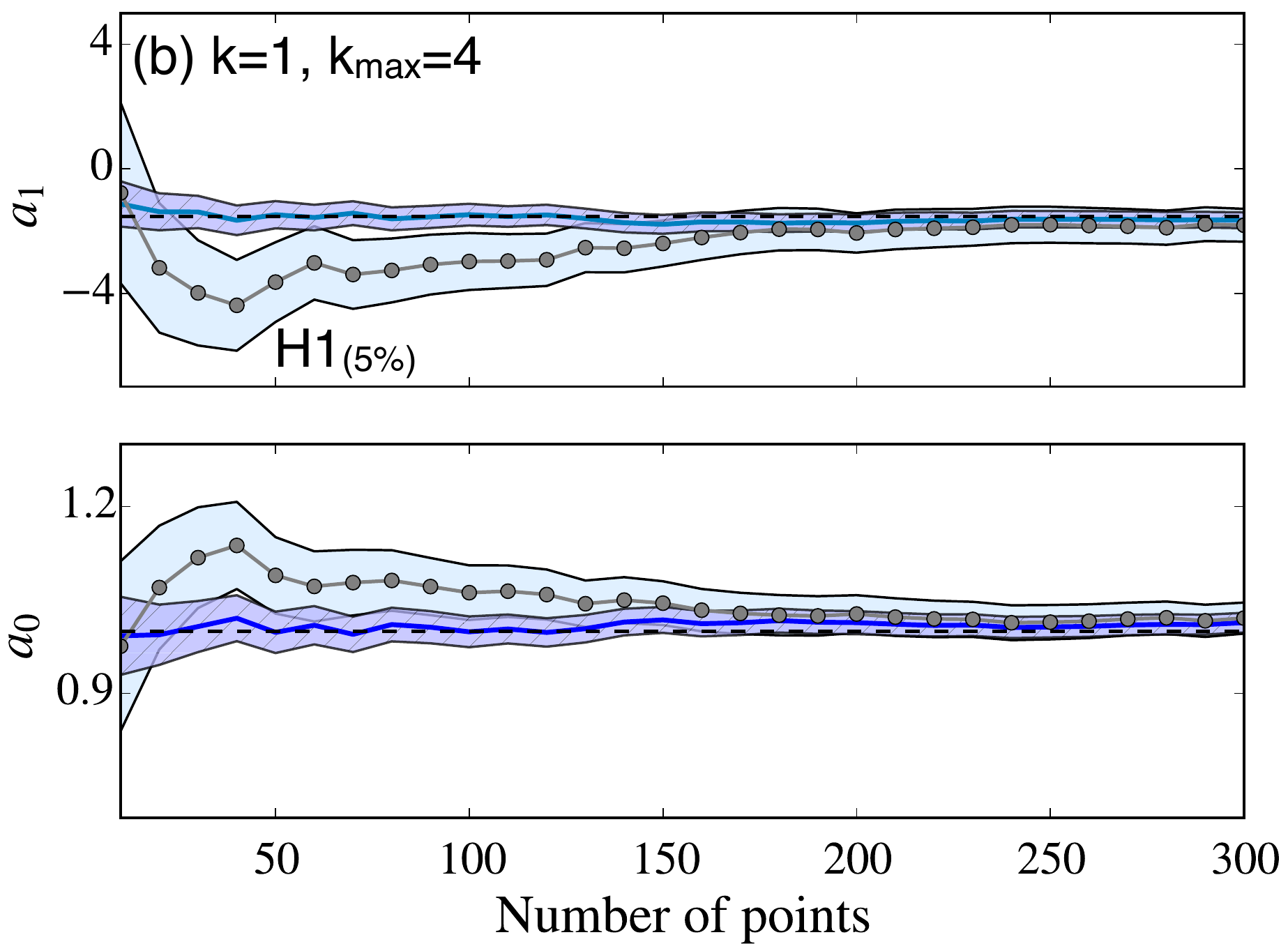}
   }
   \caption{Multi-set (a) and accumulation plots (b)
   calculated at $\kord=1$, $\kmax=4$ for data set \dataset{H}{1}{5}. The description
   is otherwise the same as Fig.~\ref{fig:datasets-H01}.
   }
\end{figure*}

\begin{figure*}[tbh]
  \subfloat{%
    \label{fig:H25_datasets_k1_kmax4}%
  \includegraphics[width=0.45\textwidth]{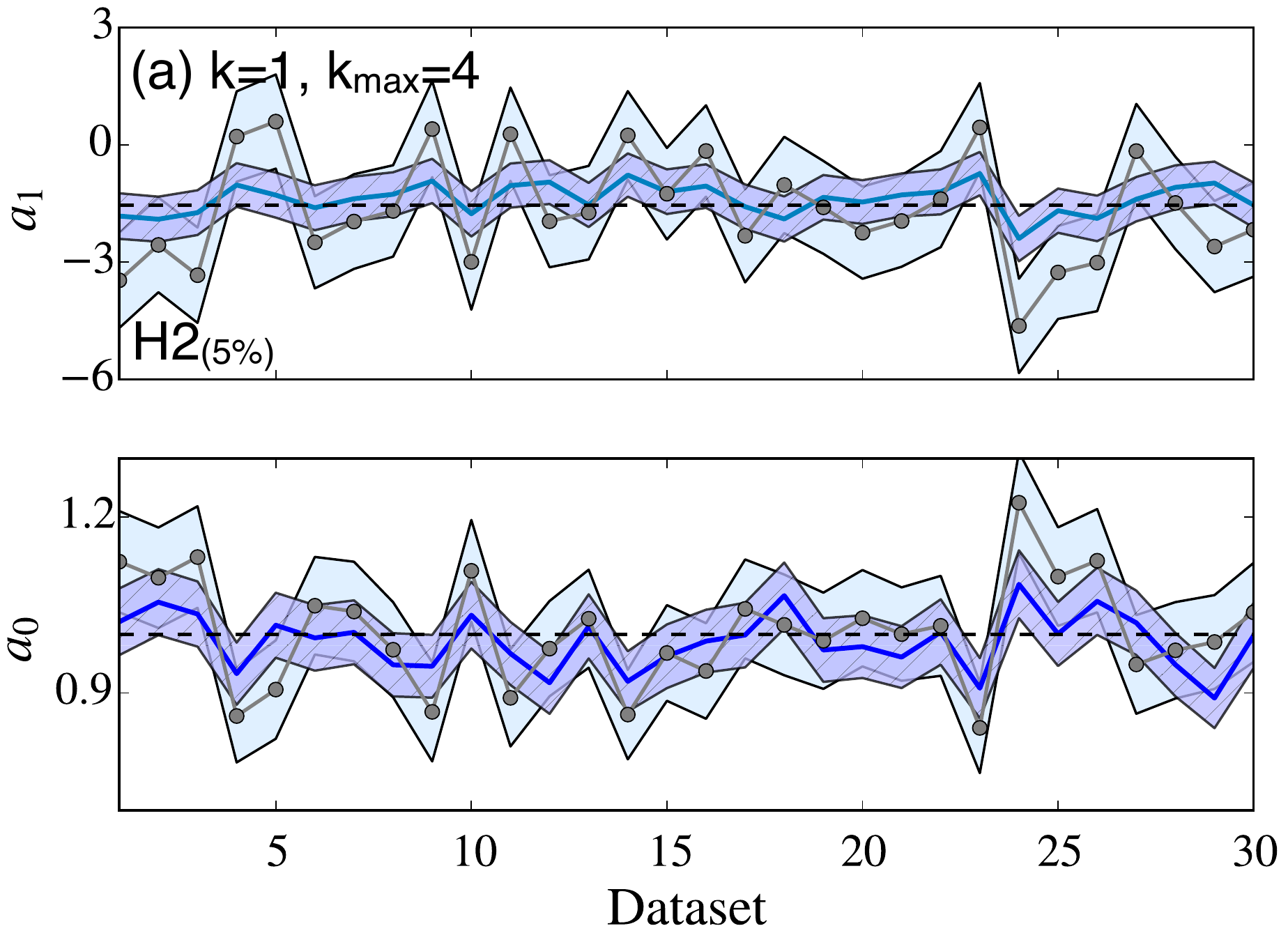}%
  }
  ~~~%
   \subfloat{%
     \label{fig:H25_accumulated_k1_kmax4}%
     \includegraphics[width=0.45\textwidth]{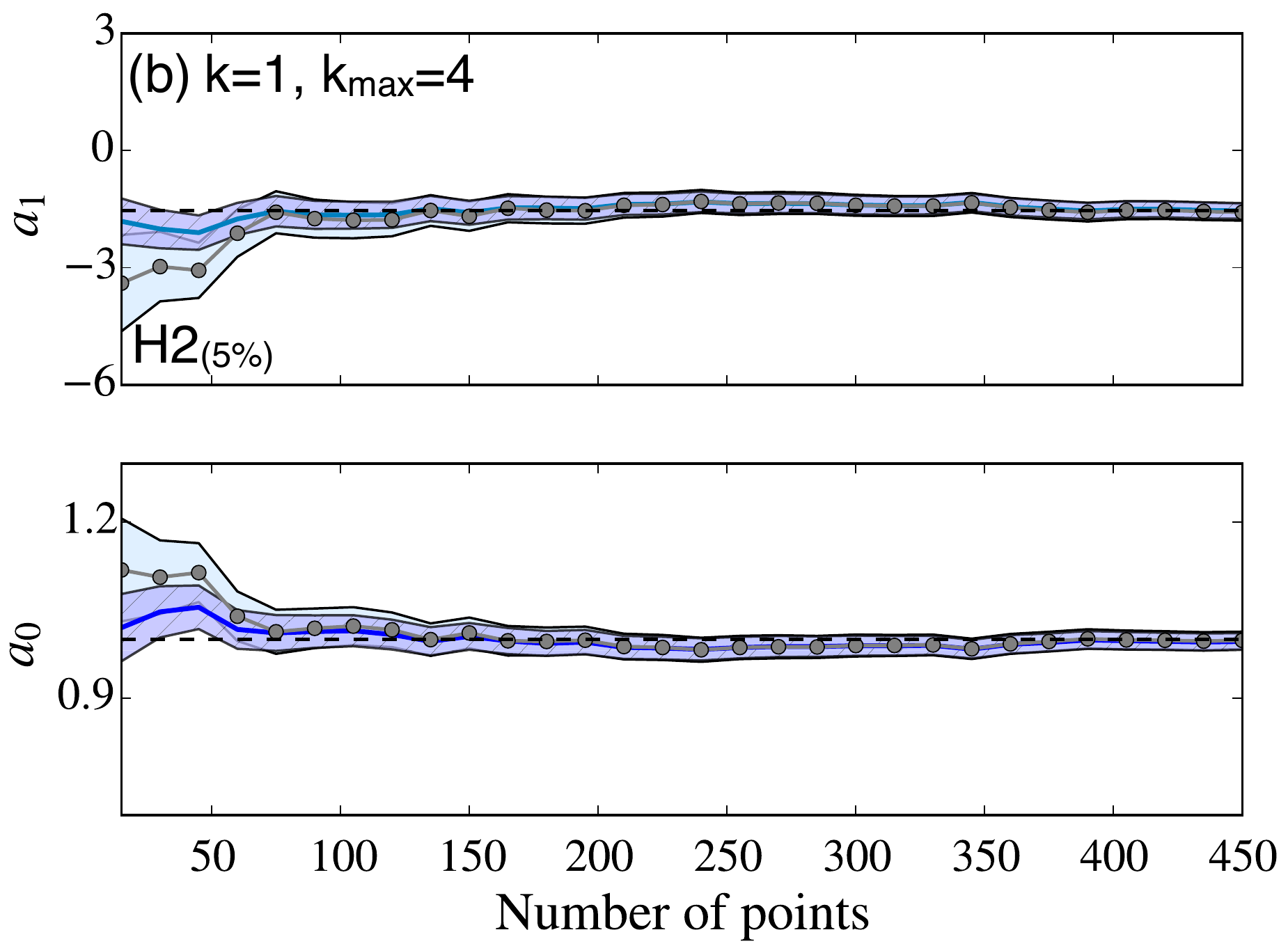}
   }
   \caption{Multi-set (a) and accumulation plots (b)
   calculated at $\kord=1$, $\kmax=4$ for data set \dataset{H}{2}{5}. The description
   is otherwise the same as Fig.~\ref{fig:datasets-H01}.
   }
\end{figure*}

\begin{figure*}[tbh]
  \subfloat{%
    \label{fig:H31_datasets_k1_kmax4}%
  \includegraphics[width=0.45\textwidth]{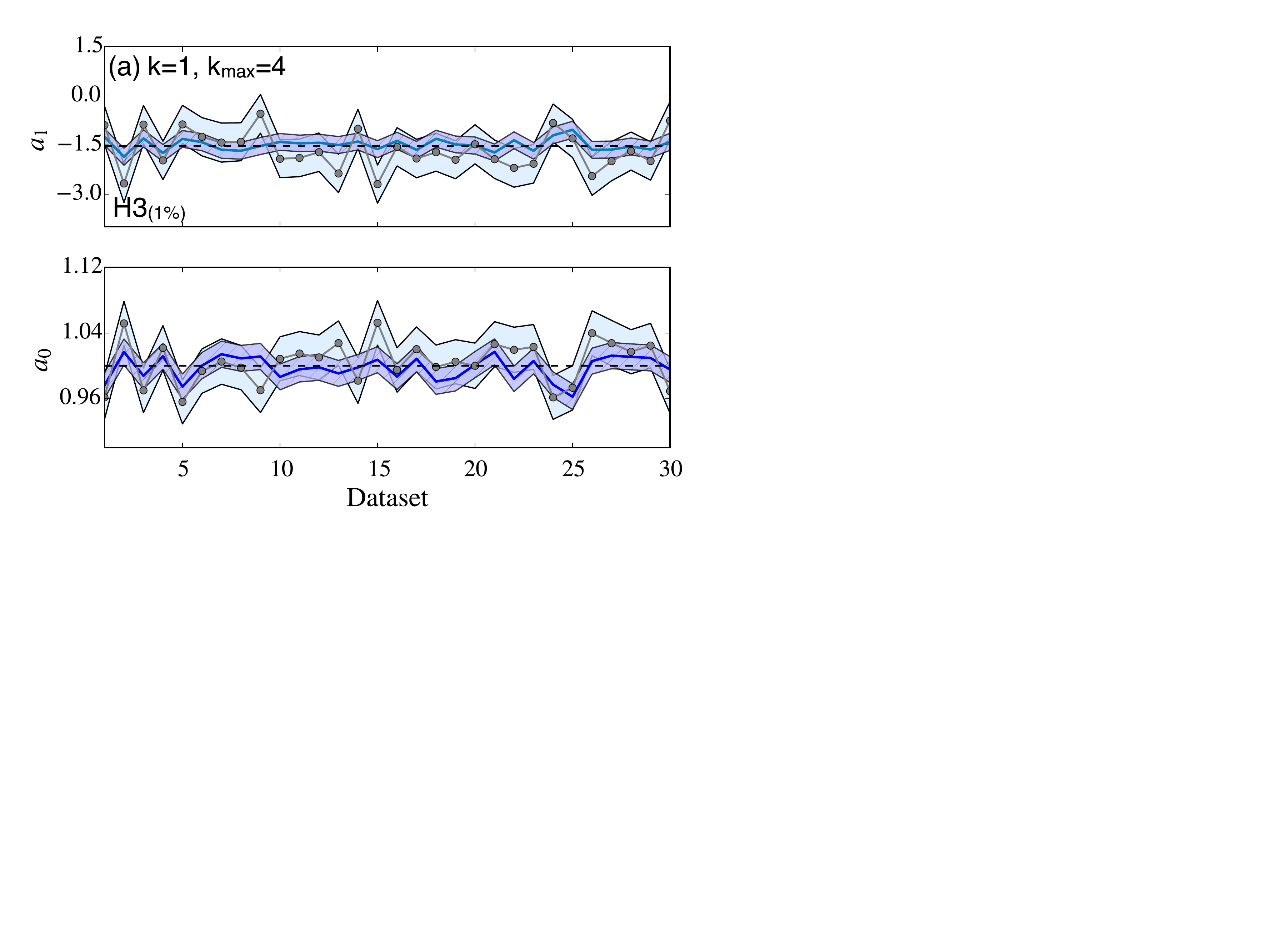}%
  }
  ~~~%
   \subfloat{%
     \label{fig:H31_accumulated_k1_kmax4}%
     \includegraphics[width=0.45\textwidth]{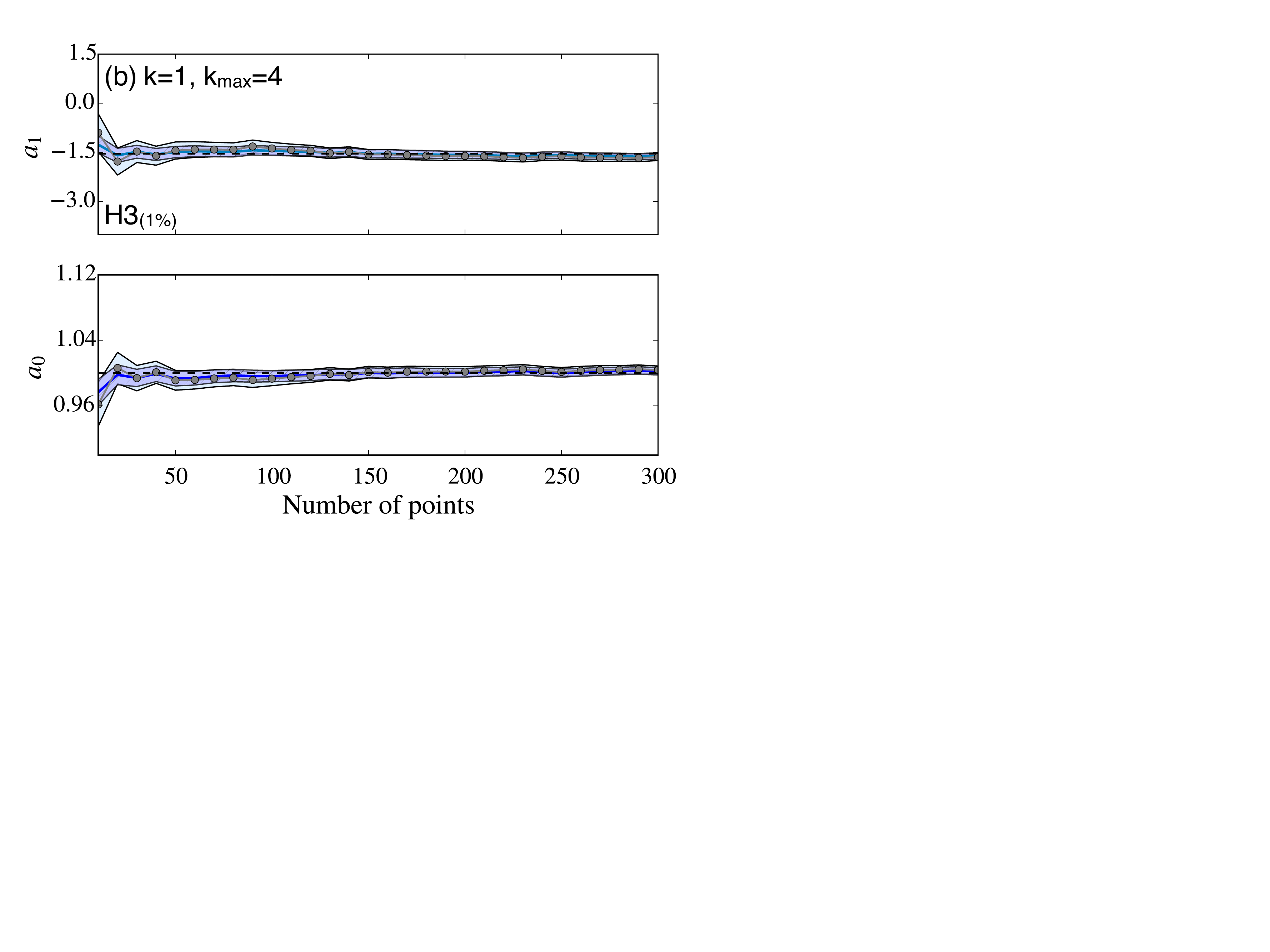}
   }
   \caption{Multi-set (a) and accumulation plots (b)
   calculated at $\kord=1$, $\kmax=4$ for data set \dataset{H}{3}{1}. The description
   is otherwise the same as Fig.~\ref{fig:datasets-H01}.}
\end{figure*}

\begin{figure}[tbh]
    \includegraphics[width=0.46\textwidth]{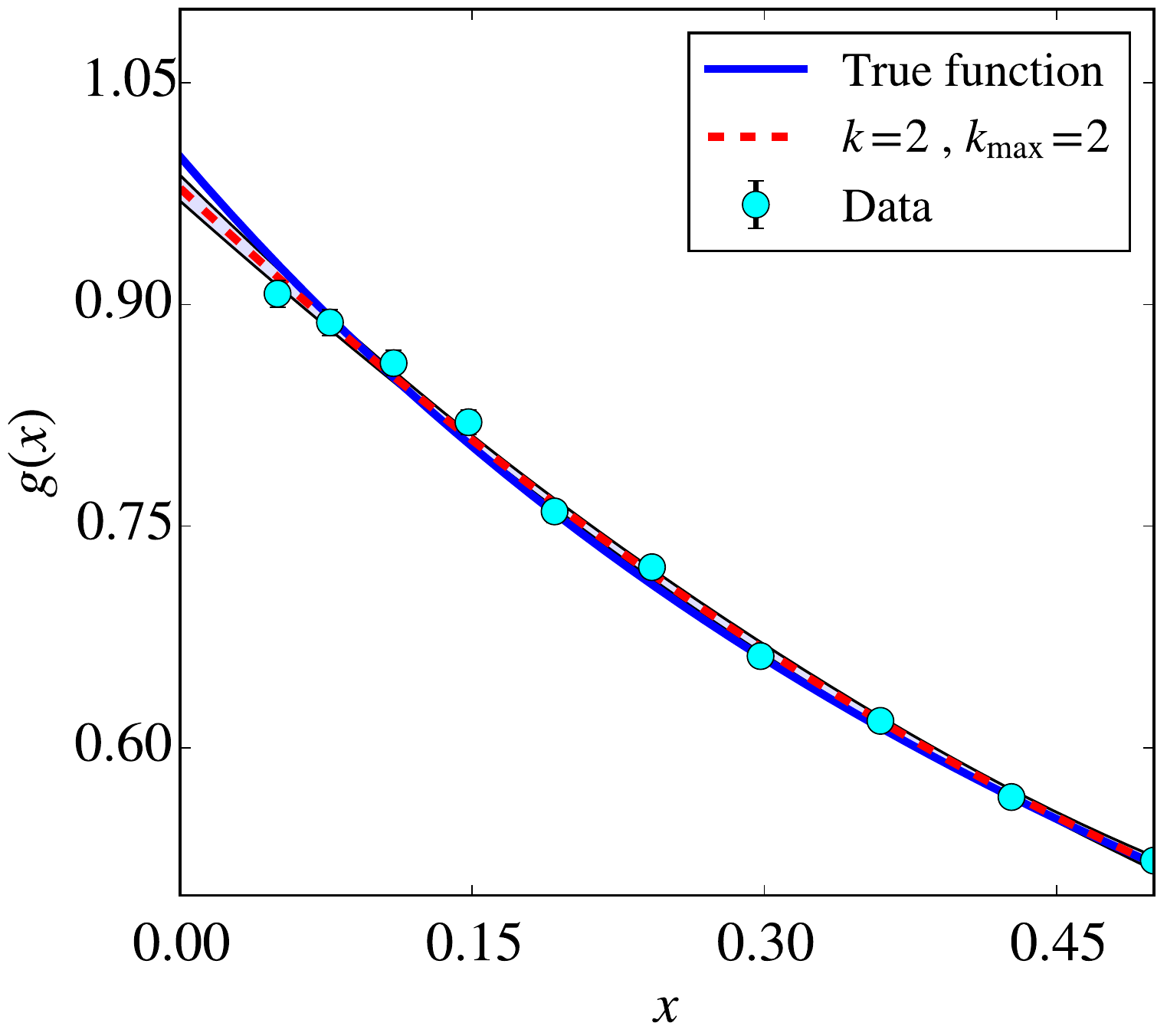}
    \caption{Comparison of data set \dataset{H}{3}{1} (corresponding to the fourth row of
    Table~\ref{tab:model-H-labels}), the underlying function for Model~H from
    Eq.~\eqref{eq:model-H}, and a Bayesian prediction using prior Set \Cprime\ with
    $\abarzero = 5$ calculated at order
    $\kord=2$, $\kmax=2$ from that data set. The error bands represent 68\% DoBs, which
    in this case are 1-$\sigma$ bands in the Gaussian approximation. \label{fig:H31-pred-abar5-k2} }
\end{figure}

\subsection{Validation: Multi-set analysis}

Finally, for the Validation step we focus on
how fluctuations in data affect our results as well as how the
errors in the available data limit parameter estimation. We posit that
some larger set of measurements for our model is available, which we
divide into smaller subsets and use to perform the parameter estimation
to see the effects of statistical fluctuations (multi-set plots outlined 
in Sec.~\ref{sec:multiset-accum}). We also take these data sets
and build them up sequentially to see how much data at that error level
is needed to produce precise estimates (accumulation plots outlined in
Sec.~\ref{sec:multiset-accum}).

\textbf{Small error, very small range.}
Figure~\subref*{fig:H01_datasets_k1_kmax3} shows a multi-set plot for
\dataset{H}{0}{1} at $\kord=1$ with $\kmax=3$, which is well into the evidence saturation
region. As overfitting occurs with the uniform prior, data fluctuations lead to 
fluctuations in the coefficients with large errors.
The results using the naturalness prior have a smaller spread about the central value
and smaller error bands.
As $\kmax$ is increased, the overfitting and coefficient fluctuations in the uniform prior
case become even more severe, while the results with the naturalness prior are insensitive
to increasing $\kmax$. Figure~\subref*{fig:H01_accumulated_k1_kmax3} shows how
the estimates improve as more data are accumulated (in sets of 10 sampled
the same way as \dataset{H}{0}{1}). Even with 300 points, the uniform prior
results are not as precise as the naturalness prior results. The advantage of extra prior
information is more in evidence in the top panel with $a_1$, where the uniform prior error
bands are much larger.

\textbf{Large error, small range.}
The multi-set plot for \dataset{H}{1}{5} is shown in Fig.~\subref*{fig:H15_datasets_k1_kmax4},
this time for $\kord=1$ with $\kmax=4$, which is in the saturation region.
Compared to \dataset{H}{0}{1}, the fluctuations of estimates are much larger,
due to the larger data error and sparser mesh, while the uniform-prior results
are fluctuating even more due to overfitting. The accumulated results for the uniform prior in 
Fig.~\subref*{fig:H15_accumulated_k1_kmax4}
are not consistent at the 1-$\sigma$ level
until about 150 points---and even there they still have errors that are 
larger due to different input information. The naturalness prior results are 
consistent with the true value even for only 10 points, but the error in the
estimates does not decrease much as more data are added. 
Any further increase in precision is limited by the range of the data and the fluctuations in it.

\textbf{Large error, larger range.}
The multi-set and accumulated results for \dataset{H}{2}{5} are shown in 
Figs.~\subref*{fig:H25_datasets_k1_kmax4} and
\subref*{fig:H25_accumulated_k1_kmax4} for $\kord=1$, $\kmax = 4$. Compared to
\dataset{H}{1}{5}, the fluctuations between data sets in Fig.~\subref*{fig:H25_datasets_k1_kmax4}
are not as large for either prior, but the naturalness prior reduces the spread.
The accumulated results in Fig.~\subref*{fig:H25_accumulated_k1_kmax4} show that the prior
improves the results until there are about 75 data points available and the results
with different priors become the same. However, for larger values of $\kmax$ the uniform
prior result will not converge to the Bayesian one as well as it does for $\kmax=4$, as
we saw for Model~D in Fig.~\ref{fig:D15-accum-k1-kmax6}.

\textbf{Small error, small range.}
The multi-set plot in Fig.~\subref*{fig:H31_datasets_k1_kmax4} 
for the first two parameters given \dataset{H}{3}{1},
 marginalized to $\kmax = 4$,
show tight error bands compared to the previous results using 5\% error data sets,
as would be expected because of the smaller 1\% relative error in the data.
The estimates in Fig.~\subref*{fig:H31_datasets_k1_kmax4} for the naturalness prior
fluctuate statistically about the true value.
Before, the results in the right-hand-side of Table~\ref{tab:H3-results-both}
showed that the 68\% interval missed the true parameter values, and a plot of results from that
estimation in Fig.~\ref{fig:H31-pred-abar5-k2} at $\kord=2,\kmax=2$ (in
the saturation region) shows a resulting prediction from that data set.
Fig.~\subref*{fig:H31_datasets_k1_kmax4} confirms that this was merely a result of
the random error in that data set. 
This is important to keep in mind: the available physical data could lead to 
this type of estimate in an EFT. Thus statistically meaningful error bars are important so that resulting
predictions have a true significance based on the data. The accumulated results in 
Fig.~\subref*{fig:H31_accumulated_k1_kmax4} show that even with just two data sets at this
value of $\kmax$, the naturalness and uniform-prior results become the same.

\textbf{Summary.}
The multi-set plots let us compare the impact of fluctuations on results using
different prior information. In some cases the accumulated results can provide a sanity check that
sufficient data will overwhelm the prior. The accumulated results
also illustrate that data errors can severely limit what can be
extracted from even a large amount of data when overfitting occurs. Using prior
naturalness information about the coefficients can help even when many data are available.

\subsection{Predictions}

After the Validation stage of Fig.~\ref{fig:flowchart}, the posteriors for
the LECs are available to make predictions of observables, with consistent
propagation of data and theory uncertainties.
Additional full loops in the flowchart may be appropriate to explore the impact on these
predictions of different priors.
In addition, we emphasize that truncation errors for predicted observables
must be included for a full uncertainty quantification.
We will not consider the Predictions stage further here, but reserve it for
applications to actual EFTs in future work.

It should be evident that the Bayesian framework we advocate is far from
automatic; there are many pitfalls along the way, but this is the nature
of the problem.
The simple case studies considered here
illustrate some general truths about parameter estimation for EFTs: 
the size of data errors and the range
of available data make a big difference; fluctuations happen; only so
much information is available and the Bayesian evidence is necessary to 
quantify it.
We have given examples of how
the multiple tools we have outlined can be used together to optimize the
extraction of the LECs and the uncertainty quantification.
But we recognize the additional burden on the practitioner, which can introduce a significant
and potentially prohibitive computational load.
An important topic for further study is to what degree one can lighten this
load while still working within the Bayes framework to provide robust
statistical error bars (e.g., as in Ref.~\cite{Higdon:2014tva}). 
We caution that taking too many shortcuts may lose information from the data 
that could improve your LECs.

\section{Additional Case Studies}  \label{sec:case_studies}

In this section we consider an additional set of parameter estimation case studies,
which highlight some particular difficulties that may be
encountered in EFT applications.  Our philosophy is
that validating the procedures and testing the diagnostics with model problems
(which at the least means analyzing the EFT with synthetic data),
where we can explore a full range of issues in fitting LECs with known answers,
is essential before turning to optimizations from real data.  The following examples
serve as prototypes for behavior we expect to see in ``real-world'' cases.

\subsection{Blind test: detecting unnaturalness}

\label{subsec:blind}

We performed several different blind tests of the procedure of 
Sec.~\ref{sec:model-problems} to verify that already knowing the answer 
has not been causing confirmation bias in our analyses. 
For example, one might worry that our use 
of a naturalness prior by construction could guarantee
that we will always find natural LECs unless we were already aware
of unnatural coefficients. The corresponding blind test, described below, 
included an unnatural
coefficient at a particular order, but the only information provided
to the tester was a set of data points and a specification of the basis
functions (the model EFT). The procedure set up in Sec.~\ref{sec:model-problems}
was applied to estimate the parameters of the model problem. 

The first signals of unnaturalness should be detected in the Guidance stage,
where the pdf of the naturalness parameter for the given prior choice is 
explored in the posterior pdf of $\abar$: $\pr(\abar|D,\kord,\kmax)$. During
the Parameter estimation stage, an observation of unnaturalness should be
confirmed by the evolution of the parameters with fixed $\abarzero$ (using
prior Set \Cprime\ or \Aprime), that is, the estimates will vary rapidly for
expected natural values of $\abarzero$, but will stabilize once $\abarzero$
is larger.

The case study we present here is a simple modification of Model~D from
Sec.~\ref{sec:procedures}. 
In particular, an unnatural coefficient was added to the 
model of Eq.~\eqref{eq:toy_func_SP}:
\beq
  g(x) = \left( \frac{1}{2} + \tan \left(\frac{\pi}{2} x\right) \right)^2 + 20 x^3 \;,
\eeq
which therefore has the same Taylor series as in Eq.~\eqref{eq:toy_func_SP} except
for an unnatural coefficient at third order in $x$, altering the convergence rate---but not
the convergence radius---of the expansion. 
Note that this is a somewhat
extreme example because the unnaturalness is very large and occurs at a
high order in the expansion.
We refer to this as Model~\Dtilde. We consider a data set \dataset{\Dtilde}{1}{5}
that consists of 10 points sampled in the range $0<x \leq 1/\pi$.

\begin{table*}[thb]
\caption{
  Coefficient estimates from sampling of $\pr(\avec|\dataset{\Dtilde}{1}{5},\kord,\kmax)$
   given the expansion from Eq.~\eqref{eq:model-th-expansion}  
   (these results are controlled by $\kmax$ only, see Sec.~\ref{subsec:setup}).
   The left side of the table is for a uniform prior, 
   which is equivalent to a least-squares fit, and includes the $\chi^2$/dof values.
   The right side of the table is using prior Set~\Cprime\ 
   from Table~\ref{tab:priors} with $\abarzero = 15$, and includes the evidence.
   For both priors the posterior pdf is a multi-dimensional
   Gaussian.
 \label{tab:Dtil-blind-both}}
  \begin{tabular}{|c|c||c|c|c|c||c|c|c|c|}
    \hline
     \multicolumn{2}{|c||}{} & \multicolumn{4}{|c||}{Uniform prior} &  \multicolumn{4}{|c|}{Gaussian prior} \\
    \hline
      $\kord$ & $\kmax$ & $\chi^2/$dof & $a_0$ & $a_1$ & $a_2$ 
      & Evidence & $a_0$ & $a_1$ & $a_2$ \\ 
    \hline
        0 & 0  & 100    & 0.47$\pm$0.01 &               &               & $\sim$ 0 & 0.47$\pm$0.01 &  &  \\
        1 & 1  & 9.9    & 0.13$\pm$0.01 & 3.8$\pm$0.1 &                 & 7.5$\times10^{-13}$& 0.13$\pm$0.01 & 3.8$\pm$0.1 &  \\
        2 & 2  & 2.2    & 0.28$\pm$0.02 & 0.30$\pm$0.5 & 13$\pm$2       & 5.1$\times10^{0}$& 0.28$\pm$0.02 & 0.35$\pm$0.4 & 13$\pm$2 \\ 
        2 & 3  & 2.4    & 0.25$\pm$0.04 & 1.3$\pm$1 & 4.1$\pm$9         & 5.4$\times10^{0}$& 0.27$\pm$0.03 & 0.81$\pm$0.7 & 8.6$\pm$5 \\ 
        2 & 4  & 2.4    & 0.35$\pm$0.1 & $-$3.1$\pm$3 & 63$\pm$40       & 5.4$\times10^{0}$& 0.26$\pm$0.03 & 0.88$\pm$0.8 & 8.2$\pm$5 \\ 
        2 & 5  & 2.9    & 0.36$\pm$0.1 & $-$3.7$\pm$8 & 75$\pm$100      & 5.4$\times10^{0}$& 0.26$\pm$0.03 & 0.92$\pm$0.8 & 8.1$\pm$5 \\ 
        2 & 6  & 0.73    & $-$0.54$\pm$0.3 & 55$\pm$20 & $-$1300$\pm$400 & 5.4$\times10^{0}$& 0.26$\pm$0.03 & 0.88$\pm$0.8 & 8.3$\pm$5 \\ 
\hline
        \multicolumn{3}{|c|}{True values}  &  0.25  &  1.57  &  2.47  &   &  0.25  &  1.57  &  2.47   \\
    \hline
\end{tabular}
\end{table*}

Beginning with the Guidance stage, we compute the evidence of \dataset{\Dtilde}{1}{5}
at several different orders and compute the posterior for $\abar$. Looking at
the evidence on the right-hand side of Table~\ref{tab:Dtil-blind-both} for Set~\Cprime, we see that
saturation occurs at $\kmax=3$, but that the ratio to $\kmax=2$ is not much 
greater than unity. Therefore, we expect that
the first three coefficients---$a_0$, $a_1$, and $a_2$---should be determined.
It is likely that the unnatural coefficient's value can {\it not} be obtained with any 
real precision, making this a bastard
of a blind test. 

\begin{figure}[tbh]
  \includegraphics[width=0.45\textwidth]{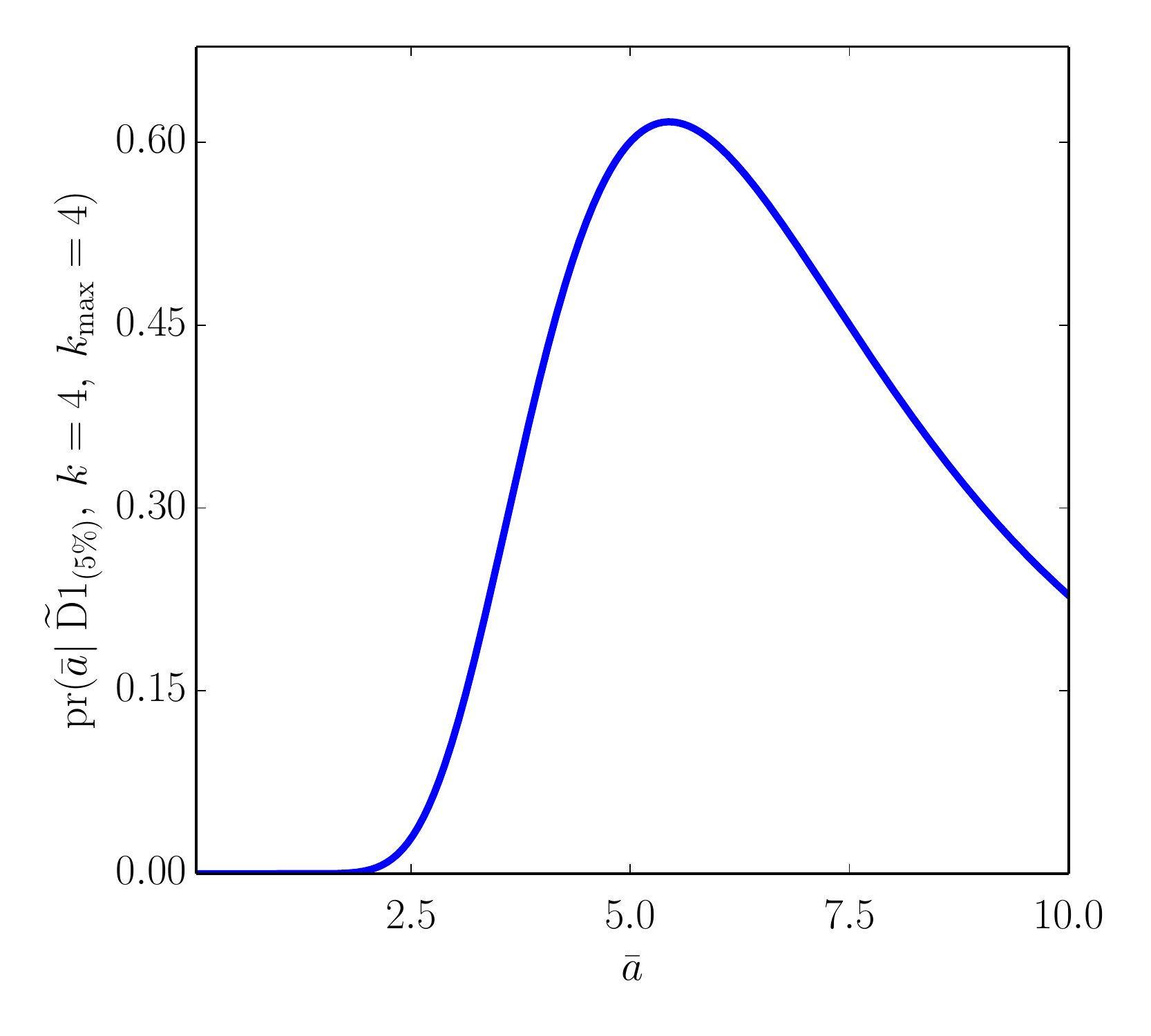}
  \caption{The posterior pdf $\pr(\abar|D,\kord,\kmax)$ assuming prior Set~C from Table~\ref{tab:priors} with
  $\abarmin=0.05$ and $\abarmax=20$, given
  data set \dataset{\Dtilde}{1}{5} calculated at $\kord=4$, $\kmax=4$.
   \label{fig:blind-abar-post-k4-kmax4}}
\end{figure}

The posterior for $\abar$ using prior Set~C is shown in 
Fig.~\ref{fig:blind-abar-post-k4-kmax4} for $\kmax=4$. This posterior
indicates that the most probable values for $\abar$ lie in
a wide range peaked around $\abar=5$, with a long tail extending to
much larger values. 
This is a clear signature of unnaturalness.
Choosing $\abar < 5$ will likely
cause the parameter estimates to be distorted by overly restricting
the prior. We therefore choose $\abarzero = 15$ in the extractions of
Table~\ref{tab:Dtil-blind-both} to avoid restricting the parameter space too much.

\begin{figure}[tbh]
\includegraphics[width=0.48\textwidth]{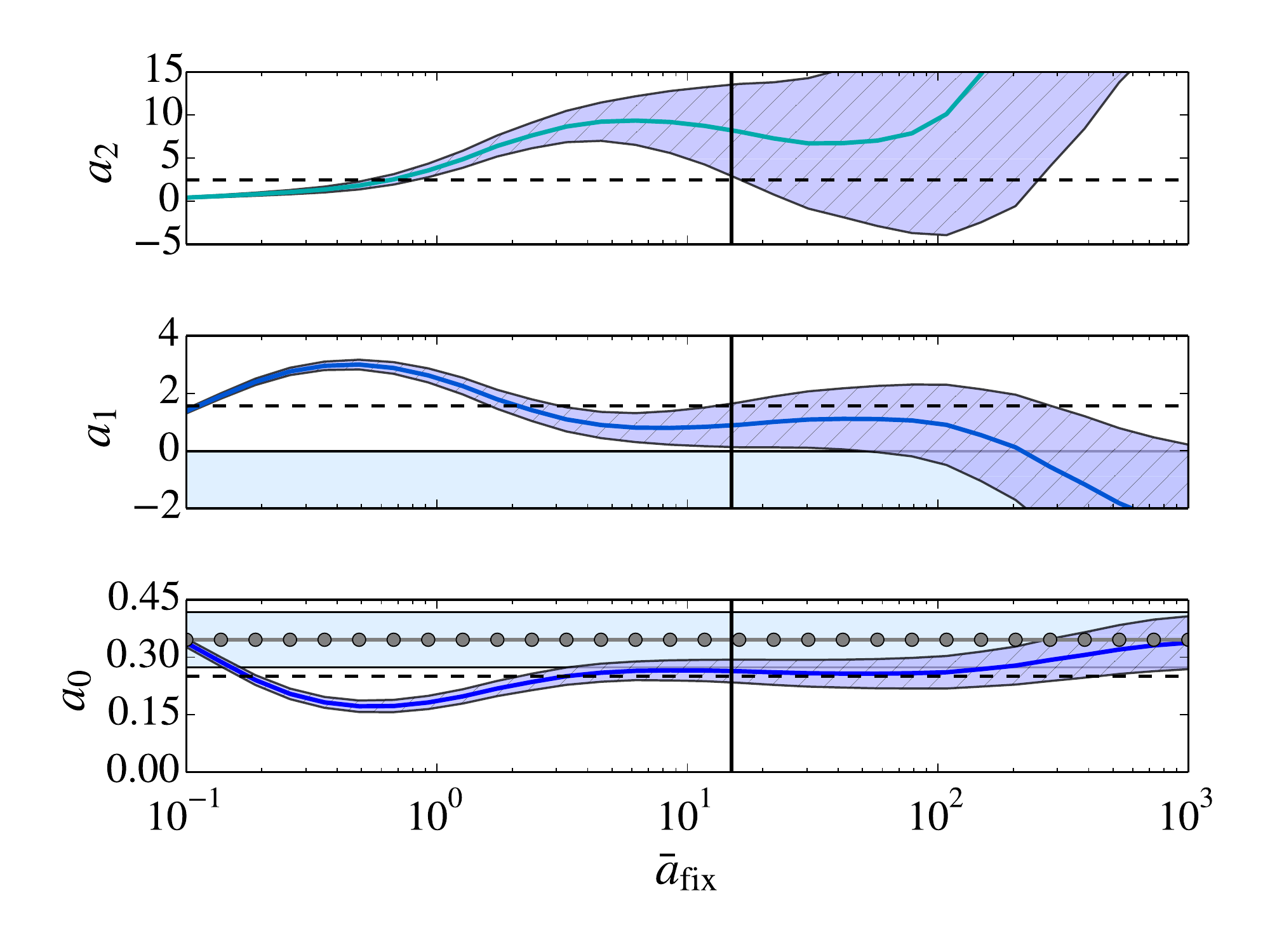}
\caption{(color online)
    Bayesian coefficient estimates (lines with darker hatched error bands)
    calculated at $\kord=2$, $\kmax=4$ as a function of $\abarzero$ using prior Set~\Cprime\
    given \dataset{\Dtilde}{1}{5}. The constant line with circles (when
    visible on the chosen y-scale) with lighter solid error bands is the least-squares estimate,
    which is independent of $\abarzero$. The error bands represent 68\% DoBs (1-$\sigma$ errors).
    The vertical lines in each panel at $\abarzero=15$ indicate the value chosen for Set~\Cprime\
    in the analysis.
\label{fig:abar-evol-blind-k2-kmax4}}
\end{figure}

Unnaturalness is further confirmed during the Parameter estimation stage.
Figure~\ref{fig:abar-evol-blind-k2-kmax4} shows how the first three
coefficients, which should be rather well constrained by the data, change
as a function of $\abarzero$ using prior Set \Cprime. Without the guidance
of the dashed lines indicating the true values of the underlying expansion,
there is a strong indication that $a_2$ is about 10, while the first two coefficients
are approximately natural. The evolution plateaus between $\abarzero$ of about 5 to 100,
after which it rapidly transitions to the least-squares result.

It is also difficult to see where the unnaturalness is present in the expansion.
In fact, if we did not marginalize over $a_3$ and $a_4$ with $\kmax=4$, the
results for these coefficients corresponding to that row in Table~\ref{tab:Dtil-blind-both}
are $a_3 = 8.02 \pm 12$ and $a_4 = 5.32 \pm 14$. The unnaturalness is present in
the estimate of $a_3$, which is, however, rather poorly determined. The correlations inherent in a 
(naive) polynomial fit result in the unnaturalness being shared between $a_3$ and its neighbors. 
I.e., the presence of
unnaturalness can distort the fit result of other coefficients which are natural. 
In this example it is---absent high-quality data in an appropriate region of $x$---difficult
to disentangle the specific order at which the one unnatural coefficient occurs. 

In spite of this, unnaturalness in the EFT fit can be identified using 
the procedure of Sec.~\ref{sec:model-problems}: there are
clear signals in the Guidance and Parameter estimation stages.
Disentangling at which order the unnatural LEC appears in the expansion
is difficult, and the presence of an unnatural LEC can distort
LECs which are in fact natural. With more information,
such as knowing which orders have natural and unnatural contributions,
we could update the prior pdfs for $\avec$ accordingly. 

\subsection{Data past breakdown of theory} \label{subsec:pastbreakdown}

An EFT expansion parameter will typically take the form of a ratio of scales that becomes one
at the breakdown scale, where dynamics not explicitly included in the EFT appear.
The analogous theory breakdown for our models is at the radius of convergence of the expansion.
We explore estimating the parameters of Model~H from Sec.~\ref{sec:model-problems}
using a data set sampled past the theory breakdown. In this
case we have data set \dataset{H}{4}{5}, which is described in Table~\ref{tab:model-H-labels}.
We have now three points sampled at $x\geq 1.3$ where the expansion
does not converge. Since the pole for Model~H is at negative $x$, the
function does not exhibit singular behavior at $x=1.3$ and so the breakdown
is ``hidden''.

Our aim is to determine signatures of the EFT breakdown using this data set, i.e., 
determine whether the procedure of Sec.~\ref{sec:model-problems} is sensitive to it.
We compute the evidence and $\abar$ posteriors to
complete the Guidance stage. The $\abar$ posteriors do not indicate unnaturalness
for any order, and so we compute the evidence for
\dataset{H}{4}{5} using $\abarzero=5$.

\begin{figure}[tbh]
    \includegraphics[width=0.45\textwidth]{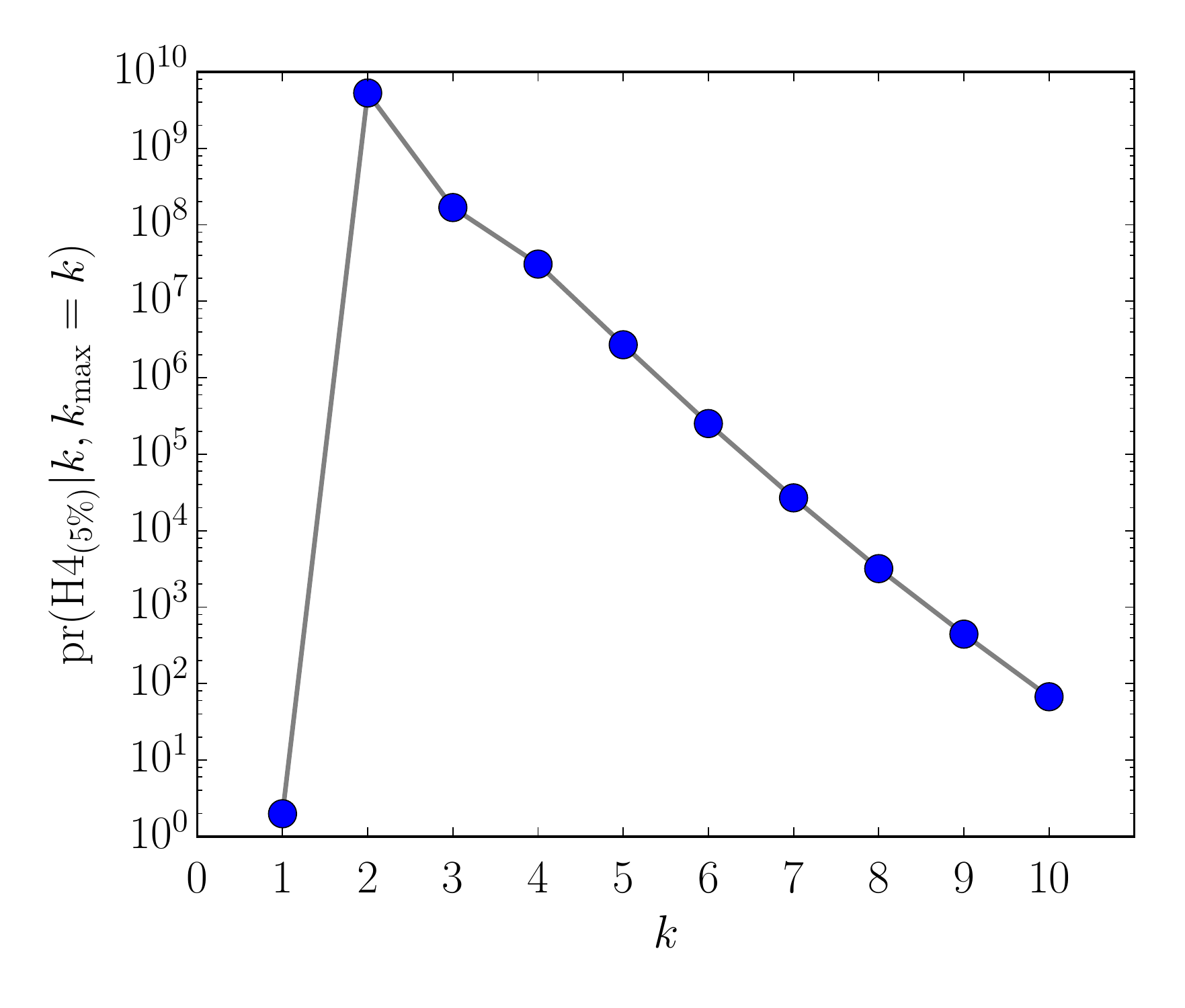}
    \caption{Evidence $\pr(\dataset{H}{4}{5}|\kord,\kmax=\kord)$ using prior
    Set \Cprime\ with $\abarzero=5$ for several values of $\kord$ with $\kmax = \kord$.
    (The evidence for $\kord = 0$ is not shown since it is nearly zero). 
    \label{fig:H45_evi_Cp5} }
\end{figure}

\begin{table*}[tbh]
\caption{
 Coefficient estimates from sampling of $\pr(\avec|\dataset{H}{4}{5},\kord,\kmax)$
   given the expansion from Eq.~\eqref{eq:model-th-expansion}, and the evidence
   at each order  (these results are controlled by $\kmax$ only, see Sec.~\ref{subsec:setup}).
   The prior used in these estimates was Set~\Cprime\ 
   from Table~\ref{tab:priors} with $\abarzero = 5$. 
 \label{tab:H45-abar-5}}
\begin{tabular}{|c|c||c|c|c|c|c|c|}
\hline
     $\kord$ & $\kmax$ & Evidence & $a_0$ & $a_1$ & $a_2$ & $a_3$ & $a_4$ \\ 
     \hline
        0 & 0 & $\sim$0                & 0.31$\pm$0.00 &  &  &  &  \\
        1 & 1 & 1.5$\times 10^1$       & 0.69$\pm$0.01 & $-$0.34$\pm$0.01 &  &  &  \\
        2 & 2 & 8.8$\times 10^{10}$    & 0.86$\pm$0.03 & $-$0.78$\pm$0.06 & 0.23$\pm$0.03 &  &  \\
        3 & 3 & 1.1$\times 10^{10}$    & 0.90$\pm$0.04 & $-$0.98$\pm$0.2  & 0.49$\pm$0.2 & $-$0.10$\pm$0.1 &  \\
        4 & 4 & 2.1$\times 10^{9}$     & 0.98$\pm$0.07 & $-$1.6$\pm$0.4   & 2.0$\pm$0.9 & $-$1.4$\pm$0.8 & 0.38$\pm$0.2 \\ 
        4 & 5 & 5.2$\times 10^{8}$     & 0.97$\pm$0.08 & $-$1.5$\pm$0.7   & 1.5$\pm$2 & $-$0.59$\pm$3 & $-$0.15$\pm$2 \\ 
        4 & 6 & 1.4$\times 10^{8}$     & 0.97$\pm$0.08 & $-$1.5$\pm$0.7   & 1.5$\pm$2 & $-$0.31$\pm$3 & $-$0.70$\pm$3 \\ 
  \hline
  \multicolumn{3}{|c|}{True values}  & 1.0 & $-1.54$ & 1.78 & $-1.82$ & $1.75$ \\
  \hline
  \end{tabular}
\end{table*}

\begin{figure}[tbh]
    \includegraphics[width=0.48\textwidth]{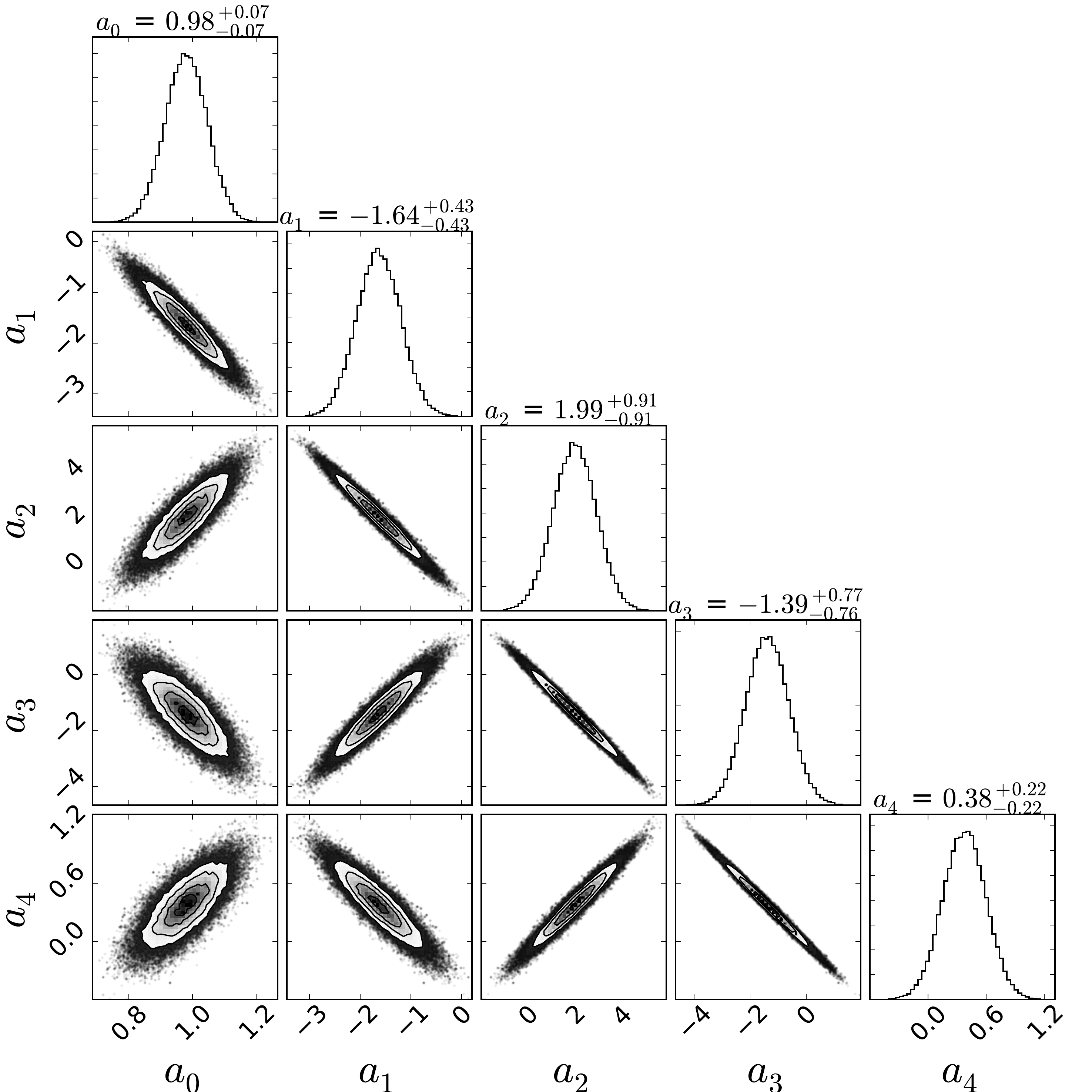}
    \caption{Projected posterior plot (see Fig.~\ref{fig:triangle_D1}) calculated at order $\kord=4$, $\kmax=4$ given
    data set \dataset{H}{4}{5} using prior Set~\Cprime\ with $\abarzero=5$.
    \label{fig:H45_gauss_abar_5_post}}
\end{figure}

Figure~\ref{fig:H45_evi_Cp5} shows the evidence values. Compared to the
saturation behavior using data sets sampled at smaller $x$ values, the behavior
of the evidence is strikingly different. The evidence decreases exponentially with
increasing order of the expansion, and the pattern continues without saturating past the highest order shown in
Fig.~\ref{fig:H45_evi_Cp5}. We turn to the Parameter estimation stage to
further examine the estimates with increasing $\kmax$. Table~\ref{tab:H45-abar-5} shows the
parameter estimates for the coefficients up to $\kmax=6$. Although the results for
$a_0$ and $a_1$ are fairly stable, the results for $a_2$ and beyond are not stable
as we saw in other test cases. Examining the projected posterior plot in Fig.~\ref{fig:H45_gauss_abar_5_post}
at $\kord=\kmax=4$ gives us a clear picture of the interaction of the likelihood with the prior
in this case. We see tight correlations between coefficients remain in spite of the
naturalness prior, unlike before where higher-order coefficients became almost
uncorrelated with leading-order coefficients.

\begin{figure*}[bht]
  \subfloat{%
    \label{fig:H45_xmax_k1}%
  \includegraphics[width=0.45\textwidth]{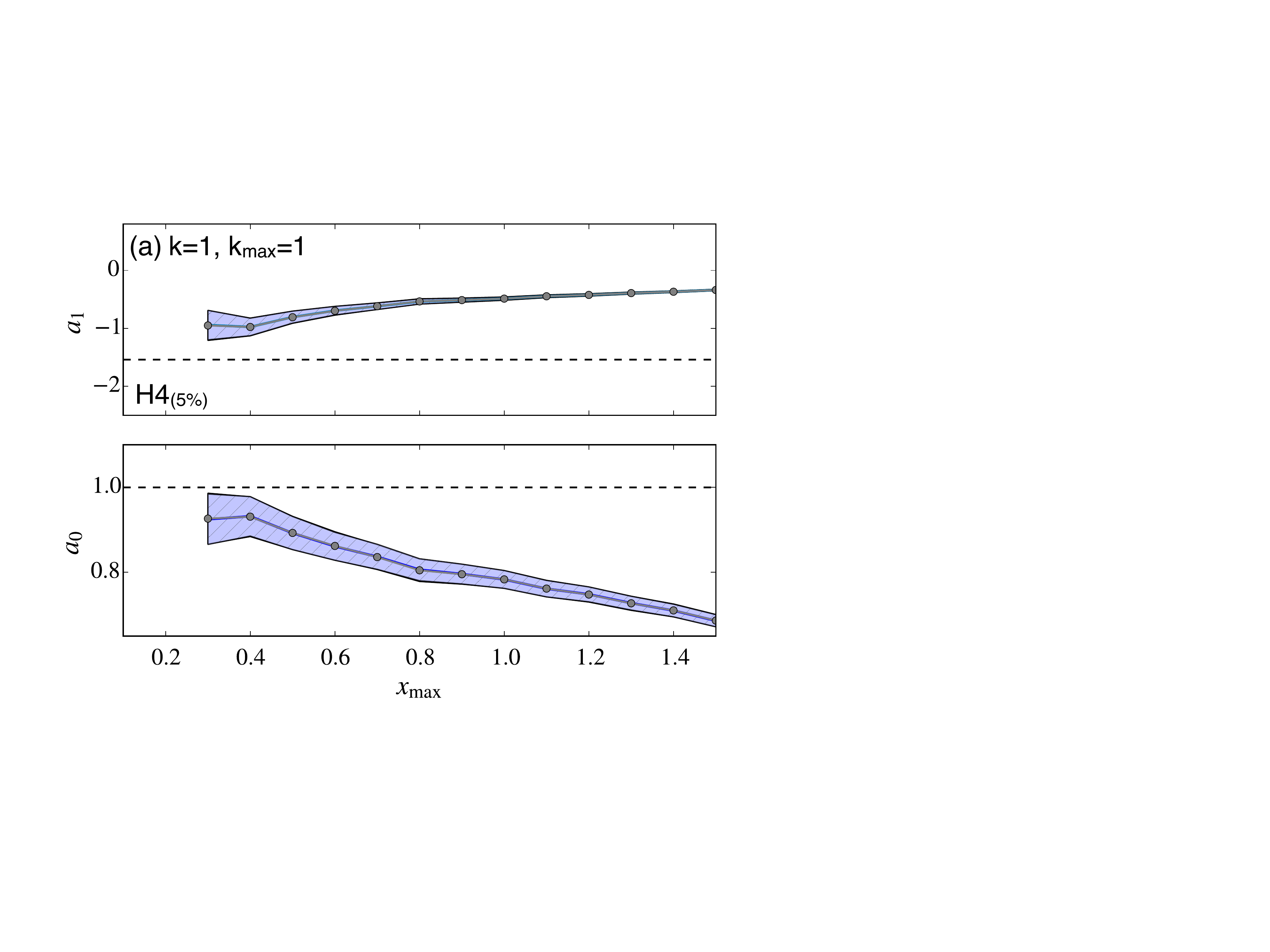}%
  }
  ~~~%
   \subfloat{%
     \label{fig:fig:H45_xmax_k2}%
     \includegraphics[width=0.45\textwidth]{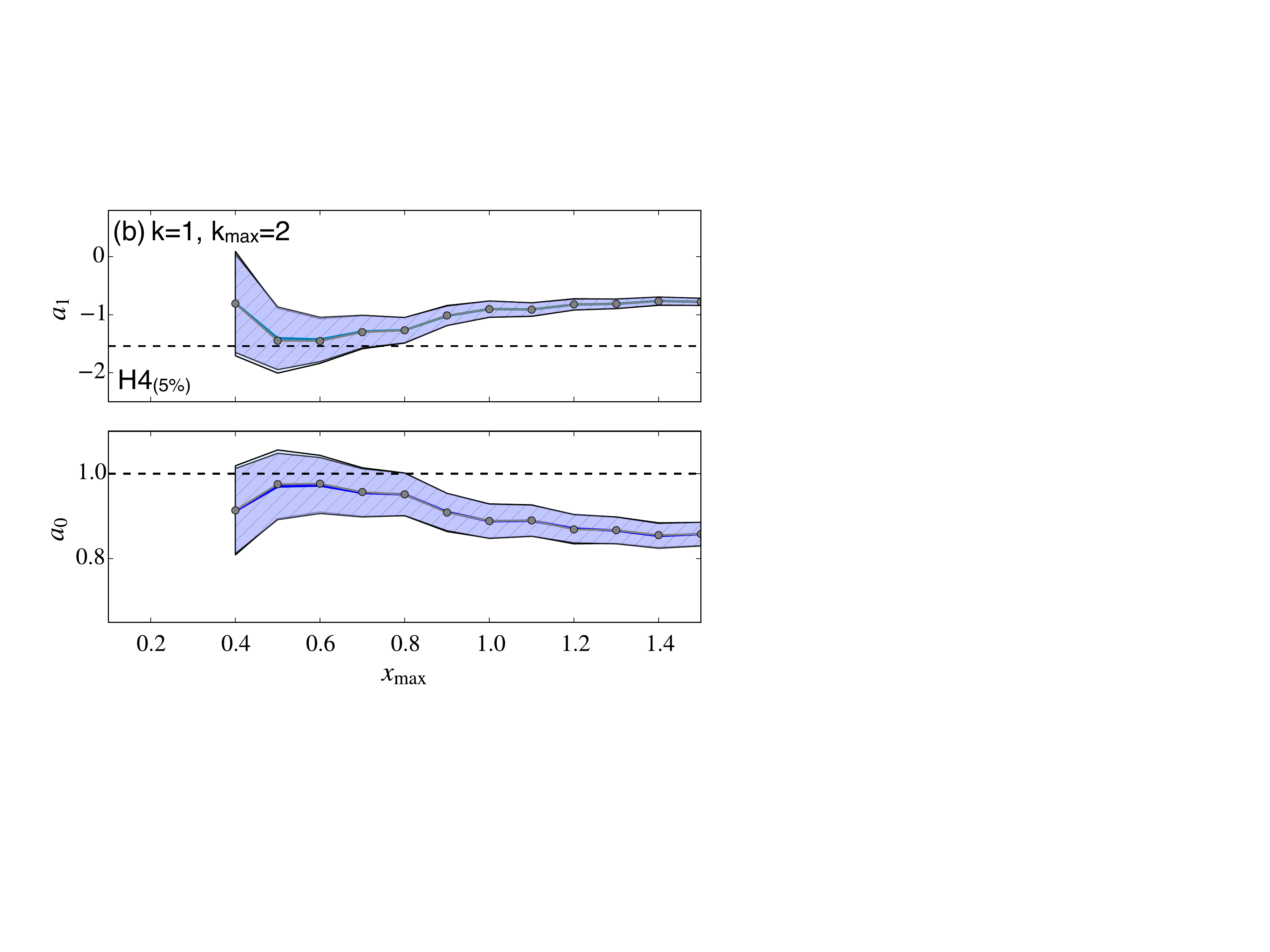}
   }
   \\%
   \subfloat{%
    \label{fig:H45_xmax_k3}%
  \includegraphics[width=0.45\textwidth]{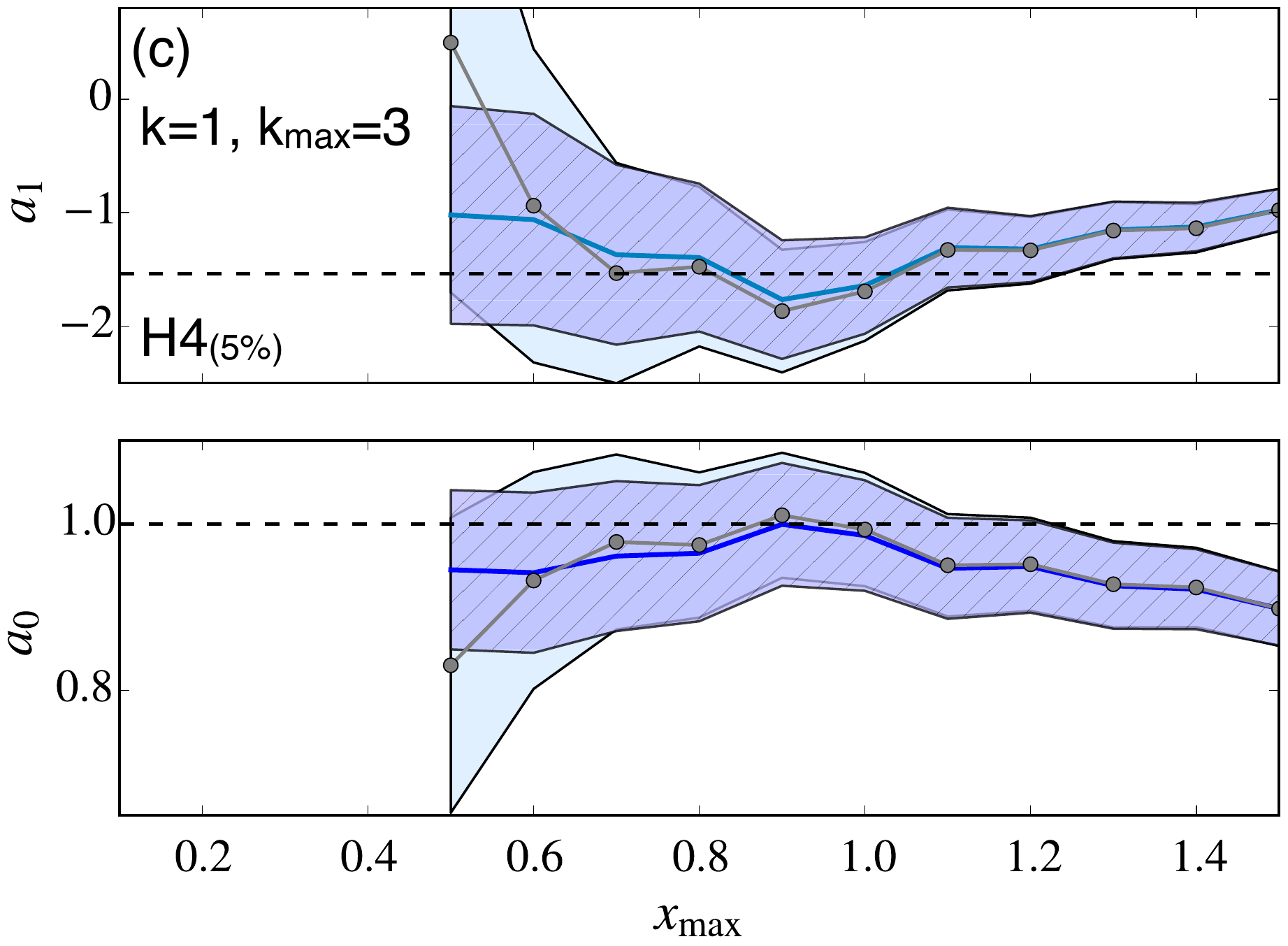}%
  }
  ~~~%
   \subfloat{%
     \label{fig:fig:H45_xmax_k4}%
     \includegraphics[width=0.45\textwidth]{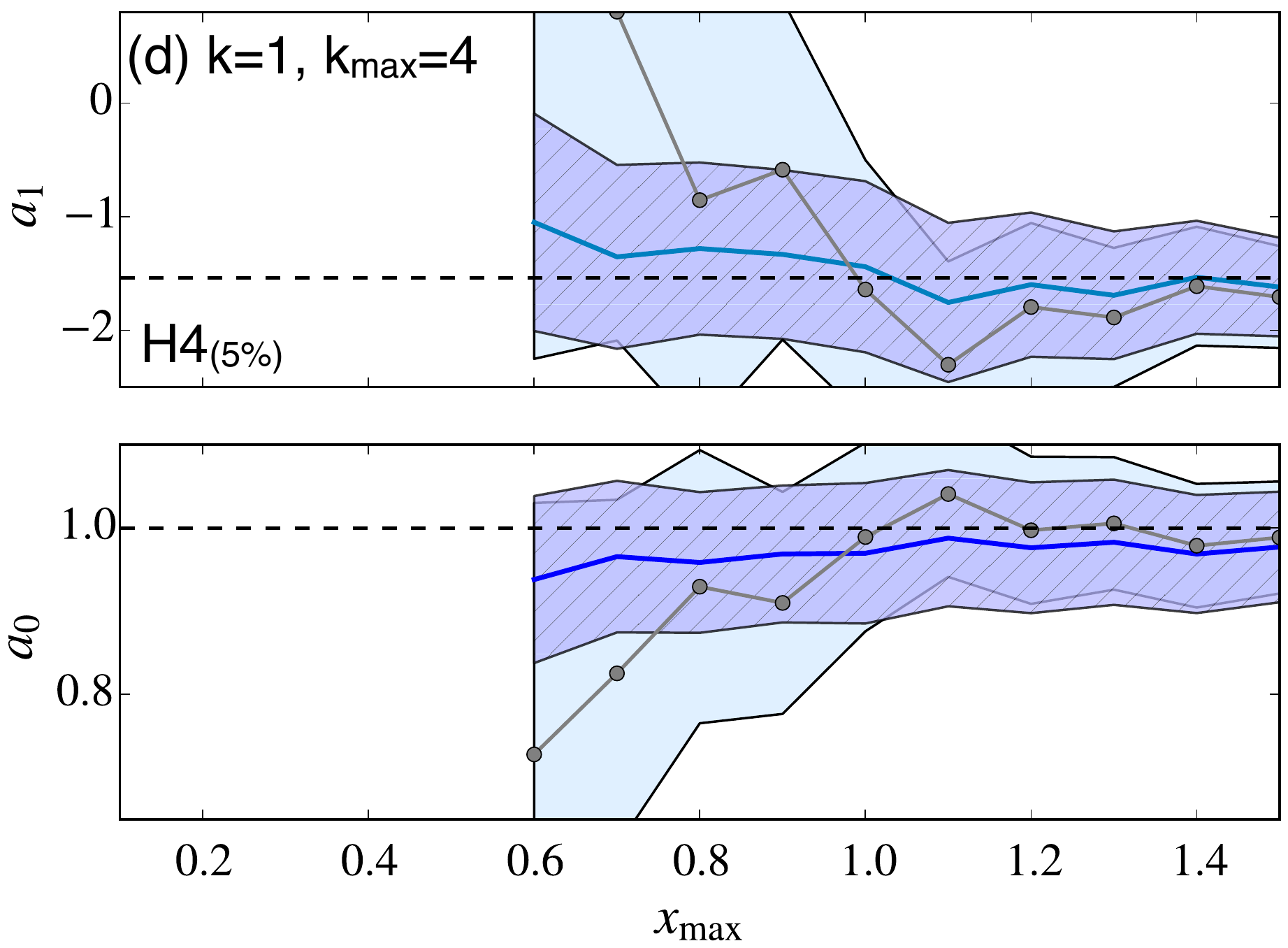}
   }
   \caption{Bayesian coefficient estimates
    as data from data set \dataset{H}{4}{5} are sequentially added at the
    high-$x$ end. The largest $x$-value in the set is denoted as $\xmax$.
    The solid lines with darker hatched error bands represent estimates
    using prior Set \Cprime\ with $\abarzero=5$, and the lines with circles
    with lighter solid error bands represent the least-squares estimates. 
    The error bands are 68\% DoBs (1-$\sigma$ errors), which coincide
    in (a) and (b).
   }
\end{figure*}

We continue with the Parameter estimation stage, examining the $\xmax$ plots
for \dataset{H}{4}{5}, which compare the prior Set~\Cprime\ results with the
uniform-prior results.  Figure~\subref*{fig:H45_xmax_k1} shows
the $\xmax$ plot for $\kord=\kmax=1$, and we continue up to
$\kord=1,\ \kmax=4$ in Fig.~\subref*{fig:fig:H45_xmax_k4}. In Fig.~\subref*{fig:fig:H45_xmax_k2},
which corresponds to the evidence peak in Fig.~\ref{fig:H45_evi_Cp5}, the results
are not stable with respect to $\xmax$, and without relying on the dashed
lines of the true values, no conclusive statement can be made. However, once
we go to order $\kmax=4$ in Fig.~\subref*{fig:fig:H45_xmax_k4}, the results are
fairly stable with respect to $\xmax$, but $a_0$ and $a_1$
remain significantly correlated with the higher-order coefficients.

When we have data over a large range in the expansion parameter, the correlations
between lower-order and higher-order coefficients may not be washed out by
the naturalness prior. Since we have data at values of $x$ where terms at different
orders have similar contributions, the higher-order coefficients are not
uniquely determined, resulting in large correlations and delicate cancellations.
The likelihood should be further analyzed using singular value decomposition (SVD)
to see the effects of the data. The example here shows that with enough low-$x$ data,
the leading behavior can still be extracted, but that further analysis is
necessary to find the best way to utilize the available data.


\subsection{Nucleon mass in $\chi$PT} \label{sec:nucleonmass}

Lattice calculations of observables near the physical
pion mass are computationally expensive. Therefore, there is a
need for extrapolations to the physical point, and chiral perturbation
theory (\chiPT) should be an ideal tool for extrapolations \cite{Bernard:2015wda}. However, 
the lack of lattice data near (or below) the physical point means
that extracting the coefficients in an expansion about zero pion mass can be difficult.
It often turns out that we need to disentangle contributions of different orders 
in the $\chi$PT expansion at a pion mass where the expansion
parameter is not particularly small. In this section we confront such difficulties, by applying our generic EFT tools
for parameter estimation to chiral extrapolations from synthetic
lattice data for the nucleon mass as a function of the pion mass, $m$. 

At a technical level these $\chi$PT expansions differ from the model
problems described above since there can be several LECs at a single order,
due to the possibility of different non-analytic functions multiplying the same
overall power of $m$. 
Orders can also be skipped, e.g., there is
no term $\propto m$ in the $\chi$PT expansion of the nucleon mass.
Therefore in this section we no longer refer to $\kord$ and $\kmax$, replacing them
by $\pord$ and $\pmax$, which indicate $\chi$PT orders, rather than numbers of LECs.

Following previous work in Ref.~\cite{Schindler:2008fh}, we fit the LECs of the expansion 
for the nucleon mass from SU(2) \chiPT. We are particularly interested in the
first two low-order coefficients. The leading coefficient is the nucleon
mass in the chiral limit, and the term at chiral order $\pord=2$ is related
to the pion-nucleon sigma term \cite{Bali20131}. 

The nucleon mass expansion in terms of the intrinsic
scale $\Lambda$ is given by \cite{Schindler:2008fh}:
\beq
  \begin{split}
  \frac{M_{N}(m)}{\Lambda} & = \frac{M_0}{\Lambda} + 
  \ktilde_1 \left(\frac{m}{\Lambda}\right)^2
   + \ktilde_2 \left(\frac{m}{\Lambda}\right)^3 \\
  & + \ktilde_3 \left(\frac{m}{\Lambda}\right)^4 \log\left(\frac{m}{\mu}\right)
   + \ktilde_4 \left(\frac{m}{\Lambda}\right)^4 \\
  & + \ktilde_5 \left(\frac{m}{\Lambda}\right)^5 \log\left(\frac{m}{\mu}\right)
    + \ktilde_6 \left(\frac{m}{\Lambda}\right)^5 \\
  & + \ktilde_7 \left(\frac{m}{\Lambda}\right)^6 \log\left(\frac{m}{\mu}\right)^2 
  + \ktilde_8 \left(\frac{m}{\Lambda}\right)^6 \log\left(\frac{m}{\mu}\right) \\
  & + \ktilde_9 \left(\frac{m}{\Lambda}\right)^6 + \ldots \;,
  \end{split}
  \label{eq:mn-expansion}
\eeq
where $m$ is the pion mass, $\mu$ is the renormalization scale for the loop contributions,
and the expansion is
up to sixth order in \chiPT. Since the expansion is about zero pion mass,
ideally we would determine the free parameters from data sampled
at small quark masses. We explore synthetic lattice data sampled at
various ranges of $m$, investigating the feasibility of chiral extrapolations.

We can define scale-invariant LECs in this expansion by expressing the
basis functions in terms of $m/\Lambda$, absorbing the $\mu$ dependence
into the coefficients. If we wish to take into account the contributions at that order
using marginalization, it is sufficient to account for the dominant
power $m^n$. 
However, we do not make this approximation here, instead accounting for all terms in the 
chiral expansion up to sixth order, as written in Eq.~\eqref{eq:mn-expansion}.

Following Ref.~\cite{Schindler:2008fh}, we generate synthetic data by 
taking $M_0 = 880\,$MeV and computing
values for the LECs of Eq.~\eqref{eq:mn-expansion} at
the renormalization scale $\mu = M_0$ with $\Lambda = 500\,$MeV.
These constants are given by (with $\ktilde_0 \equiv M_0/\Lambda$ and the relevant dimensions
of [GeV]$^{-n}$)~\cite{Schindler:2008fh, Schindler:2008fherratum}
\beq
  \begin{split}
    & \ktilde_0 = 1.76, \quad \ktilde_1 = 1.92, \quad \ktilde_2 = -1.41, \quad \ktilde_3 = 0.81, \\
    & \ktilde_4 = 1.03, \quad \ktilde_5 = 2.97, \quad \ktilde_6 = 4.41,\\
    & \ktilde_7 = 0.4, \quad \ktilde_8 = 0.31, \quad \ktilde_9 = -3.13.
  \end{split}
  \label{eq:mn-lec-vals}
\eeq
At this value of $\Lambda$ these LECs
are approximately natural, and henceforth we use it
in our calculations. 

\begin{figure}[tbh]
  \includegraphics*[width=0.98\columnwidth]{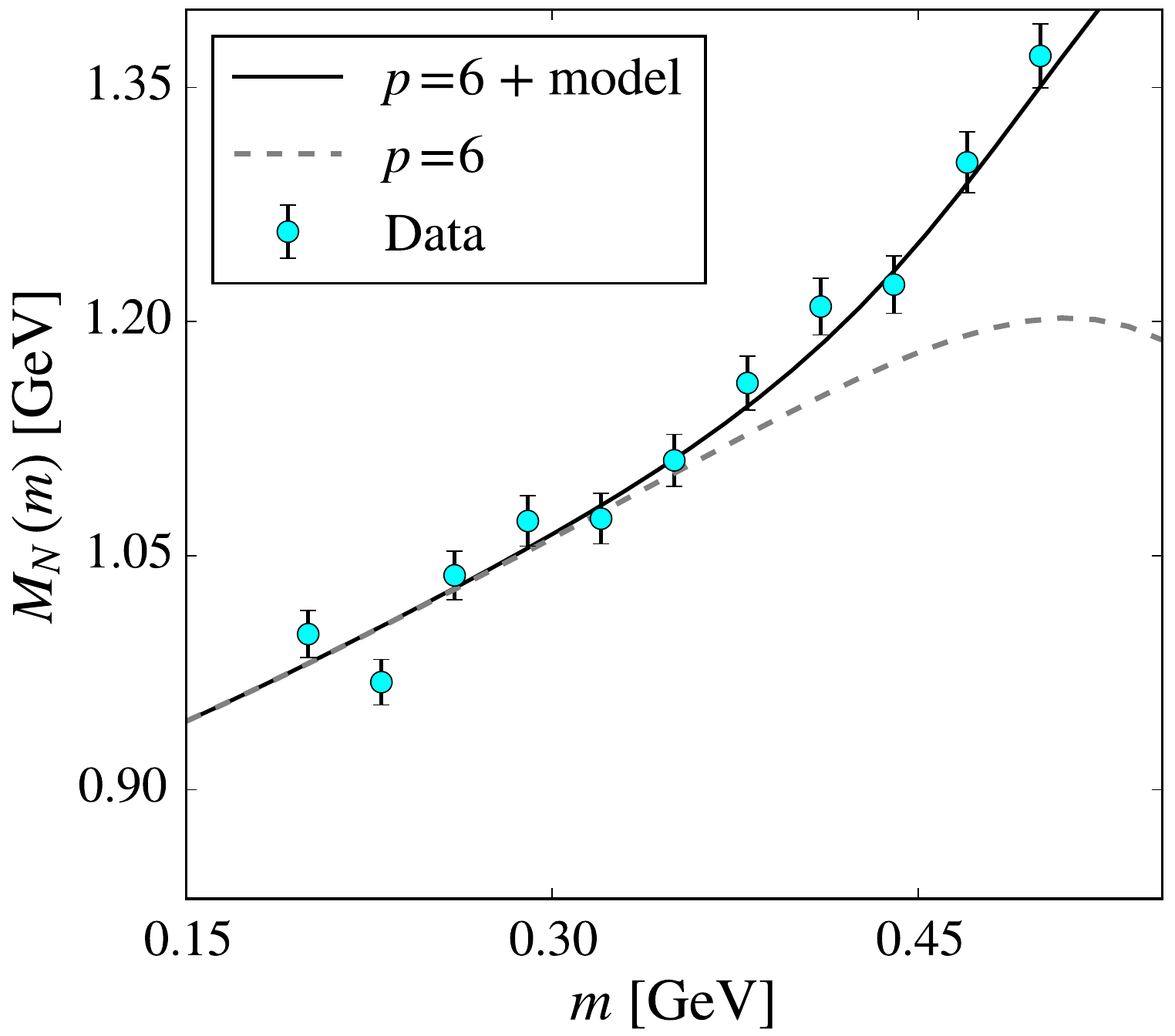}
  \caption{(color online)
  Comparison of data set \datasetmn{M}{N}{2}{1.5} (corresponding to the third row of
    Table~\ref{tab:MN-labels}, the underlying expansion to order $\pord=6$ from
    Eq.~\eqref{eq:mn-expansion}, and the function of Eq.~\eqref{eq:mn-full},
     from which the data set was sampled.
  \label{fig:MN2_data}}
\end{figure}

With these constants chosen, we add to the chiral expansion a 
model-dependent term that turns on slowly and becomes strong
around $\Lambda=500$ MeV which simultaneously turns off the expansion. The
synthetic data are generated from the function \cite{Schindler:2008fh}
\beq
  M(m) = M_N(m)\left(1 - g\left(\frac{m}{\Lambda} \right) \right) + M_{\text{model}}(m) \, g\left(\frac{m}{\Lambda}\right) \;.
  \label{eq:mn-full}
\eeq
This model-dependent term has the form
\beq
  M_{\text{model}}(m) = \alpha + \beta m \;,
  \label{eq:model-dep-mpi}
\eeq
where the values $\alpha = 1\,$GeV and $\beta = 1$ are selected following
Ref.~\cite{Schindler:2008fh}. The function that controls the dominance of each
term in Eq.~(\ref{eq:mn-full}) is \cite{Schindler:2008fh}
\beq
  g(x) = \frac{2}{\pi}\arctan (x^8) \;.
\eeq
Figure~\ref{fig:MN2_data} shows a plot of the function of Eq.~\eqref{eq:mn-full}
from which synthetic data are sampled compared to the chosen chiral expansion alone and a
data set \datasetmn{M}{N}{2}{1.5}.

We consider three data sets to represent different scenarios for
lattice simulations.
The pion mass meshes at which each was sampled are enumerated in
Table~\ref{tab:MN-labels}. We first examine two realistic cases of synthetic
data, \datasetmn{M}{N}{2}{1.5} and \datasetmn{M}{N}{1}{1.5}, which are sampled at and
above the physical pion mass with a 1.5\% relative error. After seeing the
limitations of data that simulates the current limitations of
lattice calculations
we explore an idealized set, 
\datasetmn{M}{N}{0}{1.5}, where the sampling begins at $m = 50\,$MeV.

\begin{table}[tb]
    \caption{Nucleon mass data set labels for sampling
    grid ranges and number of points, where $m$ is the pion mass.
     \label{tab:MN-labels}}
    \begin{tabular}{|c|c|c|c|}
        \hline
        Label & \# of pts. & Grid & Spacing \\
        \hline
        MN0    & 11 &$0.05 \leq m \leq 0.15$ GeV & linear \\
        MN1    & 11 &$0.14 \leq m \leq 0.44$ GeV & linear \\
        MN2    & 11 &$0.2 \leq m \leq 0.5$ GeV & linear \\
        \hline
    \end{tabular}
\end{table}

\begin{table*}[thb]
\caption{
   Coefficient estimates from sampling of $\pr(\ktildevec|\datasetmn{M}{N}{1}{1.5},\pord,\pmax)$
   given the nucleon mass expansion from Eq.~\eqref{eq:mn-expansion}
    (these results are controlled by $\pmax$ only, see Sec.~\ref{subsec:setup}).
   The left side of the table is for a uniform prior, 
   which is equivalent to a least-squares fit, and includes the $\chi^2$/dof values.
   The right side of the table is using prior Set~\Cprime\ 
   from Table~\ref{tab:priors} with $\abarzero = 5$, and includes the evidence. 
   For both priors the posterior pdf is a multi-dimensional Gaussian.
 \label{tab:MN1-results-both}}
  \begin{tabular}{|c|c||c|c|c|c||c|c|c|c|}
    \hline
     \multicolumn{2}{|c||}{} & \multicolumn{4}{|c||}{Uniform prior} &  \multicolumn{4}{|c|}{Gaussian prior} \\
    \hline
      $\pord$ & $\pmax$ & $\chi^2/$dof & $\ktilde_0$ & $\ktilde_1$ & $\ktilde_2$ 
      & Evidence & $\ktilde_0$ & $\ktilde_1$ & $\ktilde_2$ \\ 
    \hline
        0 & 0  & 35    & 2.1$\pm$0.01  &                &                           & $\sim$0           & 2.1$\pm$0.01 &  &  \\
        2 & 2  & 0.85    & 1.8$\pm$0.02 & 0.82$\pm$0.04 &                           & 4.1$\times10^5$ & 1.8$\pm$0.0 & 0.82$\pm$0.04 &  \\
        3 & 3  & 0.96    & 1.8$\pm$0.04 & 0.82$\pm$0.4  & 0.002$\pm$0.4             & 3.0$\times10^4$ & 1.8$\pm$0.0 & 0.82$\pm$0.3 & 0.00$\pm$0.4 \\ 
        3 & 4  & 0.91    & 1.6$\pm$0.3 & 7.7$\pm$10     & $-$27$\pm$60              & 1.4$\times10^4$ & 1.8$\pm$0.1 & 2.1$\pm$1 & $-$1.2$\pm$4 \\ 
        3 & 5  & 0.98    & 5.1$\pm$4 & $-$490$\pm$600   & 5800$\pm$6000             & 6.9$\times10^3$ & 1.8$\pm$0.1 & 1.9$\pm$2 & $-$0.97$\pm$4 \\ 
        3 & 6  & 0.53   & $-$2000$\pm$1000 & 8.8$\times10^5$$\pm$5$\times10^5$ & $-$2.4$\times10^7$$\pm$1$\times10^7$
                                                & 3.9$\times10^3$ & 1.8$\pm$0.1 & 1.8$\pm$2 & $-$0.83$\pm$4 \\ 
\hline
        \multicolumn{3}{|c|}{True values}  &  1.76 &  1.92 &  $-$1.41 &  &  1.76 &  1.92 &  $-$1.41    \\
    \hline
\end{tabular}
\end{table*}
\begin{table*}[thb]
\caption{
   Same as Table~\ref{tab:MN1-results-both} except sampling from 
   $\pr(\ktildevec|\datasetmn{M}{N}{2}{1.5},\pord,\pmax)$.
    \label{tab:MN2-results-both}}
  \begin{tabular}{|c|c||c|c|c|c||c|c|c|c|}
    \hline
     \multicolumn{2}{|c||}{} & \multicolumn{4}{|c||}{Uniform prior} &  \multicolumn{4}{|c|}{Gaussian prior} \\
    \hline
      $\pord$ & $\pmax$ & $\chi^2/$dof & $\ktilde_0$ & $\ktilde_1$ & $\ktilde_2$ 
      & Evidence & $\ktilde_0$ & $\ktilde_1$ & $\ktilde_2$ \\ 
    \hline
        0 & 0  & 51    & 2.2$\pm$0.01  &               &                                               & $\sim$0           & 2.2$\pm$0.01 &  &  \\
        2 & 2  & 1.4    & 1.8$\pm$0.02 & 0.89$\pm$0.04 &                                               & 1.4$\times10^4$ & 1.8$\pm$0.02 & 0.89$\pm$0.04 &  \\
        3 & 3  & 1.4    & 1.9$\pm$0.1 & 0.49$\pm$0.4 & 0.37$\pm$0.3                                    & 1.8$\times10^3$ & 1.9$\pm$0.1 & 0.47$\pm$0.4 & 0.38$\pm$0.3 \\ 
        3 & 4  & 1.7    & 2.6$\pm$0.7 & $-$21$\pm$20 & 89$\pm$80                                       & 3.1$\times10^2$ & 1.8$\pm$0.1 & 0.98$\pm$2 & 0.10$\pm$4\\ 
        3 & 5  & 1.2    & 37$\pm$20 & $-$2700$\pm$1000 & 2.5$\times10^4$$\pm$1$\times10^4$             & 1.2$\times10^2$ & 1.8$\pm$0.1 & 0.61$\pm$2 & 0.49$\pm$4 \\ 
        3 & 6  & 0.20    & 450$\pm$8000 & $-$1.3$\times10^5$$\pm$2$\times10^6$ & 3.1$\times10^6$$\pm$5$\times10^7$
                                                                                                       & 6.0$\times10^1$ & 1.9$\pm$0.2 & 0.14$\pm$2 & 0.87$\pm$4 \\ 
\hline
        \multicolumn{3}{|c|}{True values}  &  1.76 &  1.92 &  $-$1.41 &  &  1.76 &  1.92 &  $-$1.41    \\
    \hline
\end{tabular}
\end{table*}

\begin{table*}[thb]
\caption{Same as Table~\ref{tab:MN1-results-both} except sampling from 
   $\pr(\ktildevec|\datasetmn{M}{N}{0}{1.5},\pord,\pmax)$.\label{tab:MN0-results-both}}
  \begin{tabular}{|c|c||c|c|c|c||c|c|c|c|}
    \hline
     \multicolumn{2}{|c||}{} & \multicolumn{4}{|c||}{Uniform prior} &  \multicolumn{4}{|c|}{Gaussian prior} \\
    \hline
      $\pord$ & $\pmax$ & $\chi^2/$dof & $\ktilde_0$ & $\ktilde_1$ & $\ktilde_2$ 
      & Evidence & $\ktilde_0$ & $\ktilde_1$ & $\ktilde_2$ \\ 
    \hline
        0 & 0  & 5.4    & 1.82$\pm$0.01  &               &                          & 0.015          & 1.82$\pm$0.01 &  &  \\
        2 & 2  & 1.1    & 1.73$\pm$0.02 & 2.2$\pm$0.33 &                           & 4.0$\times10^6$ & 1.73$\pm$0.02 & 2.2$\pm$0.3 &  \\
        3 & 3  & 1.2    & 1.71$\pm$0.03 & 3.7$\pm$2.7 & $-$4.7$\pm$8              & 3.3$\times10^6$  & 1.72$\pm$0.02 & 2.4$\pm$1.4 & $-0.72\pm$4 \\ 
        3 & 4  & 1.3    & 2.1$\pm$0.3 & $-$140$\pm$110 & 2000$\pm$2000             & 3.1$\times10^6$ & 1.72$\pm$0.02 & 2.5$\pm$1.4 & $-0.66\pm$4 \\ 
        3 & 5  & 1.9    & 1.5$\pm$5 & 450$\pm$5000 & $-$16000$\pm$2$\times 10^5$ & 3.1$\times10^6$   & 1.72$\pm$0.02 & 2.5$\pm$1.4 & $-$0.60$\pm$4 \\ 
        3 & 6  & 0.25    & 1500$\pm$1000 & $-$4.9$\times10^6$ $\pm$5$\times10^6$ &     
        3.7$\times10^7$ $\pm$4$\times10^7$                                        & 3.1$\times10^6$  & 1.72$\pm$0.02 & 2.5$\pm$1.4 & $-0.61\pm$4 \\ 
\hline
        \multicolumn{3}{|c|}{True values}  &  1.76 &  1.92 &  $-$1.41 &  &  1.76 &  1.92 &  $-$1.41    \\
    \hline
\end{tabular}
\end{table*}

We first analyze data sets that emulate currently available lattice
calculations. \datasetmn{M}{N}{1}{1.5} is sampled with the lowest $m$-value near the
physical point $m=140$ MeV up to $m = 440$ MeV, where there is noticeable contribution to the generated
data from the model-dependent part. In the
Guidance stage, we consider the evidence in Table~\ref{tab:MN1-results-both}, which
peaks at $\pmax=2$ but decreases by two orders of magnitude from $\pmax=2$
to $\pmax=6$. There is no indication of unnaturalness in the $\abar$ posteriors at
any order. 

We proceed with the Parameter estimation, looking for stability with increasing orders
of the extraction in accordance with the evidence calculations from the Guidance
stage. Even though the evidence peaks at $\pmax=2$, the parameter estimates in
Table~\ref{tab:MN1-results-both} do not stabilize at that order. In 
fact, they do not stabilize until $\pmax=4$. The result for $\ktilde_0$
is well-determined and consistent with the true value at the 1-$\sigma$ level, but $\ktilde_1$ is rather
poorly determined, although the estimate is consistent with the true value to within the quoted
error estimates. With the data currently available to practitioners, lattice extrapolations will likely be
difficult past the leading order.

\begin{figure*}[tbh!]
  \subfloat{%
    \label{fig:MN2_xmax_k2_kmax4}%
  \includegraphics[width=0.48\textwidth]{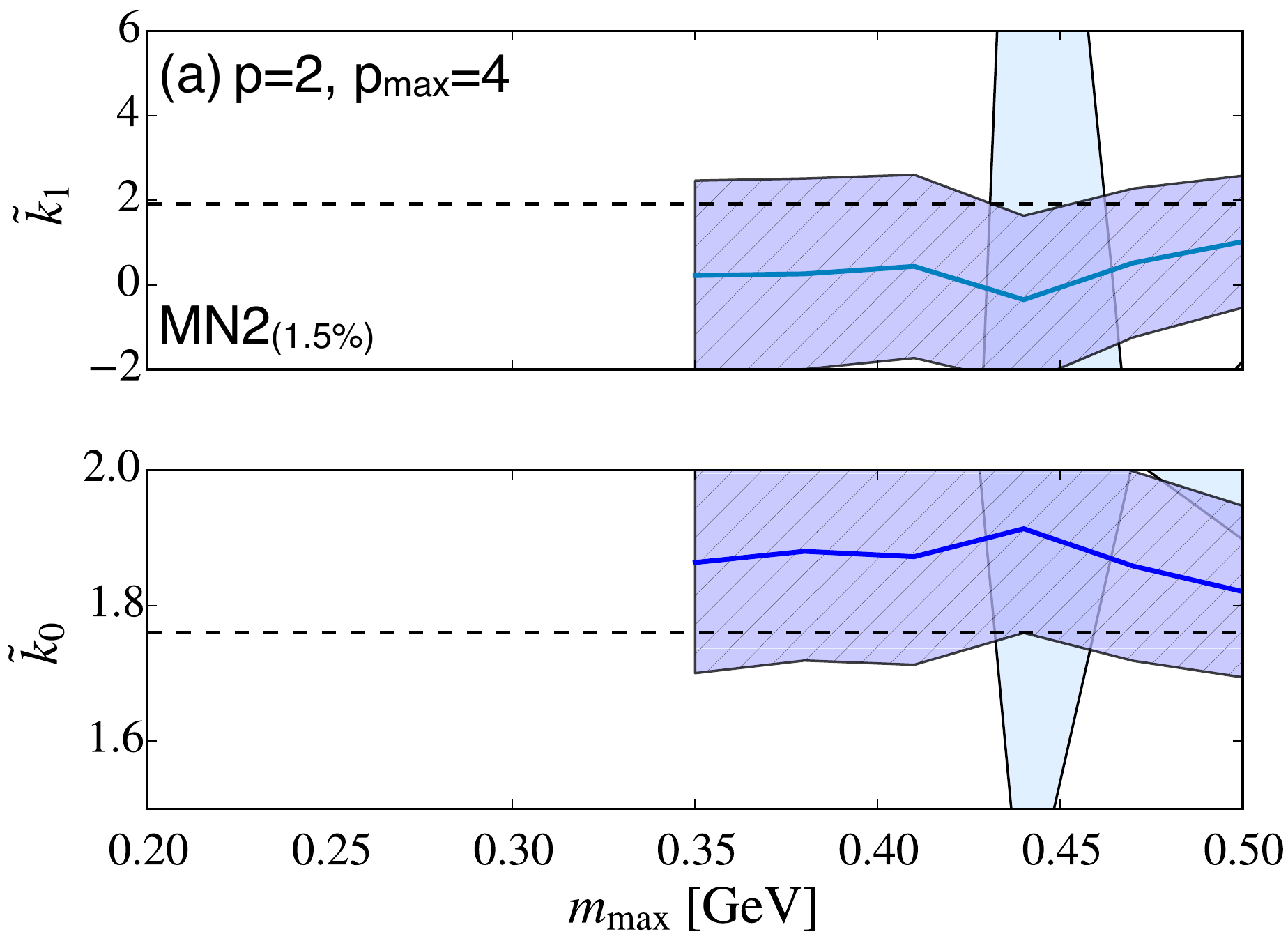}%
  }
  ~~~%
   \subfloat{%
     \label{fig:MN2_xmax_k2_kmax5}%
     \includegraphics[width=0.48\textwidth]{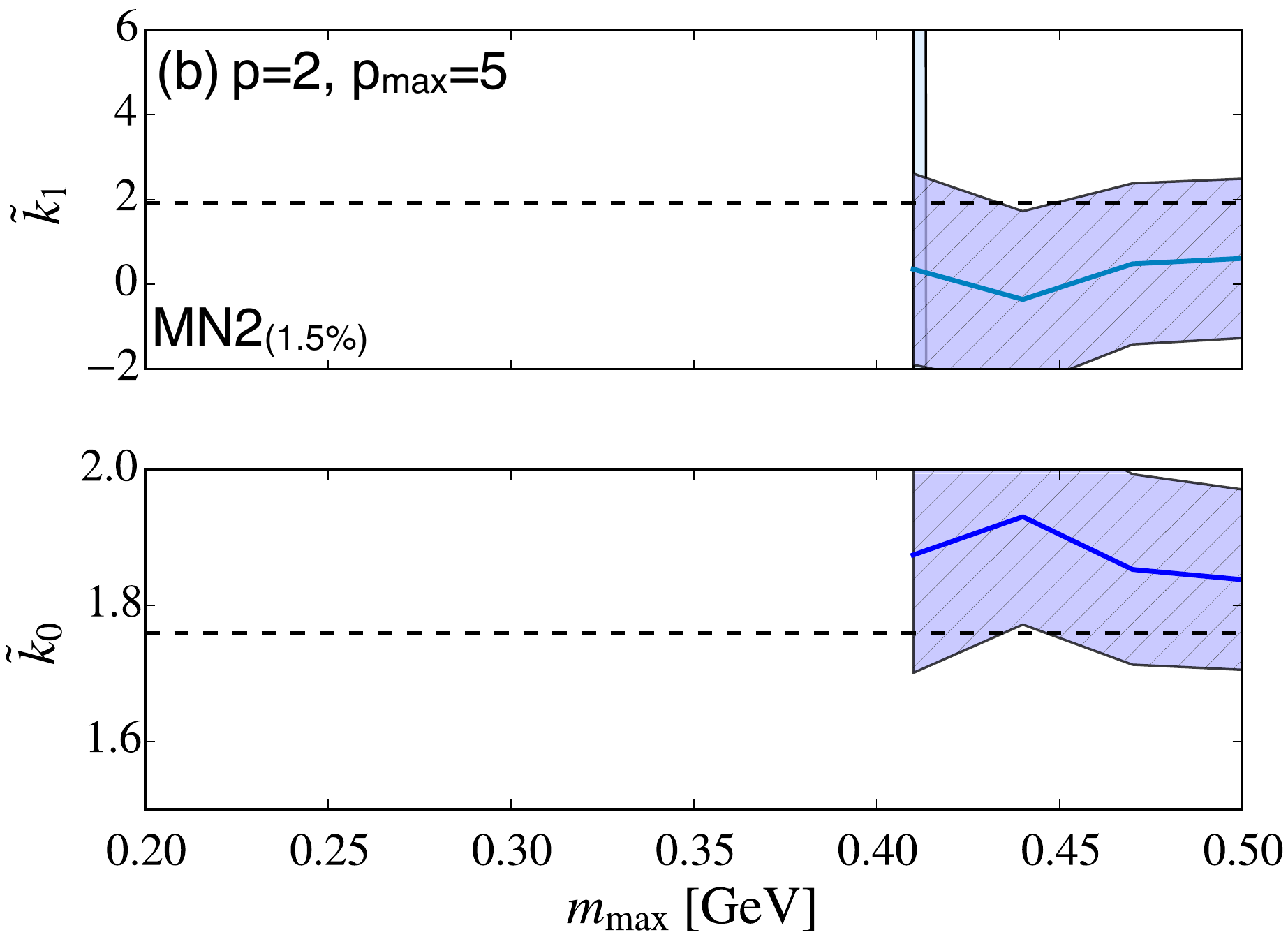}
   }
   \caption{(color online) Bayesian coefficient estimates
    as data from data set \datasetmn{M}{N}{2}{1.5} are sequentially added at the
    high-$m$ end. The largest $m$-value in the set is denoted as $\mmax$.
    The solid lines with darker hatched error bands represent estimates
    using prior Set \Cprime\ with $\abarzero=5$, and the line with circles
    (off-scale in these panels)
    with lighter solid error bands represents the least-squares estimates. 
    The error bands represent 68\% DoBs (1-$\sigma$ errors).}
\end{figure*}

We also examine the results of parameter estimation from \datasetmn{M}{N}{2}{1.5}. A plot of this
data set is shown in Fig.~\ref{fig:MN2_data} compared to the underlying nucleon mass function used to generate
the data and the chiral expansion up to sixth order. The data are sampled up to a
region where the model-dependent term becomes significant. Table~\ref{tab:MN2-results-both} shows the
evidence and parameter estimates using prior Set~\Cprime\ compared to the uniform-prior results. 
The evidence shows no pleasing saturation behavior for this data set, perhaps indicating 
that we are sampling too near the breakdown of the expansion, as in
Sec.~\ref{subsec:pastbreakdown}. However, 
there is no indication of unnaturalness in the $\abar$ posterior at any order.
It appears that the results are not stabilizing as the order
is increased, but comparing the $\xmax$ plots of Fig.~\subref*{fig:MN2_xmax_k2_kmax4}
and Fig.~\subref*{fig:MN2_xmax_k2_kmax5} at $\pmax = 4$ and $\pmax=5$ respectively, the
final results look consistent and stable with $\xmax$. The coefficients are reproduced at
the 1-$\sigma$ level with rather large errors. Even extracting the leading coefficient in
this case is very difficult, with the next-order coefficient mostly undetermined. As
in Sec.~\ref{subsec:pastbreakdown}, it is not possible to extract conclusive results where
the expansion cannot describe the data well.

\begin{figure}[tbh]
  \includegraphics*[width=0.98\columnwidth]{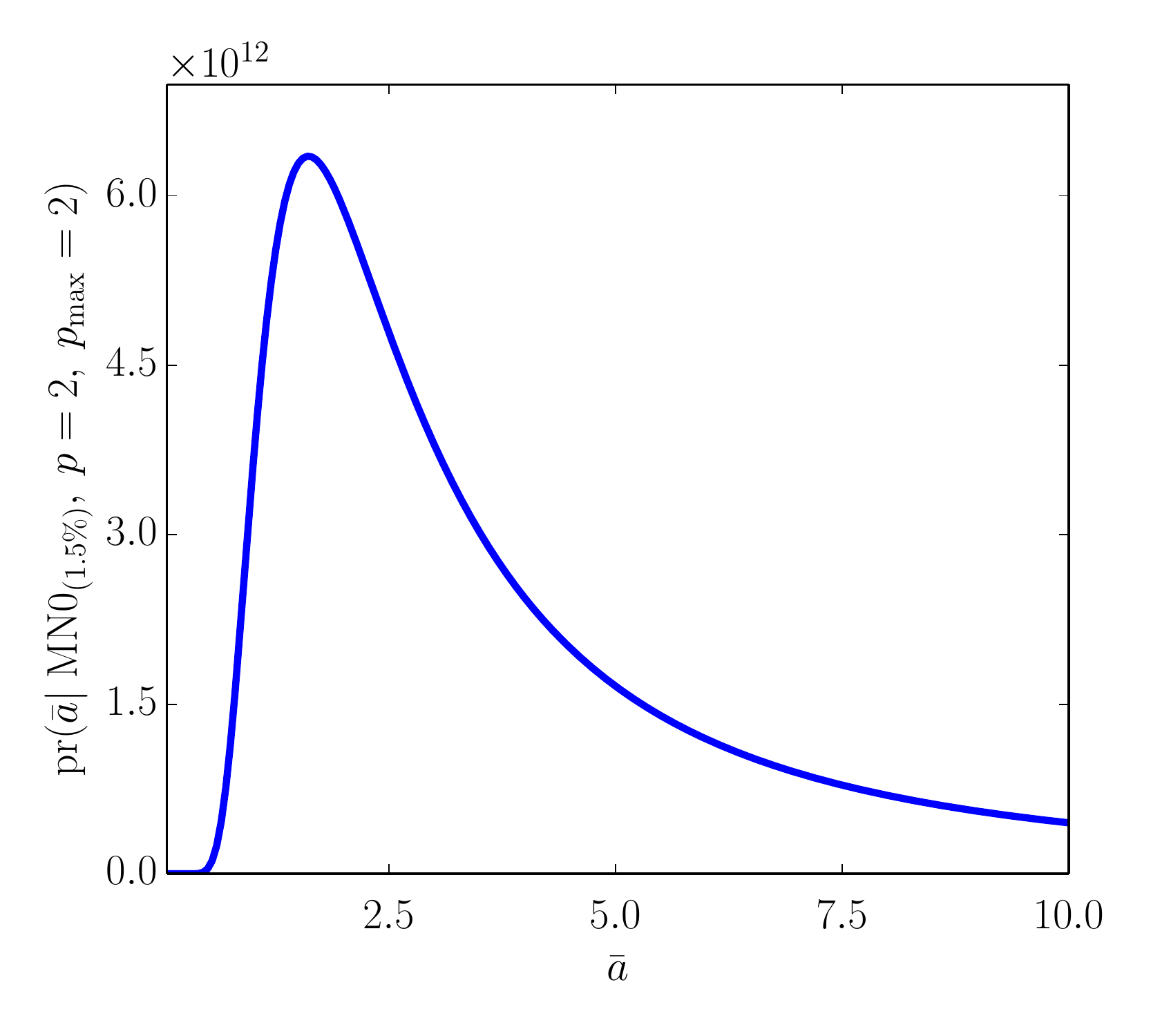}
  \caption{(color online) 
  The posterior pdf $\pr(\abar|D,\pord,\pmax)$ calculated at $\pord=2$, $\pmax=2$ 
  using prior Set~C from Table~\ref{tab:priors} with
  $\abarmin~=~0.05$ and $\abarmax~=~20$, given
  data set \datasetmn{M}{N}{0}{1.5}.
  \label{fig:MN0_abar_post}}
\end{figure}

We finally examine \datasetmn{M}{N}{0}{1.5}, which is an idealized set
sampled at significantly smaller pion masses than the previous two sets.
We first compute the evidence and naturalness parameter posteriors to complete the
Guidance stage for \datasetmn{M}{N}{0}{1.5}. Figure~\ref{fig:MN0_abar_post}
shows the posterior for $\abar$ assuming prior Set~C at chiral order $\pord=2$,
$\pmax=2$. We see no indications of unnaturalness in this data set, and again
choose $\abarzero=5$ for prior Set~\Cprime. Computing the evidence using
this choice, the results in Table~\ref{tab:MN0-results-both} show that the evidence reaches
saturation at about $\pmax=2$, where the values peak slightly, and then levels off at
higher orders. The expectation is that only two coefficients will be determined
by this data. Even in the unrealistic case where we have high-quality data below
the physical pion mass, we do not expect to determine more than two LECs
in the chiral expansion.

\begin{figure}[tbh]
  \includegraphics*[width=0.98\columnwidth]{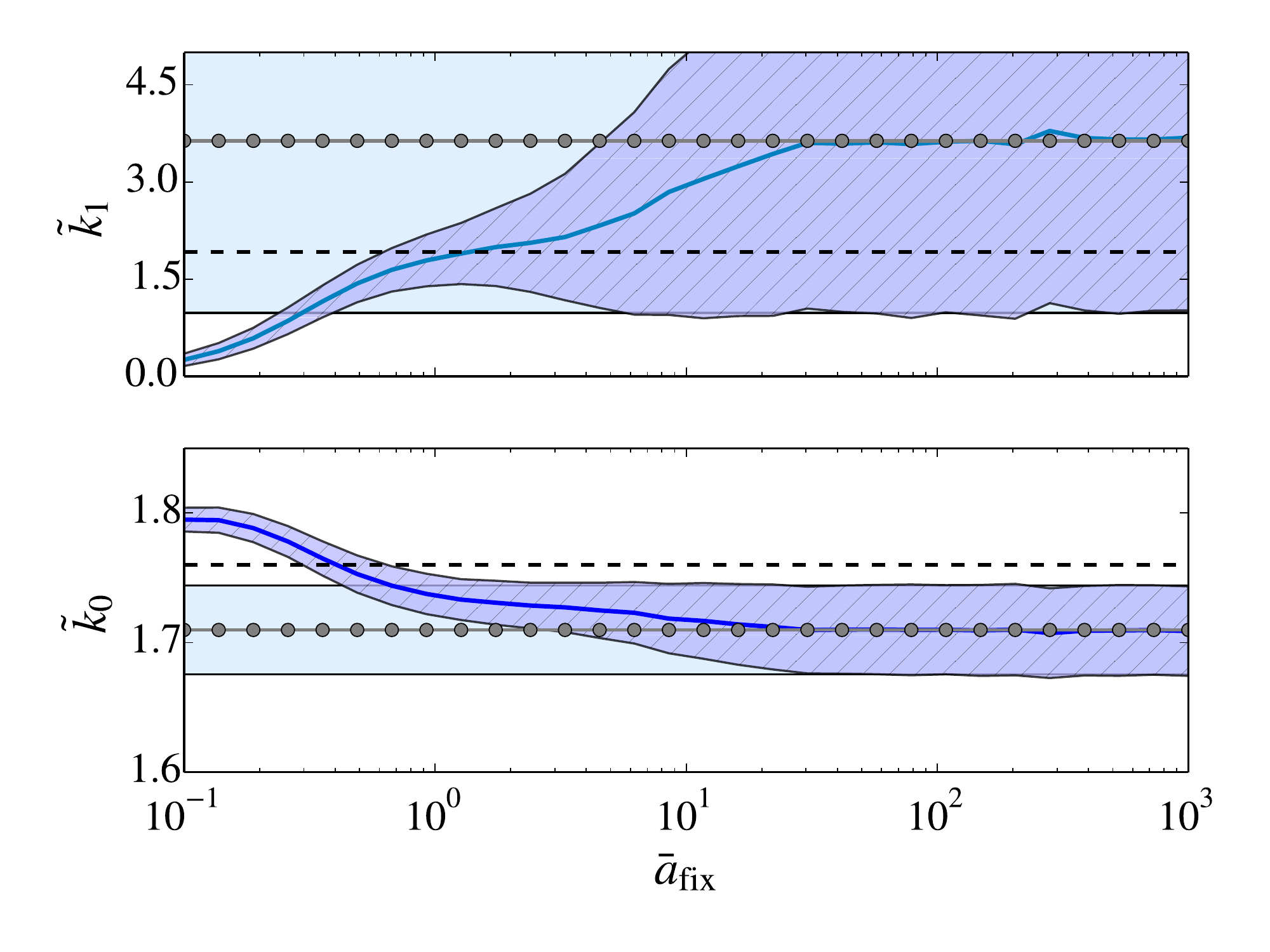}
  \caption{(color online)  Bayesian coefficient estimates (lines with darker hatched error bands)
    calculated at $\pord=2$, $\pmax=3$
    as a function of $\abarzero$ using prior Set~\Cprime\
    given \datasetmn{M}{N}{0}{1.5}. The constant line with circles
    with lighter solid error bands is the least-squares estimate, which is independent of $\abarzero$. 
    The error bands represent 68\% DoBs (1-$\sigma$ errors).
  \label{fig:MN0_abar_evol}}
\end{figure}

\begin{figure*}[tbh!]
  \subfloat{%
    \label{fig:MN0_xmax_k2_kmax2}%
  \includegraphics[width=0.48\textwidth]{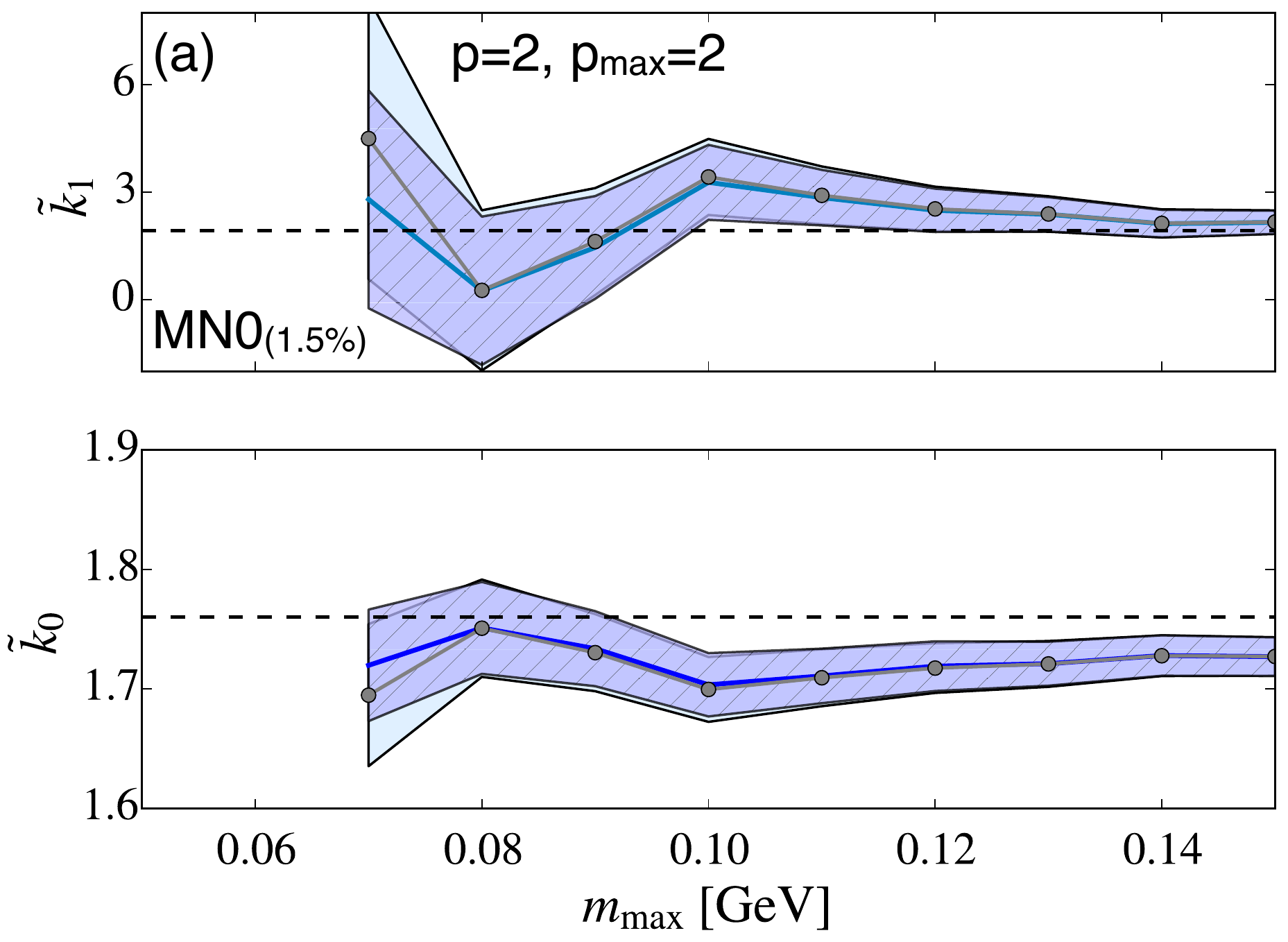}%
  }
  ~~~%
   \subfloat{%
     \label{fig:MN0_xmax_k2_kmax3}%
     \includegraphics[width=0.48\textwidth]{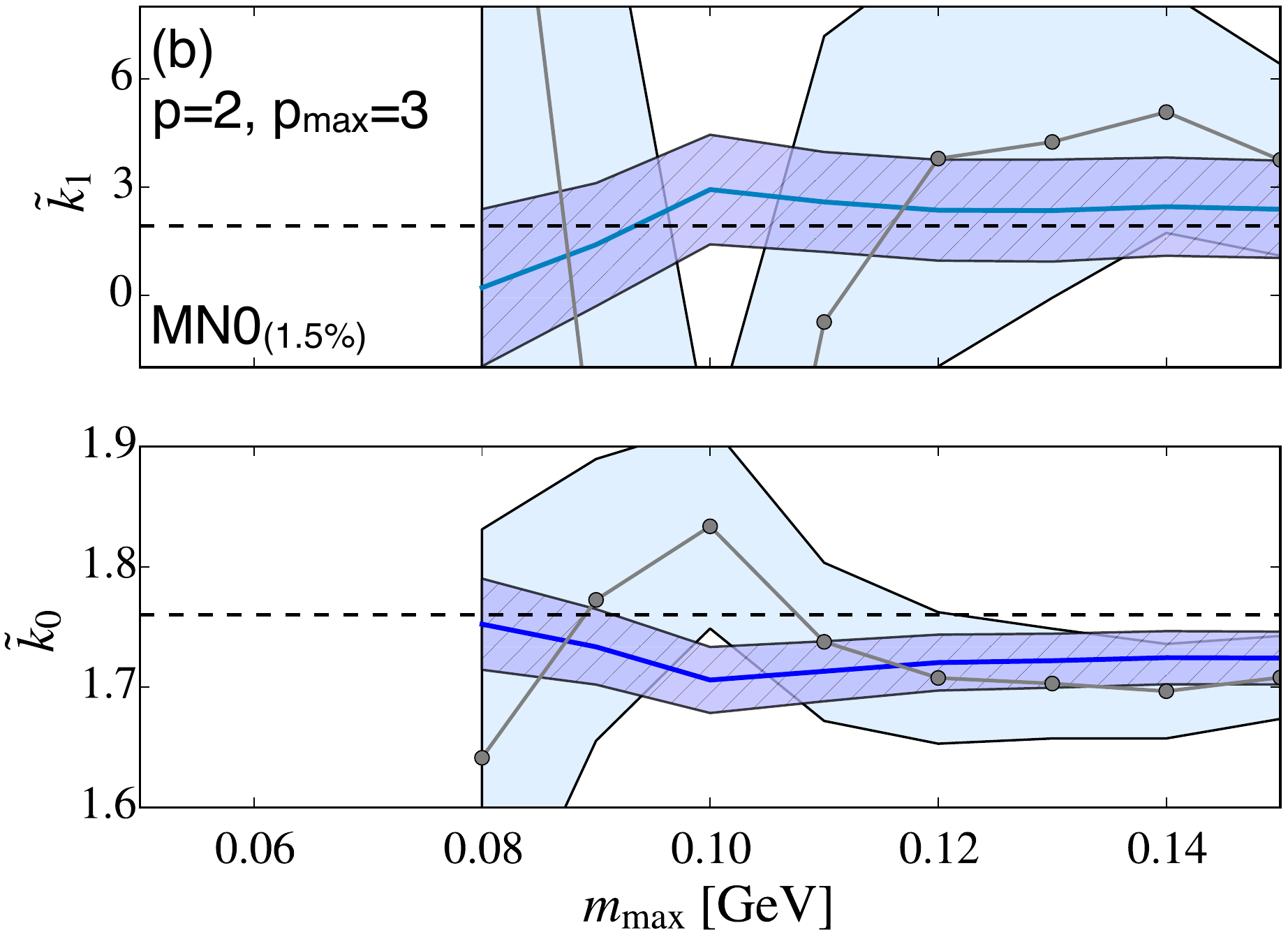}
   }
   \caption{(color online) Bayesian coefficient estimates
    as data from data set \datasetmn{M}{N}{0}{1.5} are sequentially added at the
    high-$m$ end. The largest $m$-value in the set is denoted as $\mmax$.
    The lines with darker hatched error bands represent estimates
    using prior Set \Cprime\ with $\abarzero=5$, and the line with circles
    with lighter solid error bands represents the least-squares estimates. 
    The error bands represent 68\% DoBs (1-$\sigma$ errors).}
\end{figure*}

Turning to the Parameter estimation, the results in Table~\ref{tab:MN0-results-both}
show that the data set MN0$_{(1.5\%)}$ leads to an underestimate of the parameter $\ktilde_0$ 
compared to the true value, while $\ktilde_1$ has a better central value than was obtained
from \datasetmn{M}{N}{2}{1.5}, albeit still with a sizeable uncertainty. 
The underestimation
of $\ktilde_0$ is another example of the effect of fluctuations in the data. The sizeable error on $\ktilde_1$ 
indicates that, although the coefficient is fixed by the data, the precision we can obtain is quite 
limited. At order $\pmax=2$, which was the saturation order from the Guidance stage,
the naturalness-prior parameter estimate for $\ktilde_1$ is spuriously precise---likely a result of underfitting of \datasetmn{M}{N}{2}{1.5}
at this order. This illustrates the importance
of going well into the region of $\pmax$ where the evidence flattens out, in order to eliminate the possibility
of underfitting. 
Examining the results using a uniform prior in Table~\ref{tab:MN0-results-both},
we see very similar parameter estimates as for the naturalness prior at $\pord=\pmax=2$, but any attempt
to check the stability of that result by going to higher chiral order leads to extreme overfitting: by
the time we reach $\pmax=6$, where there are ten total coefficients, least squares is producing
central values and 68\% intervals of order $10^6$ or $10^7$. 
The additional information imparted to the fit by the naturalness prior is
essential to meaningful parameter estimation once we go to $\pmax=3$. And going to
$\pmax=3$ is necessary in order to achieve stability of the extraction with chiral order.

The $\xmax$ plots in Figs.~\subref*{fig:MN0_xmax_k2_kmax3}
and \subref*{fig:MN0_xmax_k2_kmax2} also show that 
stability with respect to the pion-mass interval over which data is fit  is only achieved
at order $\pmax=3$---in accordance with the inference from 
Table~\ref{tab:MN0-results-both} that there is underfitting at $\pmax=2$.
The $\abar$ relaxation plot in Fig.~\ref{fig:MN0_abar_evol}
confirms a naturalness parameter $\abarzero$ between 1 and about 5; this is consistent
with the guidance from Fig.~\ref{fig:MN0_abar_post}.
\begin{figure*}[tbh]
  \subfloat{%
    \label{fig:MN0_datasets_k2_kmax4}%
  \includegraphics[width=0.48\textwidth]{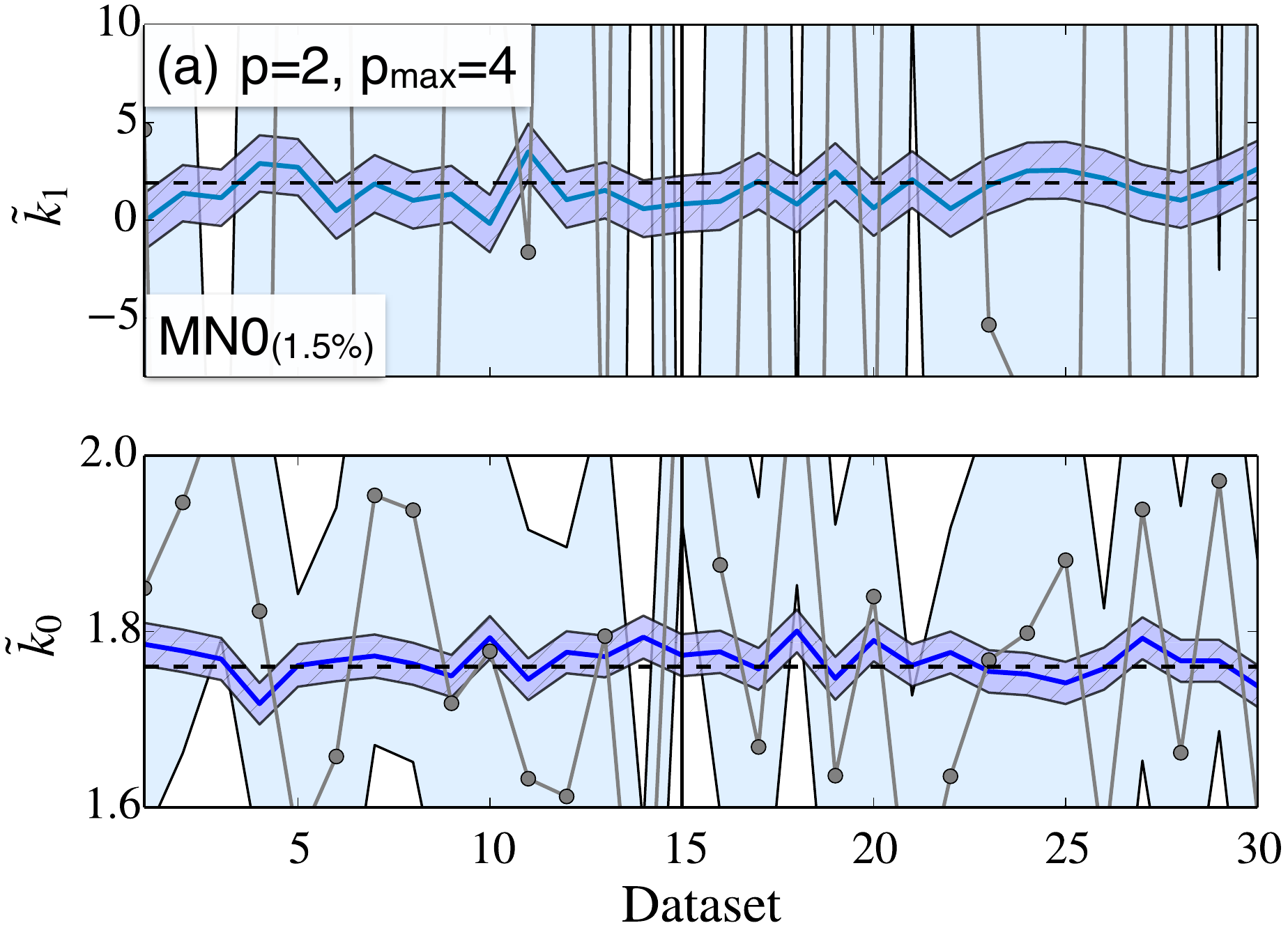}%
  }
  ~~~%
   \subfloat{%
     \label{fig:MN0_accumulated_k2_kmax4}%
     \includegraphics[width=0.48\textwidth]{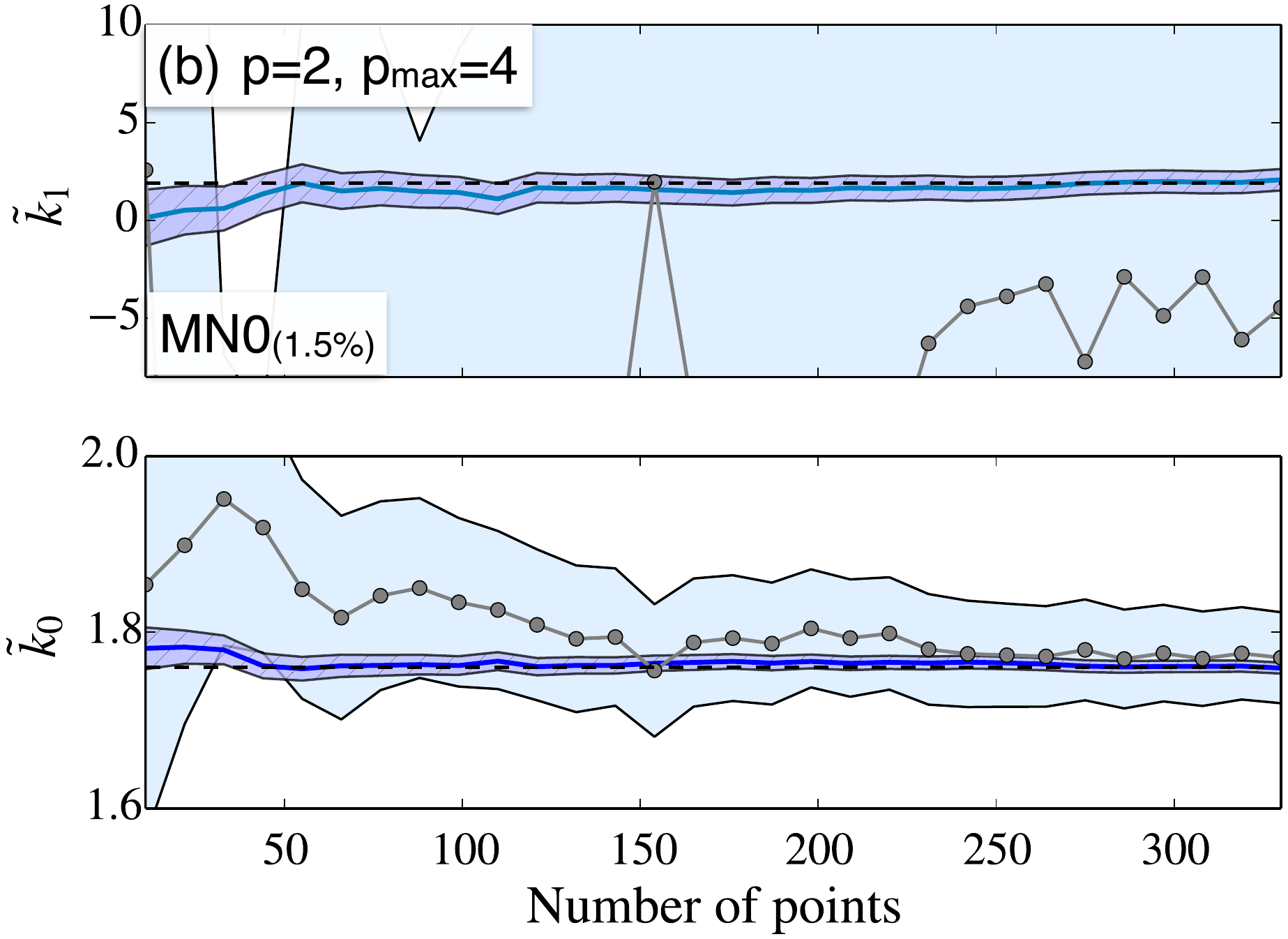}
   }
   \caption{(color online)
   Multi-set (a) and accumulation plots (b)
   calculated at $\pord=2$, $\pmax=4$. 
   The shaded regions denote 68\% error bands for the uniform (line with circles with lighter solid band)
   and naturalness prior (line with darker hatched band).
   The data sets used in (a) are 30 samples on the \datasetmn{M}{N}{0}{1.5} mesh from Table~\ref{tab:MN-labels}. 
   The same data are accumulated set by set to generate (b).
   The naturalness prior used was Set~\Cprime\ with $\abarzero = 5$.}
\end{figure*}

Finally, the Validation stage gives some indication of the effects of fluctuations
in the data set, where we compare many independent samplings on the \datasetmn{M}{N}{0}{1.5}
grid in the multi-set plot of Fig.~\subref*{fig:MN0_datasets_k2_kmax4}, where we choose $\pmax=4$.
The fluctuations in the results using the uniform prior
are severe, and the results using prior Set \Cprime\ fluctuate much less. Comparing the results for
the first two coefficients at $\pmax=4$ with these 30 independent data sets to the previous
estimates with our single data set \datasetmn{M}{N}{0}{1.5}, we conclude that the naturalness prior
result for $\ktilde_0$ in Table~\ref{tab:MN0-results-both} is consistent with the fluctuations seen
in Fig.~\subref*{fig:MN0_datasets_k2_kmax4}.
At the same
order we examine the accumulation plot with these data sets in Fig.~\subref*{fig:MN0_accumulated_k2_kmax4}.
The results for $\ktilde_0$ are consistent to within 1-$\sigma$ when we have about 50 data points,
but the error reduces significantly once we have more than 300 points. The error in $\ktilde_1$
reduces slightly once we have over 300 data, but at this level of error in the
data it is still difficult to extract $\ktilde_1$ with small uncertainty. The uniform-prior results for
$\ktilde_1$ in Fig.~\subref*{fig:MN0_accumulated_k2_kmax4} are much less precise than the
results using naturalness information, even though $\ktilde_0$ is well determined by these data
and only a uniform prior. 

In summary, extrapolating the nucleon mass in the chiral limit from lattice data is a difficult
problem due to the lack of data at low pion masses, the level of error in
available lattice data, and the contributions of several coefficients at the
same order. Attempting to estimate LECs of the chiral expansion of the nucleon
mass past chiral order $\pmax=2$ will be very difficult for these reasons.
Constraints on terms non-analytic in the quark mass will certainly assist
in the extraction, but extracting $\ktilde_0$ and $\ktilde_1$ is still problematic. The
use of the Bayesian parameter estimation procedure laid out in this work
will assist practitioners in obtaining chiral extrapolations with credible
uncertainty estimates, which can be improved as new data becomes available.

\subsection{Nonlinear models}

The models considered so far are linear in the coefficients that represent
LECs.  However, in many EFTs of interest, such as chiral EFT, the observables
are nonlinear functions of the LECs and can be more complicated and/or expensive to compute. Here we
give a brief example to show the effectiveness of our procedures for cases in
which the observable is nonlinear in the coefficients. We will address in future work cases where calculating
an observable to order $\kmax$ is not practical, and how to circumvent this issue
using the naturalness of the expansion of the observable used in the fit, as discussed in
Sec.~\ref{subsec:setup}. This will be
relevant in applications to real EFTs such as chiral EFT. We will now briefly
discuss one model example to demonstrate the utility of the diagnostics to observables
nonlinear in the parameters.

We define a model observable (called Model~T) for this example by
\beq
  \mbox{T}[g(x)] = \frac{1}{1 + g(x)} \;, 
\eeq
where $g(x)$ can represent any of the linear expansions we have considered. We
will consider the model linear observable $g(x)$ from Sec.~\ref{sec:model-problems}, Model~H
defined in Eq.~\eqref{eq:model-H}. This model will be referred to by Model~T[H].
The theoretical prediction for the observable is
\beq
  \mbox{T}[\gth(x)] = \frac{1}{1 + \gth(x)} \;,
\eeq 
where $\gth(x)$ is defined in Eq.~\eqref{eq:model-th-expansion}, and this observable
can be calculated to order $\kord$ using the vector of coefficients $\avec$ up to
that order. Using
Eq.~\eqref{eq:avec-post-general-exp}, we can compute the posterior for the coefficients
$\avec$ with a simple modification to the likelihood. 
Since the observable is no longer linear in $\avec$,
the likelihood becomes significantly more complicated, and the repeated computations necessary
for sampling the posterior or computing the evidence can become cumbersome.

\begin{table*}[thb]
\caption{
   Coefficient estimates from sampling of $\pr(\avec|\datasetnonlin{T}{H}{0}{1},\kord,\kmax)$
   given the expansion from Eq.~\eqref{eq:model-th-expansion}
    (these results are controlled by $\kmax$ only, see Sec.~\ref{subsec:setup}).
   The left side of the table is for a uniform prior, 
   which is equivalent to a least-squares fit, and includes the $\chi^2$/dof values.
   The right side of the table is using prior Set~\Cprime\ 
   from Table~\ref{tab:priors} with $\abarzero = 5$, and includes the evidence.
   For both priors the posterior pdf is well-approximated by a multi-dimensional
   Gaussian. \label{tab:TH01-both}}
  \begin{tabular}{|c|c||c|c|c|c||c|c|c|c|}
    \hline
     \multicolumn{2}{|c||}{} & \multicolumn{4}{|c||}{Uniform prior} &  \multicolumn{4}{|c|}{Gaussian prior} \\
    \hline
      $\kord$ & $\kmax$ & $\chi^2/$dof & $a_0$ & $a_1$ & $a_2$ 
      & Evidence & $a_0$ & $a_1$ & $a_2$ \\ 
    \hline
        0 & 0  & 5.6    & 0.92$\pm$0.01  &                 &                       & 9.1$\times10^4$    & 0.92$\pm$0.01 &  &  \\
        1 & 1  & 1.5    & 0.99$\pm$0.01  & $-$1.3$\pm$0.21 &                       & 6.9$\times10^{11}$ & 0.99$\pm$0.01 & $-$1.3$\pm$0.2 &  \\
        2 & 2  & 1.5    & 1.0$\pm$0.02   & $-$2.6$\pm$0.93 & 11$\pm$8          & 7.4$\times10^{11}$     & 0.99$\pm$0.02 & $-$1.6$\pm$0.5 & 2.9$\pm$4 \\ 
        2 & 3  & 1.6    & 1.0$\pm$0.04   & $-$4.6$\pm$2.8  & 57$\pm$60            & 7.4$\times10^{11}$  & 0.99$\pm$0.02 & $-$1.6$\pm$0.5 & 2.8$\pm$4 \\ 
        2 & 4  & 1.2    & 1.1$\pm$0.1    & $-$18$\pm$7.3   & 540$\pm$300           & 7.4$\times10^{11}$ & 0.99$\pm$0.02 & $-$1.6$\pm$0.5 & 2.9$\pm$4 \\ 
        2 & 5  & 1.2    & 1.0$\pm$0.1    & 1.3$\pm$18      & $-$460$\pm$900        & 7.4$\times10^{11}$ & 0.99$\pm$0.02 & $-$1.6$\pm$0.5 & 2.9$\pm$4 \\ 
        2 & 6  & 1.5    & 1.1$\pm$0.3    & $-$22$\pm$47    & 1100$\pm$3000         & 7.4$\times10^{11}$ & 0.99$\pm$0.02 & $-$1.6$\pm$0.5 & 3.0$\pm$4 \\ 
\hline
        \multicolumn{3}{|c|}{True values} & 1.0 & $-1.54$ & 1.78 & & 1.0 & $-1.54$ & 1.78 \\
    \hline
\end{tabular}
\end{table*}

We will consider a single data set \datasetnonlin{T}{H}{0}{1} in this section, which is sampled
on a linear mesh in the range $0.01 \leq x \leq 0.1$, representing a high-quality
data set sampled at small expansion parameter values, similar to the analysis of
\dataset{H}{0}{1} for the linear Model~H in Sec.~\ref{sec:model-problems}. We expect to
reproduce the leading behavior of the underlying expansion well using these data.
Table~\ref{tab:TH01-both} shows results up to $\kord = 2$ for the marginalized posteriors at different
$\kmax$s, comparing results using a uniform prior and Set~\Cprime.
We see that, as in the linear cases, overfitting occurs when naturalness information is
not included, starting as early as $\kmax=2$ in the left column of Table~\ref{tab:TH01-both}.
We will now demonstrate some of the diagnostics from the flowchart in Fig.~\ref{fig:flowchart}.

\begin{figure}[tbh]
    \includegraphics[width=0.98\columnwidth]{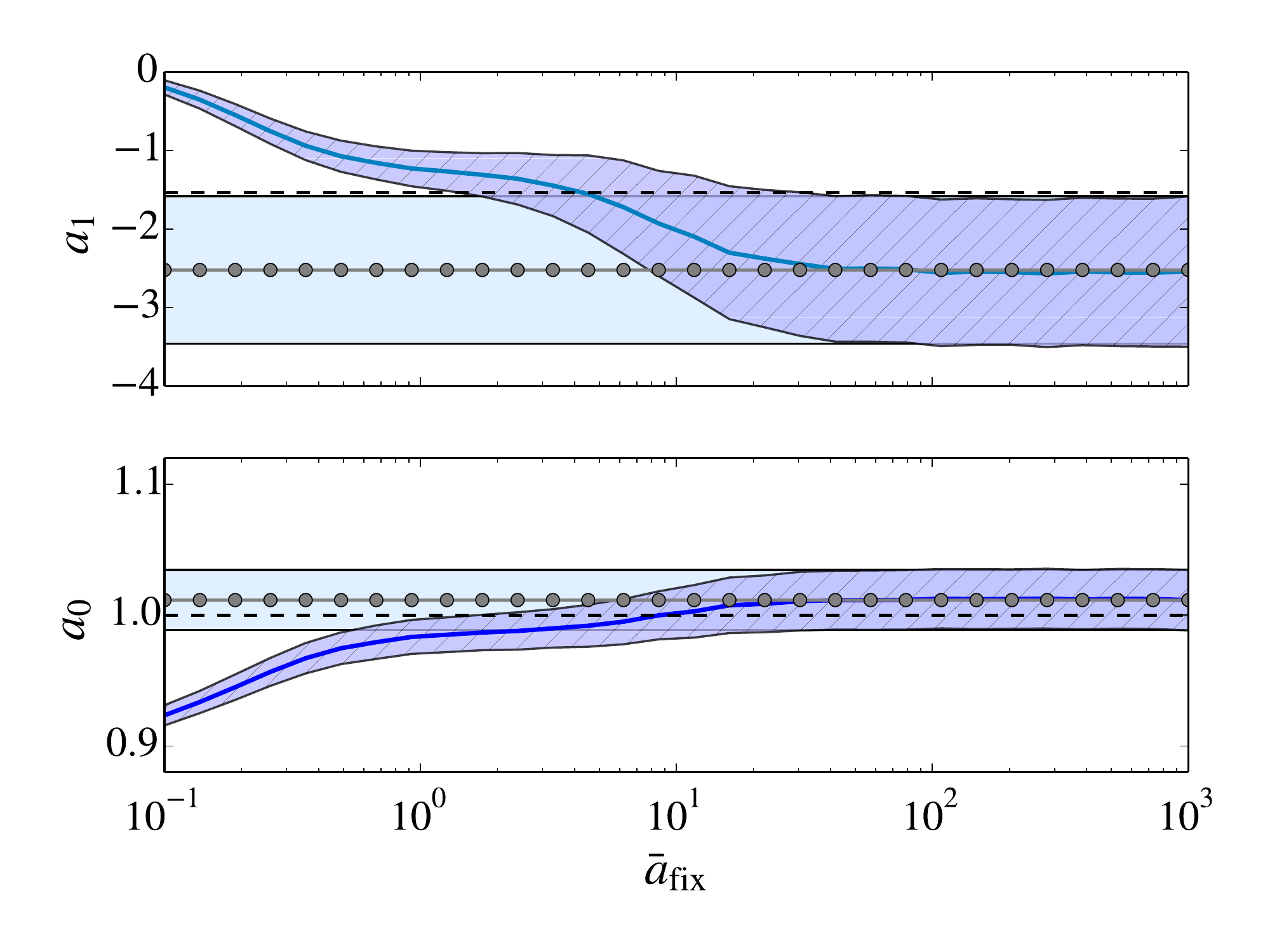}
    \caption{(color online) 
     Bayesian coefficient estimates (lines with darker hatched error bands)
     calculated at $\kord=1$, $\kmax=2$ as a function of $\abarzero$ using prior Set \Cprime\
    given \datasetnonlin{T}{H}{0}{1}. The constant line with circles
    with lighter solid error bands is the least-squares estimate, which is independent of $\abarzero$. 
    The error bands represent 68\% DoBs (1-$\sigma$ errors).
     \label{fig:TH01_abar_relaxation}}
\end{figure}

\begin{figure*}[bht!]
  \subfloat{%
    \label{fig:TH01_datasets_k1_kmax2}%
  \includegraphics[width=0.48\textwidth]{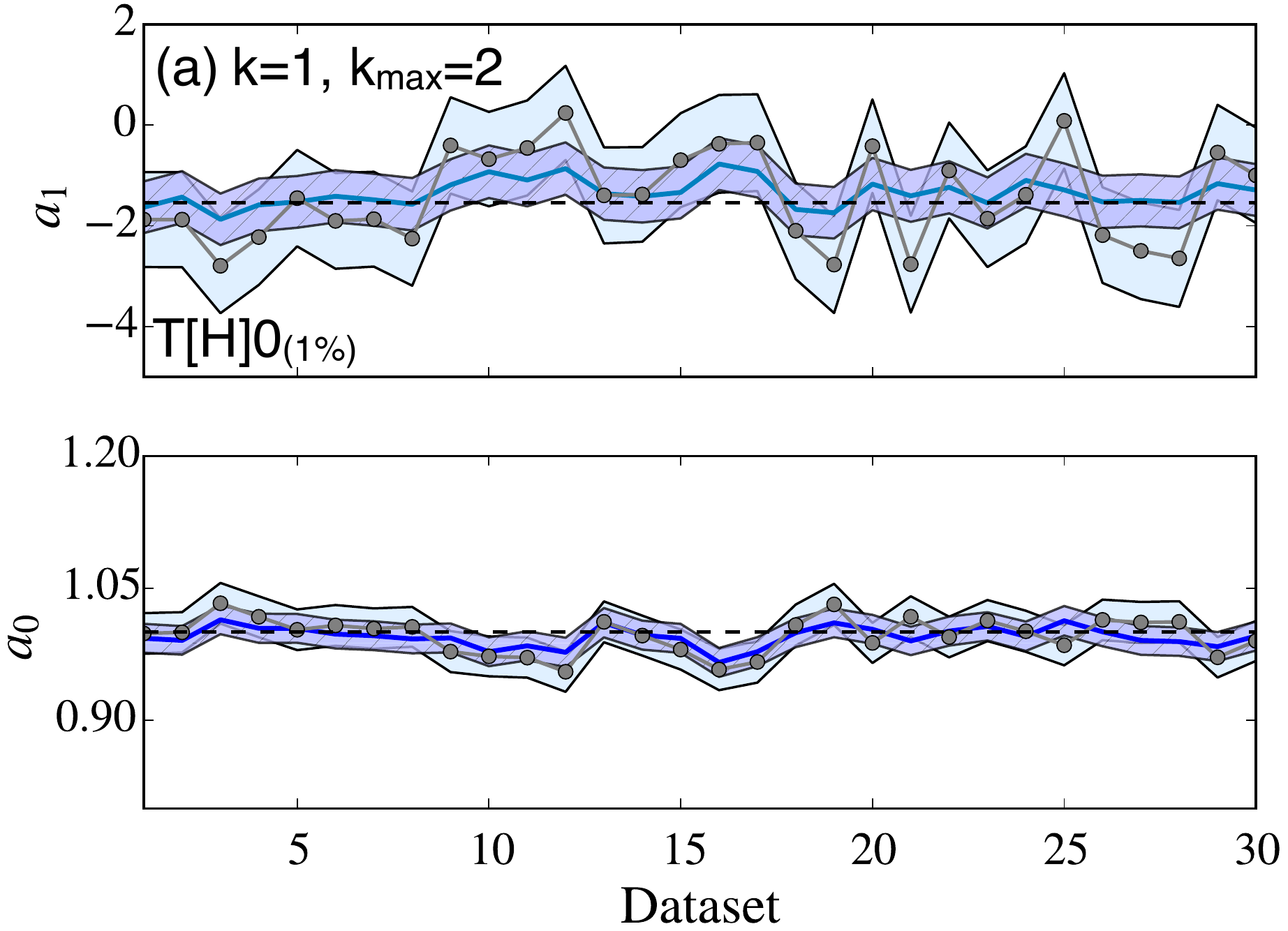}%
  }
  ~~~%
   \subfloat{%
     \label{fig:TH01_accumulated_k1_kmax2}%
     \includegraphics[width=0.48\textwidth]{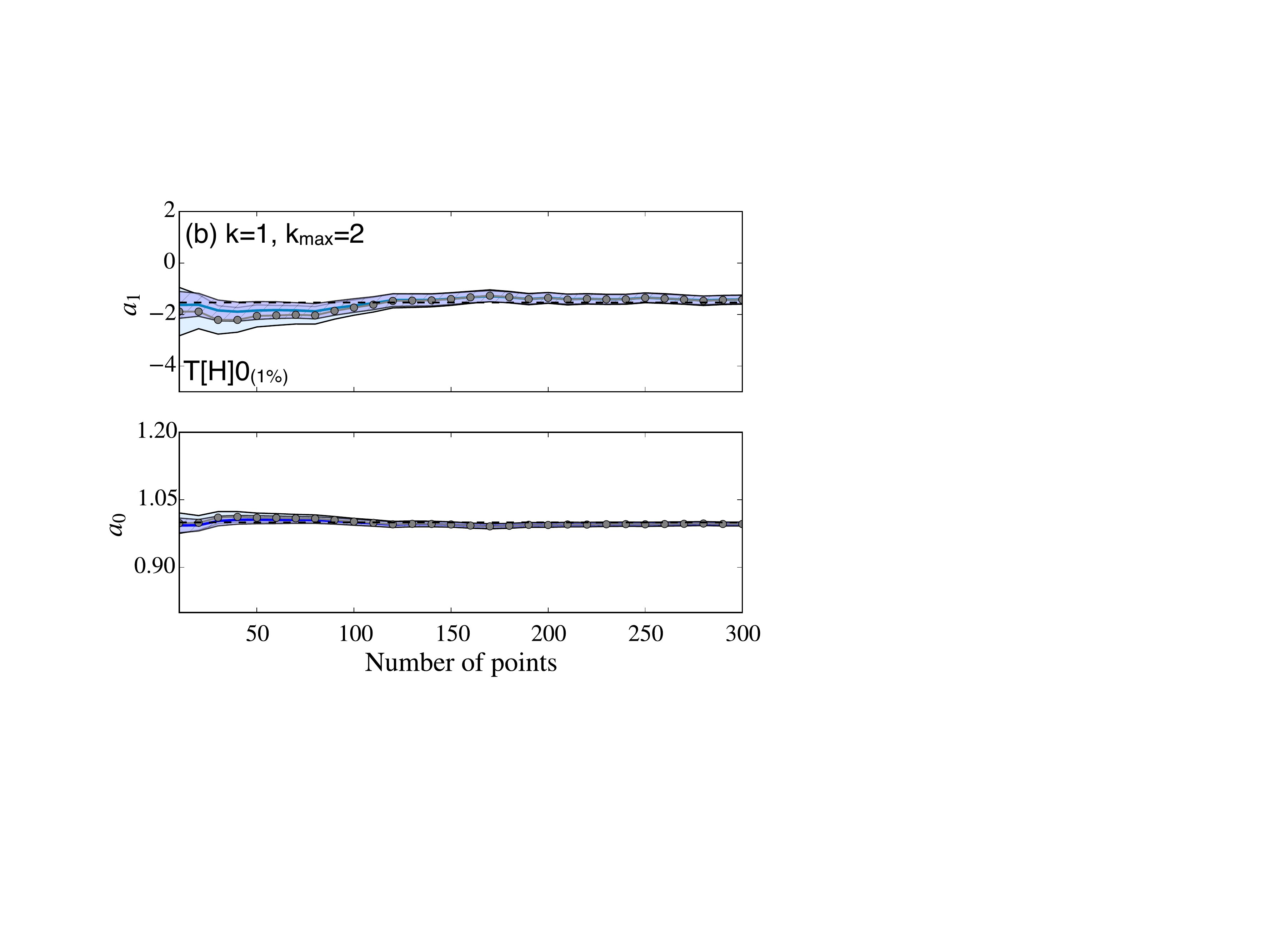}
   }
   \caption{(color online) Multi-set (a) and accumulation plots (b)
   calculated at $\kord=1$, $\kmax=2$. 
   The shaded regions denote 68\% error bands for the uniform (line with circles with lighter solid band)
   and naturalness prior (line with darker hatched band).
   The data sets used in (a) are 30 samples on the \datasetnonlin{T}{H}{0}{1} mesh which is the same
   as the \dataset{H}{0}{1} mesh from Table~\ref{tab:model-H-labels}. 
   The same data are accumulated set by set to generate (b).
   In each case the prior was Set~\Cprime\ with $\abarzero = 5$}
\end{figure*}
 
The evidence for different values of $\kmax$ is shown in Table~\ref{tab:TH01-both},
and we see that the evidence saturates around $\kmax = 2$, while the ratio of the evidence at
$\kmax=2$ to $\kmax=1$ is only about $1.1$, indicating that no more information is gained for
$a_2$ from this data set. Using prior Set \Cprime\ with $\abarzero=5$ results in excellent agreement
of $a_0$ and $a_1$ with their true values, but with $a_2$ not well-constrained. The results for
the posterior maxima in the right side of Table~\ref{tab:TH01-both} confirm the guidance from the
evidence that extractions past first order will not contain much information.

In terms of choosing a prior, Fig.~\ref{fig:TH01_abar_relaxation} demonstrates that at $\kmax=2$, the
first two coefficients are consistent with the true values for $\abarzero$ between about 2 and 10. We could
also compute the $\abar$ posterior to analyze naturalness information contained in the data. Using
the simplified prior Set \Cprime\ with $\abarzero = 5$ therefore produces consistent results in this case.

The inclusion of additional information in the parameter estimation is beneficial, as 
it was in Sec.~\ref{sec:model-problems}. In this nonlinear problem, the procedure
results in the best possible parameter estimates with the available information. 
Even at $\kmax=2$ overfitting occurs when naturalness information is not included,
 and the multi-set plot of Fig.~\subref*{fig:TH01_datasets_k1_kmax2}
shows the benefit of the naturalness prior--- and that the consquences of statistical
fluctuations are still unavoidable, even with the high-quality data set \datasetnonlin{T}{H}{0}{1}.
The accumulation plot in Fig.~\subref*{fig:TH01_accumulated_k1_kmax2} shows that the
prior is beneficial even for a large data set until about 10 data sets are available,
at which point the uniform prior and naturalness prior results become the same. 

The procedures detailed in Fig.~\ref{fig:flowchart} are thus directly applicable to parameter
estimation in cases when the available data are related nonlinearly to the parameters. We have
demonstrated in a simple test case the success of the diagnostics with respect to the type of observable used. 
Of course, the limitations identified in this section and in the test cases of 
Sec.~\ref{sec:model-problems} will still be present in nonlinear applications, but the major
theme persists: the best parameter estimates are obtained by using all available
information about the problem at hand. 


\section{Summary and outlook} \label{sec:conclusion}

In this paper, we present a detailed framework for EFT parameter estimation
using Bayesian statistics, which naturally combines statistical and systematic
uncertainties.  Fitting the LECs in an EFT poses a basic dilemma: we want
to use more data to suppress the impact of data errors and avoid overfitting, 
but the accuracy decreases for larger values of the EFT expansion parameters,
so using too much can lead to underfitting.
By marginalizing over omitted higher-order terms using 
naturalness priors, we are able to use \emph{all} of the data, 
without having to decide a particular value of the EFT parameter at which to break off the fit.
In particular, a prior on naturalness suppresses overfitting by limiting how much
LECs can play off each other.
This in turn means we find stability with respect to expansion order and 
amount of data.

This success is not achieved automatically, however; there are multiple ways the
parameter estimation can go wrong.  By studying simple models reflecting the
possible behaviors of an EFT, we have identified pitfalls and ways to address
them.  
We have developed a suite of diagnostic tools to guide and analyze the fit
of LECs. These diagnostics:
\begin{itemize}
\item use the Bayes evidence ratio to find how many LECs
can be determined by the data;
\item  identify potential sensitivity to the choice of prior;
\item trace the way the prior eliminates spurious correlations between undetermined coefficients;
\item indicate the region in the EFT parameter where a
particular LEC is determined;
\item show the impact that fluctuations in the data have on the
parameter estimation.
\end{itemize}
They are summarized in Table~\ref{tab:diagnostic_list} 
and a possible flowchart for applying them in parameter estimation  
is given in Fig.~\ref{fig:flowchart}.
Multiple examples in Sec.~\ref{sec:procedures}, \ref{sec:model-problems}, 
and \ref{sec:case_studies} illustrate how to use these tools.
These examples are not exhaustive but highlight some important features
of Bayesian uncertainty quantification for EFT parameter estimation.

Our goal has been to make this framework adaptable to a wide range of EFT
applications.
We plan to test the framework in applications such 
as fitting LECs for the nucleon mass chiral expansion with genuine lattice
data and for chiral nucleon-nucleon and three-nucleon forces.
This will require us to address practical challenges such as the computational
cost of MCMC evaluations, which may require compromises in our procedures.
It will also be of interest to investigate whether augmentations of basic
least-squares fitting procedures commonly used by practitioners to address
the special features of EFTs can
be justified within a Bayesian framework.

While we have focused here on parameter estimation,
we believe that the Bayesian framework will enable practitioners to consistently 
achieve all the general goals of UQ for EFT calculations:
\bi
  \I reflect all sources of uncertainty in an EFT prediction using a
     likelihood or prior for each;
  \I compare theory predictions and experimental results statistically, with
      error bands as Bayesian degree-of-belief (DoB) intervals;
  \I distinguish uncertainties from IR (long-range) vs. UV (short-range) physics
      by avoiding overfitting;
  \I provide guidance on how to better extract LECs by exploiting how
   Bayes' theorem propagates new info (e.g., will an additional or better
   measurement or lattice calculation help and by how much?);
  \I test whether EFT is working as advertised --- do our predictions exhibit the
   anticipated systematic improvement? --- by analyzing the trends of DoB intervals
   and applying model selection.
\ei
Investigations are in progress to extend and validate our diagnostics
and procedures for each of these goals.

\acknowledgments{We thank H.~Griesshammer, D.~Higdon, J.~McGovern, E.~Lawrence, A.~Peter, M.~Schindler,
A.~Steiner, and S.~Wild for useful discussions.
Useful feedback on the manuscript was provided by A.~Dyhdalo, M.~Heinz, S.~Koenig, S.~More, R.~Perry, and 
C.~Plumberg.
This work was supported in part by the National Science Foundation
under Grant No.~PHY--1306250, the U.S. Department of Energy under
grant DE-FG02-93ER40756, the NUCLEI SciDAC Collaboration under
DOE Grant DE-SC0008533, and the Seattle Chapter of
the Achievement Rewards for College Scientists Foundation.}


\appendix

\section{Details on implementation of Markov Chain Monte Carlo via emcee}
\label{app:emcee}

MCMC sampling methods are particularly suited to efficiently sample pdfs with large parameter
spaces~\cite{Foreman_Mackey:2013aa}, and have been applied widely in many physical applications,
such as astrophysics~\cite{Trotta:2008qt,Hou:2011jz}. 
 We apply a particular MCMC sampling algorithm called \emcee~\cite{Foreman_Mackey:2013aa}, 
which is a Python implementation of Goodman and Weare's affine-invariant ensemble sampler
for MCMC. The goal of the algorithm is to improve over traditional
Metropolis-Hastings (M-H) methods~\cite{Metropolis53} and modifications that require significant burn-in
phases with the price of many computationally expensive evaluations of the pdf. 
Reference~\cite{Gelman03} contains summaries of several methods for posterior 
simulation, including the aforementioned M-H methods and Gibbs sampling.

The \emcee\ code uses an ensemble of several walkers to explore the parameter space.
Each starts in a user-specified part of the parameter space and is allowed to diffuse,
exploring different features of
the distribution in parallel. The number of walkers needed to sufficiently explore the parameter
space generally depends on the problem of interest, but Ref.~\cite{Foreman_Mackey:2013aa}
suggests using hundreds of walkers. For the simple model problems in this work
we find it
is sufficient to use a number of walkers equal to twice the number of parameters, 
and to be safe we use four times the number of parameters.

  The dangers that occur during optimization can also 
occur with MCMC walkers--- they can find unlikely parts of the parameter space and
become ``stuck''.
Assessment of the convergence of the walkers is crucial to diagnose
such problems. Sometimes walkers that are not converging can be identified graphically,
but several useful quantitative methods exist to diagnose the convergence. 
Foreman-Mackey et al.\ 
advocate using the autocorrelation time, which quantifies the number of evaluations
needed to obtain independent samples of the distribution~\cite{Foreman_Mackey:2013aa}. 
When the autocorrelation time
is large, the number of evaluations needed by the MCMC algorithm to achieve convergence
will be large. 

They also outline some proxies for assessing 
convergence~\cite{Foreman_Mackey:2013aa}.
One proxy is the fraction of proposed steps that were accepted in the sampling, called the acceptance
fraction $a_f$, which can be used to diagnose walkers that did not converge during the sampling. We
use this simple diagnostic to ensure that walkers converged in our MCMC sampling
throughout this work, using the rejection criteria $a_f < 0.2$ for each separate walker.
Low acceptance fractions are a sign that the sampler is stuck in an area of low
probability, which often occurs in multimodal distributions. Some methods to deal with
multimodality include clustering algorithms~\cite{Hou:2011jz} and parallel-tempering
ensemble MCMC, which is implemented in \emcee.

\newpage
\bibliography{./bayesian_refs}

\end{document}